\documentclass[pre,aps,twocolumn,superscriptaddress]{revtex4-2}
\usepackage{filecontents}

\usepackage{graphicx,color,setspace}
\usepackage[utf8]{inputenc}
\usepackage{mathtools}
\usepackage[normalem]{ulem}
\usepackage{tabularx}
\usepackage{multirow}
\usepackage{float}
\usepackage{amsmath}
\usepackage{xr}
\makeatletter
\newcommand*{\addFileDependency}[1]{
\typeout{(#1)}
\@addtofilelist{#1}
\IfFileExists{#1}{}{\typeout{No file #1.}}
}\makeatother

\newcommand*{\myexternaldocument}[1]{%
\externaldocument{#1}%
\addFileDependency{#1.tex}%
\addFileDependency{#1.aux}%
}

\myexternaldocument{final_supplementary}
\usepackage{siunitx}

\newcommand{\cmmnt}[1]{\ignorespaces}

\begin{document}

\title{Inverse design of a pyrochlore lattice of DNA origami\\through model-driven experiments}

\author{Hao Liu}
\affiliation{School of Molecular Sciences and Center for Molecular Design and Biomimetics, The Biodesign Institute, Arizona State University, 1001 South McAllister Avenue, Tempe, Arizona 85281, USA}
\author{Michael Matthies}
\affiliation{School of Molecular Sciences and Center for Molecular Design and Biomimetics, The Biodesign Institute, Arizona State University, 1001 South McAllister Avenue, Tempe, Arizona 85281, USA}
\author{John Russo}
\affiliation{Dipartimento di Fisica, Sapienza Universit\`{a} di Roma, P.le Aldo Moro 5, 00185 Rome, Italy}
\author{Lorenzo Rovigatti}
\affiliation{Dipartimento di Fisica, Sapienza Universit\`{a} di Roma, P.le Aldo Moro 5, 00185 Rome, Italy}

 \author{Raghu Pradeep Narayanan}
 \affiliation{School of Molecular Sciences and Center for Molecular Design and Biomimetics, The Biodesign Institute, Arizona State University, 1001 South McAllister Avenue, Tempe, Arizona 85281, USA}
 
 \author{Thong Diep}
 \affiliation{School of Molecular Sciences and Center for Molecular Design and Biomimetics, The Biodesign Institute, Arizona State University, 1001 South McAllister Avenue, Tempe, Arizona 85281, USA}
 
\author{Daniel McKeen}
\affiliation{Department of Chemical Engineering, Columbia
University, 817 SW Mudd, New York, NY 10027,
USA}

\author{Oleg Gang}
\affiliation{Department of Chemical Engineering, Columbia
University, 817 SW Mudd, New York, NY 10027,
USA}
\affiliation{Department of Applied Physics and Applied Mathematics, Columbia University, New York, NY 10027, USA}
\affiliation{Center for Functional Nanomaterials, Brookhaven National Laboratory, Upton, NY, 11973, USA}

\author{Nicholas Stephanopoulos}
\affiliation{School of Molecular Sciences and Center for Molecular Design and Biomimetics, The Biodesign Institute, Arizona State University, 1001 South McAllister Avenue, Tempe, Arizona 85281, USA}
\author{Francesco Sciortino}
\affiliation{Dipartimento di Fisica, Sapienza Universit\`{a} di Roma, P.le Aldo Moro 5, 00185 Rome, Italy}
\author{Hao Yan}
\affiliation{School of Molecular Sciences and Center for Molecular Design and Biomimetics, The Biodesign Institute, Arizona State University, 1001 South McAllister Avenue, Tempe, Arizona 85281, USA}
\author{Flavio Romano}
\affiliation{Department of Molecular Sciences and Nanosystems, Ca' Foscari University of Venice, Via Torino 155, 30171 Venezia-Mestre, Italy}
\affiliation{European Centre for Living Technology (ECLT), Ca’ Bottacin, 3911 Dorsoduro Calle Crosera, 30123 Venice, Italy}

\author{Petr \v{S}ulc}
\affiliation{School of Molecular Sciences and Center for Molecular Design and Biomimetics, The Biodesign Institute, Arizona State University, 1001 South McAllister Avenue, Tempe, Arizona 85281, USA}
\affiliation{School of Natural Sciences, Department of Bioscience, Technical University Munich, 85748 Garching, Germany}

\begin{abstract}

Sophisticated statistical mechanics approaches and human intuition have
demonstrated the possibility to self-assemble complex lattices or finite size constructs, but have mostly only been successful {\it in silico}. The proposed strategies quite often fail in experiment due to unpredicted traps associated to kinetic slowing down (gelation, glass transition), as well as to competing ordered structures. An additional challenge that theoretical predictions face is the difficulty to encode the desired inter-particle interaction potential with the currently available library of nano- and micron-sized particles. To overcome these issues, we conjugate here SAT-assembly --- a patchy-particle interaction design algorithm based on constrained optimization solvers --- with coarse-grained simulations of DNA nanotechnology to experimentally realize trap-free self-assembly pathways. As a proof of concept we investigate the assembly of the pyrochlore (also known as tetrastack) lattice, a highly coveted 3D crystal lattice due to its promise in construction of optical metamaterials. We confirm the successful assembly with two different patchy DNA origami designs via SAXS as well as SEM visualization of the silica-coated lattice.
Our approach offers a versatile modeling pipeline that starts from patchy particles designed {\it in silico} and ends with wireframe DNA origami that self-assemble into the desired structure.
\end{abstract}

\maketitle

\section{Introduction}

\begin{figure*}[t]
\centering
	\includegraphics[width=0.8\textwidth]{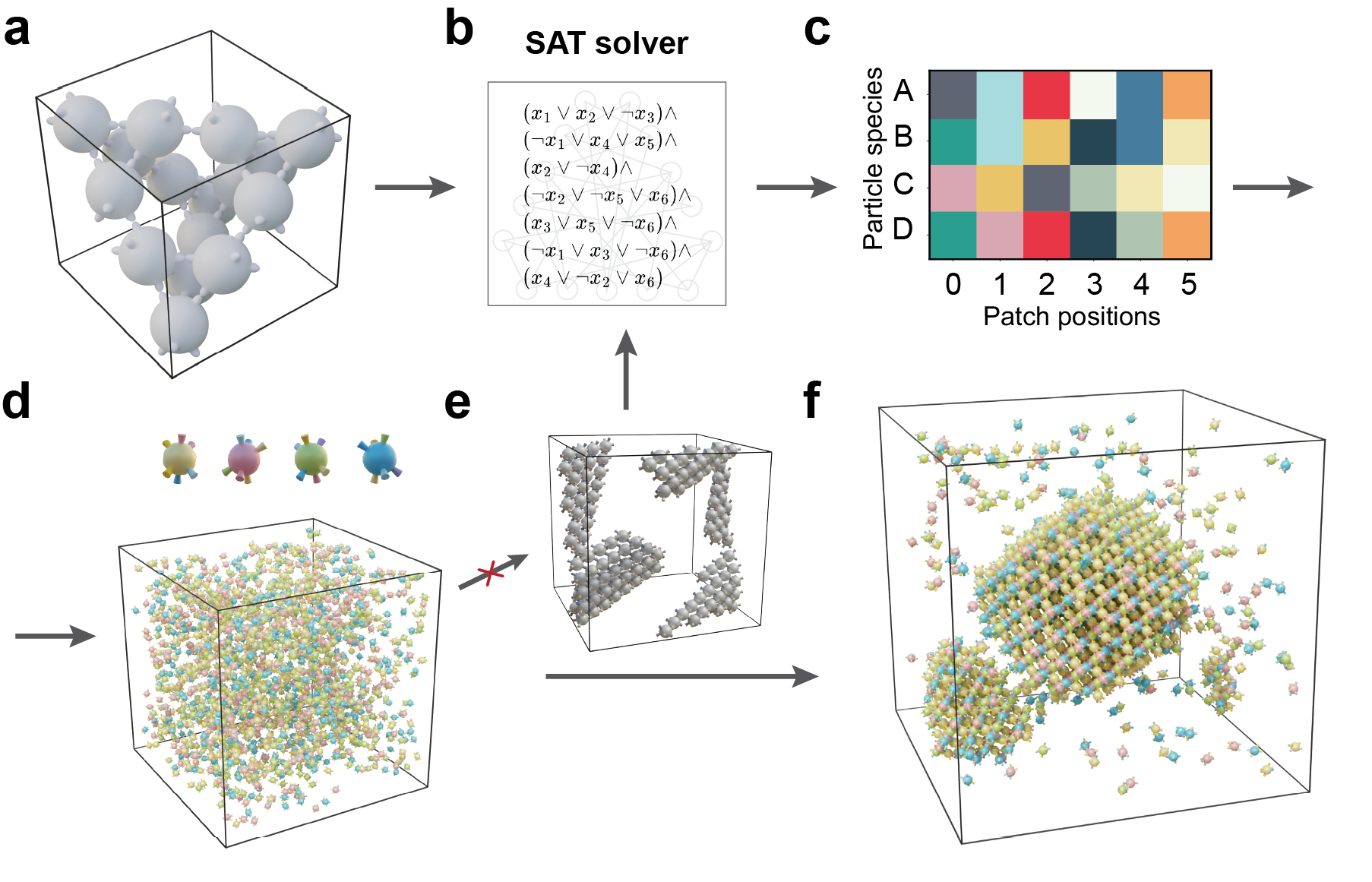}
	\caption{Workflow of the computational design with SAT-assembly: \textbf{a)} The topology of the unit cell of a pyrochlore lattice, where each particle has six neighbors. \textbf{b)} The design problem to find a fixed number of species of patchy particles that satisfy the unit cell lattice is translated into a set of Boolean clauses (shown schematically here and listed in Supp.~Mat.~Sec.~\ref{supp-sec:computational} ). \textbf{c)} From the defined clauses the SAT solver generates an interaction matrix corresponding to the assignment of colors to patches on respective species of patchy particles that can be arranged to satisfy the unit lattice interactions. \textbf{d)} We simulate the bulk assembly of patchy particles in a range of temperatures and screen for undesired assemblies that prevent the formation of the desired lattice. \textbf{e)} The identified undesired states are included as negative design in the SAT solver pipeline, explicitly banning solutions that can form these competing states. \textbf{f)} The process is iterated until we find a patchy assignment that only results in successful nucleation and assembly of the desired lattice.}
  \label{fig:overview}
 \centering
\end{figure*}

The experimental realization of nano- and mesoscopic structures with precise geometry is one of the central goals of nanotechnology. To this end, one of the most promising bottom-up approaches is self-assembly, where the building blocks are specifically designed to spontaneously aggregate into the target structure. Despite its potential impact, progress in this direction is hindered by the lack of a general framework that can discover designs that can self-assemble with high yield without encountering kinetic traps or unwanted byproducts. Some of the proposed solutions to this problem include statistical mechanical approaches \cite{halverson2013dna,miskin2016turning, kumar2019inverse,dijkstra2021predictive, rechtsman2005optimized,marcotte2011optimized,jain2014dimensionality, jacobs2016self, bupathy2021temperature, halverson2013dna,patra2017layer,mushnoori2022controlling}, machine learning-inspired protocols \cite{whitelam2020learning}, and optimization methods \cite{romano2020designing}. So far, these approaches have been mostly {\it in-silico}, with experiments relying more heavily on ingenious intuition \cite{tian2020ordered,zhang20183d,he2020colloidal,michelson2022three,macfarlane2011nanoparticle}, or painstakingly difficult trial-and-error attempts.

Several potential pitfalls typically encountered in self-assembly can considerably lower the yield of the desired structure: i) metastable states that can compete with the final product; ii) dynamically arrested states (kinetic traps); iii) low aggregation rates; iv) lack of knowledge of the underlying phase behavior of the building-blocks, especially for mixtures with many components. Additionally, from an experimental point-of-view, one has to consider the difficulty of realizing the building blocks, their size and interaction polydispersity, as well as their mechanical and molecular properties, such as softness and flexibility, that are difficult to take into account in theoretical modeling.

Here we use the synergy between theory, simulations, and experiments to address the aforementioned problems. We introduce a new modeling-driven design pipeline based on patchy particles as building blocks.
Patchy particles are a model for units that interact with addressable directional bonds~\cite{russo2021physics}, which have recently been experimentally realised at the colloidal scale~\cite{duguet2011design,he2020colloidal,swinkels2021revealing,chakraborty2022self,khalaf2023transfer}. In particular, we select as patchy particles DNA origami \cite{Rothemund2006} where the scaffold is complemented by single-stranded overhangs (sticky sequences) located in controllable positions. The possibility of controlling the interaction between different patches in a unique way (by designing their DNA sequence) serves as a key strength of DNA nanotechnology as a nanofabrication technique. We show that our approach that couples optimization methods, multiscale simulations, and DNA-origami experiments, can actively guide the assembly into the target structure.

As a test case, we focus on the pyrochlore (tetrastack) lattice,
which so far has not been successfully experimentally realized by self-assembly.
The pyrochlore lattice is of particular interest since it has been identified as having an omnidirectional photonic band gap, which is both wide and robust with respect to defects in the lattice \cite{ngo2006tetrastack,ducrot2018pyrochlore}. The self-assembly of this lattice is particularly challenging due to the presence of several competing structures with similar bonding topologies but different stacking orders which have so far prevented experimental realization \cite{romano2012patterning}.
To further demonstrate the versatility of our method we employ two different DNA origami wireframe nanostructure designs (icosahedral and octahedral shapes respectively) and show that they both robustly self-assemble into the pyrochlore superlattice, as confirmed by SAXS and SEM experiments of the silica-coated lattice.

\begin{figure}
\centering
	\includegraphics[width=0.5\textwidth]{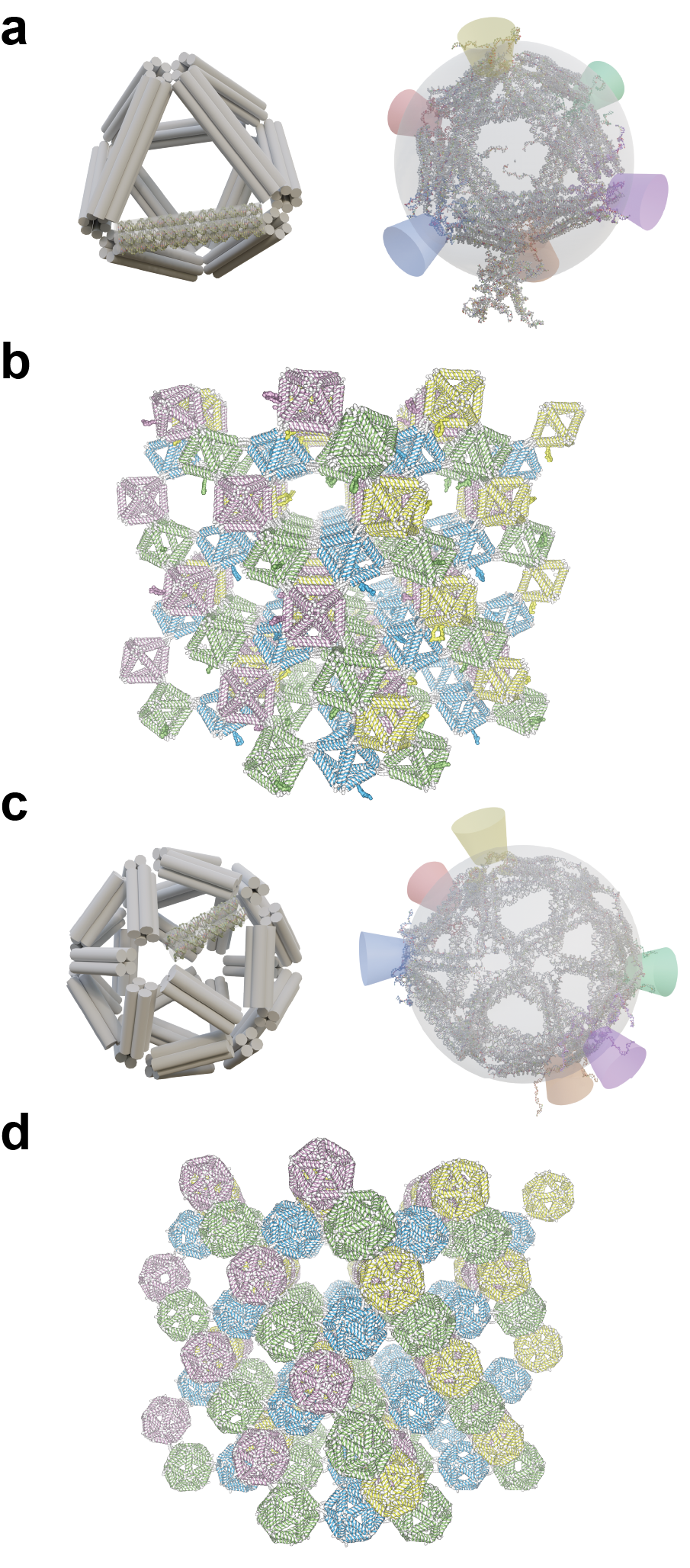}
	\caption{Transferring the patchy particle design to the sequence design of DNA nanostructures with oxDNA simulations of the assembled lattice. \textbf{a} Octahedral and \textbf{c} icosahedral DNA origamis are selected for the experimental implementation of patchy particles, with patches realized as single-stranded overhangs. Both structures have multi-helical bundle edges to ensure their structural rigidity. We ran large-scale oxDNA simulations of large clusters (consisting of 128 DNA origami) with pyrochlore lattice geometry to ensure that the lattices are mechanically stable for chosen single-stranded overhang placements and lengths. The mean structures produced from the oxDNA simulations are shown in \textbf{b} and \textbf{d} respectively.}
    \label{fig:monodlsoxdna}
\centering
\end{figure}

\section{Inverse design in silico}
We developed a multiscale approach to design DNA nanoparticles and test {\it in silico} their assembly into the pyrochlore lattice (Fig.~\ref{fig:overview}). First we use the SAT-assembly method for positive (to target the desired lattice) and negative (excluding competing structures) designs. The second step is to numerically investigate the assembly process based on a coarse-grained patchy particle model, where each nanoparticle is represented as a sphere with colored interaction sites, where only compatible colors can form a bond. The colors represent single-stranded DNA sequences, where compatible colors correspond to complementary sequences. If the simulations reveal the presence of unwanted competing structures, these are translated into additional SAT clauses that are fed back to the optimization solver to look for solutions that exclude them. 
This feedback loop can be iterated until we obtain a solution for which the simulations produce the target structure with a high yield.
The last step uses nucleotide-level modeling that translates the patchy particle design into a DNA origami amenable to experimental realization.

\subsection{SAT-assembly}
The goal of the SAT-assembly procedure is to assign an interaction matrix
that sets whether there is an attraction 
between any pair of building blocks so that the target lattice assembles without any kinetic traps or alternative free-energy minima that would lead to misassembled structures or defects in the lattice.
The number of possible ways to design interaction matrices between patchy particles explodes combinatorially with an increasing number of possible colors and particle species, which in turn makes the search of the design space very challenging. To find an interaction matrix that can avoid these trapped states, we employed the SAT-assembly framework~\cite{romano2020designing,russo2022sat} that maps the inverse design problem into a Boolean Satisfiability problem (SAT). SAT is a well studied NP-complete problem for which highly efficient solvers are available~\cite{een2005minisat}, allowing us to efficiently find solutions to the design problem; these solutions are represented as a set of binary variables that specify which color is assigned to which particle and which colors can interact. It further specifies restrictions that the interactions need to satisfy in terms of binary logic clauses composed of AND and OR and negation operations on the binary variables. These restrictions include that each color can only have one complementary color, each patch can only be assigned one color, and that the patchy particles can be arranged into the unit lattice so that all the patches on each particle are bound to a patch of complementary color (see Supp.~Mat.~Sec.~\ref{supp-sec:computational} for details). Thus, as a positive design task, we can specify in terms of binary variables and logic clauses that the target lattice is an energy minimum of the patchy particle system, and let the SAT solver find color interactions and a patch color assignment that satisfy this condition.
The first step of the SAT-assembly method is to specify the target unit cell of the lattice (Fig.~\ref{fig:overview}). In the case of the pyrochlore lattice, the unit cell is composed of 16 individual particles, where each particle has six neighbors.

\subsection{Patchy particle simulations}

 We next verify that the solutions found by SAT solver can homogeneously nucleate a pyrochlore crystal via molecular dynamics simulations (Fig.~\ref{fig:overview}). The goal of our pipeline is to realize the building blocks with DNA origami nanostructures. For simulations of DNA nanotechnology, we typically use a nucleotide-level coarse-grained model, oxDNA \cite{poppleton2023oxdna,snodin2015introducing,sulc2012sequence,ouldridge2011structural}, which has been shown to reproduce the structural, mechanical and thermodynamic properties of single- and double-stranded DNA. However, the oxDNA model is still too slow to simulate the kinetics of assembly of individual origami into the lattice. Hence, to study assembly kinetics, we used a coarser model where the individual DNA nanostructures are represented as patchy spheres (Fig.~\ref{fig:overview}d). Each patch is assigned a color, as given by the solution from SAT solver. If two colors are compatible, they are considered to correspond to complementary single-stranded DNA overhangs and, in the patchy representation, to a short-range attractive potential. The spheres interact with excluded volume interactions to prevent two particles from overlapping with each other. We have used the oxDNA model to parameterize the patchy particle model (see Supp.~Mat.~Sec.~\ref{supp-sec:computational}).

We run multiple patchy particle simulations in a range of temperatures to probe the assembly kinetics for each possible solution (see Supp.~Mat.~Sec.~\ref{supp-sec:computational} for details).
We first try a solution that uses only one species of particles, but our simulations show that it always leads to misassembled states. At high temperatures, the assembly will remain in the gas phase, whereas at low temperature the system will form a quenched glassy state; at intermediate temperatures, where one would hope to observe nucleation and assembly of the lattice, the system forms misassembled states (shown in Fig.~\ref{fig:overview}e and Supp.~Fig.~\ref{supp-fig:sup_miss}), which are stabilized by two bonds formed between two complementary pairs of patches on two patchy particles.

As a first iteration of the feedback loop, we introduce a new negative design requirement: no pair of particles can bind to each other by more than one bond, so that we explicitly prevent the system from forming the misassembled state that we previously identified in the molecular dynamics simulations.
For our pyrochlore lattice design, the SAT solver proved that no solution exists that satisfies the conditions above if we allow only one patchy particle species, forcing us to go to multicomponent systems.

In multicomponent systems it is possible to add an additional requirement that a particle cannot form any bond with a particle of the same species. In the context of DNA origami interacting via single-stranded overhangs, as we will discuss later, this will be important in order to prevent possible aggregations or blocking by unpaired staple strands when each DNA nanostructure is prepared individually.
SAT shows that this requirement suppresses all solutions with only two particle species. In fact, the SAT solver identifies that the smallest number of distinct particle species required is four (Supp.~Mat.~Sec.~\ref{supp-sec:computational}).

Our simulations show that kinetic traps are formed when a low number of colors is used. To decrease the chance of kinetic traps we then set the solver to find solutions with the maximum number of colors, 24 (that is 12 pairs of complementary colors). The resulting interaction matrix between patches is shown in Fig.~\ref{fig:overview}c. As shown in Fig.~\ref{fig:overview}f, simulations confirm the successful assembly of the 24 color solution into the desired pyrochlore lattice.

\subsection{Realization of patchy particles with DNA nanotechnology}

\begin{figure}
\centering
	\includegraphics[width=0.5\textwidth]{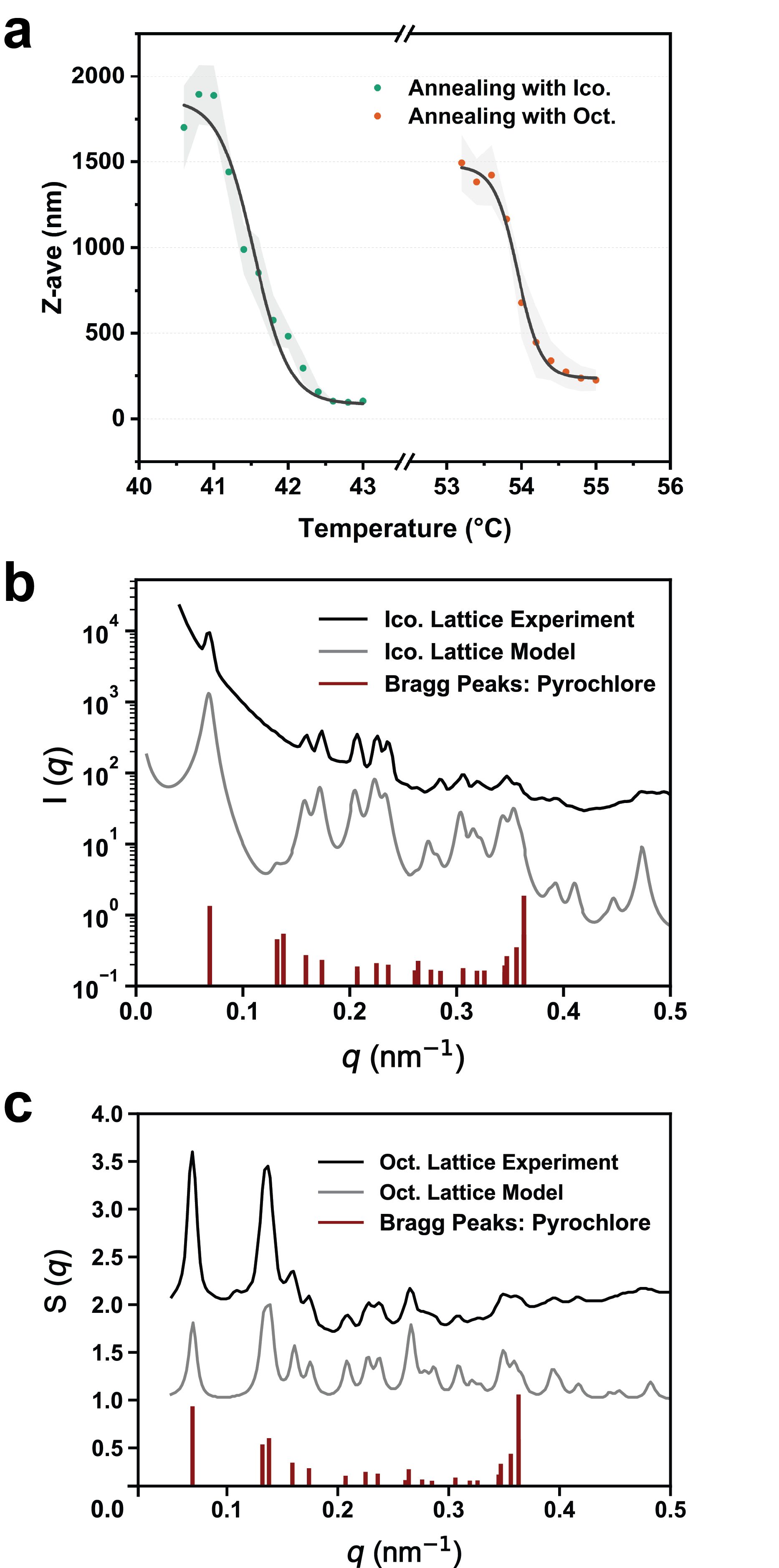}
	\caption{ \textbf{a)} Dynamic light scattering as a function of temperature during the lattice assemblies (from icosahedral and octahedral DNA origami respectively), used to identify the approximate assembly temperature. Experimental characterizations of the pyrochlore lattices with SAXS with \textbf{b)} icosahedral and \textbf{c)} octahedral building blocks. In black lines, we show the scattered intensity for lattice of icosahedral DNA origami in \textbf{b)} and structural factors are shown in \textbf{c)} for lattice of octahedral origami with encaged gold nanoparticles. The SAXS measurements are compared to the predicted scattering model for DNA origami arranged in the pyrochlore lattice (grey lines) and the Bragg peak positions (red lines).}
    \label{fig:saxs}
 \centering
\end{figure}

We next design nucleic acid nanostructures that realize the patchy particles and their interactions found by the SAT assembly and verified in patchy particle simulation (Fig.~\ref{fig:monodlsoxdna}).
We use the oxDNA model and interactive modeling tool oxView~\cite{bohlin2022design} to design DNA nanostructures to represent the patchy particles with wireframe DNA origami, where patches correspond to single-stranded overhangs with spacers (Fig.~\ref{fig:monodlsoxdna} and Supp.~Fig.~\ref{supp-fig:supmonomer}). We considered two DNA wireframe origami designs: an icosahedral shape based on origami used in Ref.~\cite{zhang2022spatially}, and a second one based on an octahedral origami from Ref.~\cite{tian2015prescribed} (Fig.~\ref{fig:monodlsoxdna}a,c). In the icosahedral origami, each ``patch'' corresponds to three single-stranded overhangs (``handles'') placed in the vertex of the DNA origami, making the overall geometry fully compatible with the corresponding patchy particle model. Each overhang has a 15-nucleotide poly-T spacer followed by the 8-nucleotide binding region at the 3' flanking end. The binding region sequences are the same on each of the three sequences in the patch, so that there is no imposed orientational control over binding of the two patches. We have optimized the assigned binding regions so that for each complementary pair, the binding free energy of complementary sequences is as close as possible for all twelve binding pairs, while making the binding between overhangs that are not supposed to interact as unfavorable as possible (see Supp.~Mat.~Sec.~\ref{supp-sec:sequences}). 

In order to demonstrate the robustness of the overhang-driven assembly and inverse-design strategy, we also employ a octahedral origami design where the vertex positions do not perfectly correspond to the patch positions. This in turn requires that each handle is sufficiently long to adapt to the imposed geometry, which is not compatible with octahedra touching their vertices. We hence design the handles with longer poly-T spacers (22 nucleotides) at each vertex to ensure that the DNA origami can arrange into the pyrochlore lattice, and use a 9-nucleotide long binding region. We verified in patchy particle simulations with patches placed in the octahedron vertex positions that the design with four particles species and 24 colors is still capable of assembling into a pyrochlore lattice (see Supp.~Sec.~\ref{supp-sec:computational}).

\begin{figure*}
\centering
	\includegraphics[width=1\textwidth]{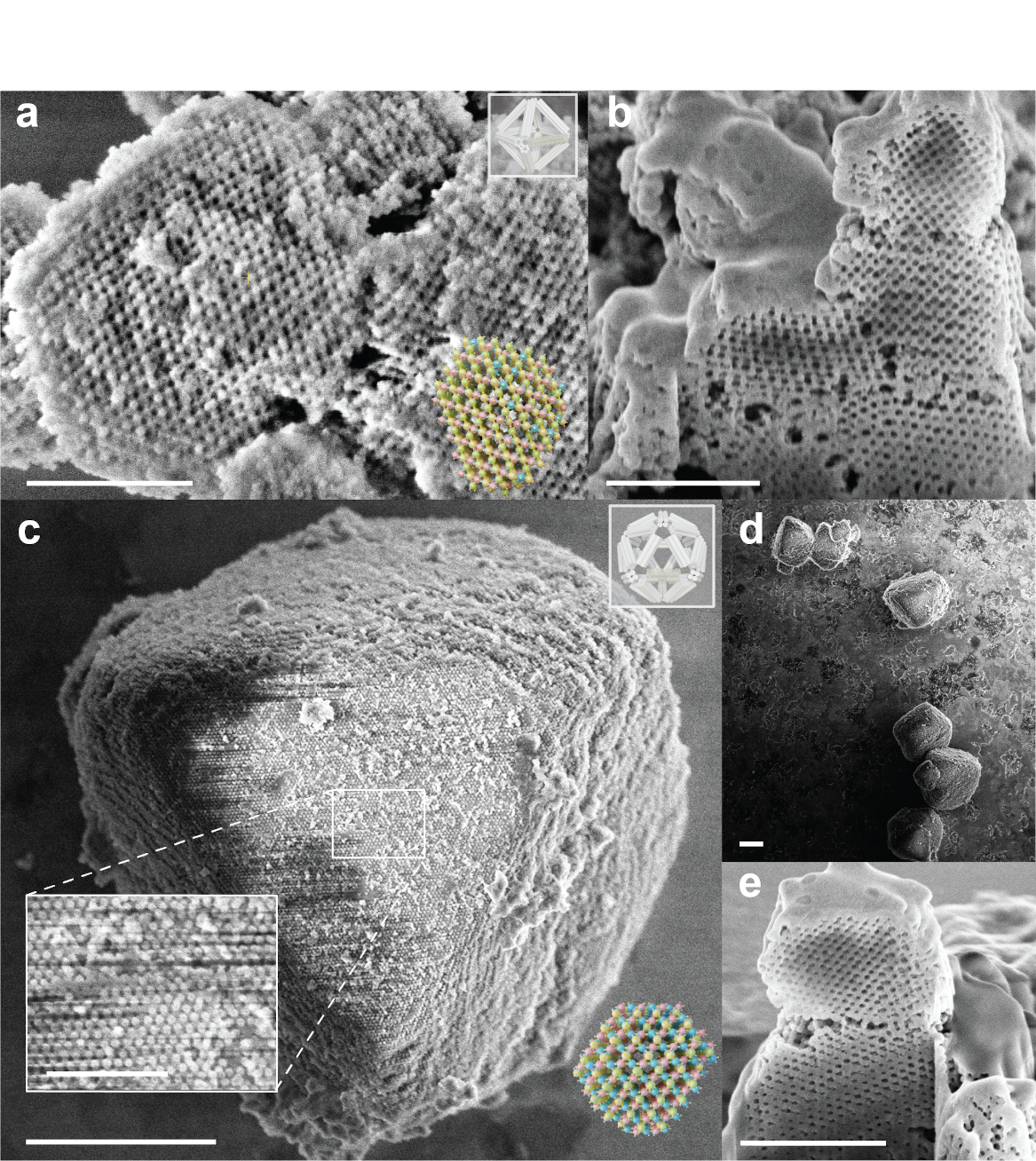}
	\caption{Experimental characterization of the fabricated pyrochlore lattice embedded with silica. Representative SEM image of the assembled pyrochlore lattice with \textbf{a} octahedral and \textbf{c} icosahedral DNA origami building blocks and \textbf{b,d} the associated cross-section of the lattice created by focused ion beam from picked smaller lattice grains. In \textbf{d}, a typical zoomed-out view of the lattice assembled from icosahedral origami is shown. Insets for \textbf{a, c} are model for monomer and assembled lattice fitted to the projected view. In \textbf{a, b}, \textbf{e}, scale bars are \SI{1}{\micro\meter} and for \textbf{c, d}, scale bars are \SI{5}{\micro\meter}.}
    \label{fig:sem1}
 \centering
\end{figure*}

Both icosahedral and octahedral designs were tested in a large scale simulation with the oxDNA model to investigate the mechanical stability and design the position and lengths of handle sequences accordingly. We therefore assembled a pyrochlore lattice cluster measuring $2\times2\times2$ unit cells in each dimension (total: 128 DNA origami, corresponding to over two million nucleotides in the simulation). Molecular dynamics simulations were conducted for both designs (icosahedron and octahedron units) at 293 K, and used to calculate the mean structure as shown in Fig.\ref{fig:monodlsoxdna}b,d. 
We assessed the movement trajectory of the center of mass (COM) for each DNA origami incorporated into the lattice and superimposed it onto the mean structure. A qualitative comparison of the relative positions of the COM trajectory and the mean structure indicates that the lattice assembled using the proposed origami design satisfies the pyrochlore geometry and is mechanically stable. We also verified that the extra single-stranded scaffold loop in the octahedral origami design (Fig.~\ref{fig:monodlsoxdna}a,b and Supp.~Fig.~\ref{supp-fig:suptetramer}) does not interfere with the desired pyrochlore geometry.

\section{Experimental lattice assembly and characterizations}
For each DNA origami design, we prepared each species in a separate PCR tube by thermal annealing, after which, depending on the folding result of the origami, we use individually determined purification method for the lattice assembly. For icosahedral DNA origami, well-folded monomeric structures, isolated using rate-zonal purification \cite{lin2013purification}, are preferentially utilized to ensure optimal lattice assembly, minimizing interference from multimeric side products. While for the octahedral DNA origami, despite further purification to exclude the influence of undesirably folded structures might produce better superlattices, we determined that removing the excess free staple strands with ultrafiltration is sufficient for the lattice to emerge, given the superior yield of the target monomeric structure. The four different origami species were then mixed together and annealed over a temperature ramp (Supp.~Mat.~Sec.~\ref{supp-secc:expmethods}).

We monitored the size change of the system with a fast annealing protocol using Dynamic Light Scattering (DLS) (Fig.~\ref{fig:saxs}a). The measured spectrum allows us to approximately identify the temperature range at which the monomers start associating ($\mathrm{T_{1}}$) and the temperature where the size of the assemblies reaches a stable size ($\mathrm{T_{2}}$). In particular, for the octahedral origami system $\mathrm{T}_1$ is 54 \textdegree{}C and $\mathrm{T}_2$ is 51 \textdegree{}C, while for the lattice assembled from icosahedral origami, it is 43 and 40 \textdegree{}C respectively. We then use an annealing protocol with a slow ramping rate from $\mathrm{T_{1}}$ to $\mathrm{T_{2}}$, after incubation of the system at a slightly higher temperature to dissociate any bonds between monomers. The annealing process, during which the mixed origami nanoparticles nucleate and further crystallize, requires at least one week for the superlattice to emerge, and both elongating and fine-tuning the annealing protocol should give rise to superlattices with enlarged sizes and improved qualities. 

The assembly of octahedral and icosahedral origami systems happens at different temperature ranges, as a result of the different particle geometry, patch distribution, and binding strength (Fig.~\ref{fig:saxs}a). We observe in the experiments that a higher binding strength is required for the octahedral system to assemble into the designed lattice, presumably because the icosahedron has more optimal patch positions, as supported by coarse-grained models (Supp.~Mat.~Sec.~\ref{supp-sec:computational}).

For characterization, we coat the annealed sample with a thin layer of silica to preserve the structural details for scanning electron microscopy (SEM) characterization \cite{wang2021dna}. We used SEM to visualize the deposited silica-DNA hybrid structure and the representative results are shown in Fig.~\ref{fig:sem1} and Supp.~Mat.~Sec.~\ref{supp-suppfigs}. We optimized the assembly conditions that affect the lattice formation, besides the annealing time, including origami concentration and ionic strength (in this case, magnesium ion concentration), based on the feedback of SEM characterization. Limited by the high binding strength, the octahedral system annealed best with the concentration of origami and magnesium being $10$ nM and $12.5$ mM respectively, to ensure the structural integrity of the nanostructure itself during lattice assembly. While with the icosahedral origami we determined (using a slow temperature ramp around the melting point) that concentrations of $10$ nM origami and $25$ mM magnesium produced the best superlattice, with increased ionic strength allowing larger lattices to emerge. For both systems, clear periodicity corresponding to the pyrochlore lattices is observed. 
The octahedral DNA origami produces on average smaller lattice grains and more poly-crystalline assemblies, with size on average \SI{1}{\micro\meter}. With icosahedral building blocks we are able to achieve larger and faceted lattices over \SI{5}{\micro\meter}. For the icosahedral system, we additionally also tried a mix-and-anneal strategy, where we provide mechanical agitation to the solutions while annealing. Through facilitating the diffusion of DNA origami particles and rocking the sedimented lattice in the bottom of the tube, we partially overcome the limitation of precipitation during thermal annealing. We find that this approach produced even larger lattices, with the largest dimension exceeding \SI{20}{\micro\meter} (Fig.~\ref{fig:sem1}c,d).

We next investigated the internal structures of the selected lattice grains with focused ion beam (FIB) cross-sectional analysis. It appears that internally associated long-range order persists without obvious assembly defects, confirming the successful experimental realization of the pyrochlore lattice (Fig.~\ref{fig:sem1}). To confirm the lattice structure, we have further performed SAXS measurements (see Supp.~Mat.~Sec.~\ref{supp-secc:saxs}) of the assembled lattices: for the octahedral design, we have additionally attached a DNA-coated gold nanoparticle inside the origami to help the SAXS characterization. By fitting the SAXS measurements with a model of diffraction pattern for pyrochlore lattice (Supp.~Mat.~Sec.~\ref{supp-secc:saxs}), we obtained lattice parameters of $156.4$ nm for the octahedron origami lattice, and $159.1$ nm for the lattice made out of icosahedral origami. The comparison of measured structure factor shows agreement with the one expected for the pyrochlore lattice (Fig.~\ref{fig:saxs}).

\section{Conclusion}

We have developed a pipeline that uses multiscale modeling and optimization algorithms to design DNA nanostructures that self-assemble into the pyrochlore lattice. Our computational tools can be generalized to also design and guide the experimental realization of other types of lattices or finite-size multicomponent assemblies \cite{bohlin2022design}. The method could also be used to design initial seeding substructures for improving the yields and resulting sizes of the seeded nucleation and growth of the lattice, as well as design other sought-after lattice geometries such as clathrates \cite{lee2023entropy}. This first successful realization of the pyrochlore lattice geometry opens a pathway towards optical metamaterials.

The reason for the generality of the procedure is that the design pipeline does not rely on the specific shape of the building blocks: the geometrical information of the local environment is only taken into account via a matrix of contacts. The nucleotide-level coarse-grained model is then used to verify that a particular DNA nanostructure realization is compatible with the designed patchy particle self-assembled system. One therefore has significant freedom in choosing the internal features, as well as the size, of the building blocks. Our approach allowed us to build upon the vast literature of successful DNA nanostructure designs, while being able to choose the size provides a handle on the lattice parameters which in turn controls the bandgap and other properties of the final lattice. The complexity of the unit cell is also not a limiting factor for the design process, since our approach is able to deal with many different building block species. Finally, the pipeline presented is not limited to long-range structures, but can be just as easily exploited to realise finite-size assemblies. The design and simulation tools are provided as free open-source software.

\section*{Acknowledgments}
We acknowledge support from the ONR Grant N000142012094 and ONR DURIP Grant N000142112876. We acknowledge the use of the Extreme
Science and Engineering Discovery Environment (XSEDE), which is supported by National Science Foundation grant number TG-BIO210009. We acknowledge the use of facilities within the Eyring Materials Center at Arizona State University supported in part by NNCI-ECCS-1542160. This material is based upon work supported by the National Science Foundation under Grant No.~2227650. Research reported in this publication was supported by The National Institute of General Medical Sciences of the National Institutes of Health under grant number DP2GM132931 to N.S. The content is solely the responsibility of the authors and does not necessarily represent the official views of the National Institutes of Health. J.R. and P.Š. further acknowledge support from the Universit`a Ca’ Foscari for a Visiting Scholarship.
J.R. acknowledges support from the European Research Council Grant DLV-759187. J.R., L.R., and F.S. acknowledge support from ICSC — Centro Nazionale di Ricerca in High Performance Computing, Big Data and Quantum Computing, funded by the European Union—NextGenerationEU. This research used the CMS beamline at the National Synchrotron Light Source II, which is a US DOE Office of Science Facilities at Brookhaven National Laboratory under contract no. DE-SC0012704. D.M and O.G were supported by the US Department of Energy, Office of Basic Energy Sciences, Grant DE-SC0008772.
This research used resources of the Advanced Light Source, which is a DOE Office of Science User Facility under contract no. DE-AC02-05CH11231. We thank Ye Tian, Yang Yang, Andreas Neophytou, Eileen Seo, Chad Simmons, Youli Li, Di Liu, Tim Liedl, and Gregor Posnjak for helpful discussions.


\bibliographystyle{old-mujstyl}
\bibliography{biblio,refmain}

\begin{thebibliography}{10}

\bibitem{halverson2013dna}
J.~D. Halverson and A.~V. Tkachenko.
\newblock DNA-programmed mesoscopic architecture.
\newblock {\em Physical Review E}, 87(6):062310, 2013.

\bibitem{miskin2016turning}
M.~Z. Miskin, G.~Khaira, J.~J. de~Pablo, and H.~M. Jaeger.
\newblock Turning statistical physics models into materials design engines.
\newblock {\em Proceedings of the National Academy of Sciences}, 113(1):34--39,
  2016.

\bibitem{kumar2019inverse}
R.~Kumar, G.~M. Coli, M.~Dijkstra, and S.~Sastry.
\newblock Inverse design of charged colloidal particle interactions for self
  assembly into specified crystal structures.
\newblock {\em The Journal of chemical physics}, 151(8):084109, 2019.

\bibitem{dijkstra2021predictive}
M.~Dijkstra and E.~Luijten.
\newblock From predictive modelling to machine learning and reverse engineering
  of colloidal self-assembly.
\newblock {\em Nature Materials}, 20(6):762--773, 2021.

\bibitem{rechtsman2005optimized}
M.~C. Rechtsman, F.~H. Stillinger, and S.~Torquato.
\newblock Optimized interactions for targeted self-assembly: application to a
  honeycomb lattice.
\newblock {\em Physical review letters}, 95(22):228301, 2005.

\bibitem{marcotte2011optimized}
E.~Marcotte, F.~H. Stillinger, and S.~Torquato.
\newblock Optimized monotonic convex pair potentials stabilize low-coordinated
  crystals.
\newblock {\em Soft Matter}, 7(6):2332--2335, 2011.

\bibitem{jain2014dimensionality}
A.~Jain, J.~R. Errington, and T.~M. Truskett.
\newblock Dimensionality and design of isotropic interactions that stabilize
  honeycomb, square, simple cubic, and diamond lattices.
\newblock {\em Physical Review X}, 4(3):031049, 2014.

\bibitem{jacobs2016self}
W.~M. Jacobs and D.~Frenkel.
\newblock Self-assembly of structures with addressable complexity.
\newblock {\em Journal of the American Chemical Society}, 138(8):2457--2467,
  2016.

\bibitem{bupathy2021temperature}
A.~Bupathy, D.~Frenkel, and S.~Sastry.
\newblock Temperature protocols to guide selective self-assembly of competing
  structures.
\newblock {\em Proceedings of the National Academy of Sciences},
  119(8):e2119315119, 2022.

\bibitem{patra2017layer}
N.~Patra and A.~V. Tkachenko.
\newblock Layer-by-layer assembly of patchy particles as a route to nontrivial
  structures.
\newblock {\em Physical Review E}, 96(2):022601, 2017.

\bibitem{mushnoori2022controlling}
S.~Mushnoori, J.~A. Logan, A.~V. Tkachenko, and M.~Dutt.
\newblock Controlling morphology in hybrid isotropic/patchy particle
  assemblies.
\newblock {\em The Journal of Chemical Physics}, 156(2):024501, 2022.

\bibitem{whitelam2020learning}
S.~Whitelam and I.~Tamblyn.
\newblock Learning to grow: Control of material self-assembly using
  evolutionary reinforcement learning.
\newblock {\em Physical Review E}, 101(5):052604, 2020.

\bibitem{romano2020designing}
F.~Romano, J.~Russo, L.~Kroc, and P.~{\v{S}}ulc.
\newblock Designing patchy interactions to self-assemble arbitrary structures.
\newblock {\em Physical Review Letters}, 125(11):118003, 2020.

\bibitem{tian2020ordered}
Y.~Tian, J.~R. Lhermitte, L.~Bai, T.~Vo, H.~L. Xin, H.~Li, R.~Li, M.~Fukuto,
  K.~G. Yager, J.~S. Kahn, et~al.
\newblock Ordered three-dimensional nanomaterials using DNA-prescribed and
  valence-controlled material voxels.
\newblock {\em Nature materials}, 19(7):789--796, 2020.

\bibitem{zhang20183d}
T.~Zhang, C.~Hartl, K.~Frank, A.~Heuer-Jungemann, S.~Fischer, P.~C. Nickels,
  B.~Nickel, and T.~Liedl.
\newblock 3D DNA origami crystals.
\newblock {\em Advanced Materials}, 30(28):1800273, 2018.

\bibitem{he2020colloidal}
M.~He, J.~P. Gales, {\'E}.~Ducrot, Z.~Gong, G.-R. Yi, S.~Sacanna, and D.~J.
  Pine.
\newblock Colloidal diamond.
\newblock {\em Nature}, 585(7826):524--529, 2020.

\bibitem{michelson2022three}
A.~Michelson, B.~Minevich, H.~Emamy, X.~Huang, Y.~S. Chu, H.~Yan, and O.~Gang.
\newblock Three-dimensional visualization of nanoparticle lattices and
  multimaterial frameworks.
\newblock {\em Science}, 376(6589):203--207, 2022.

\bibitem{macfarlane2011nanoparticle}
R.~J. Macfarlane, B.~Lee, M.~R. Jones, N.~Harris, G.~C. Schatz, and C.~A.
  Mirkin.
\newblock Nanoparticle superlattice engineering with DNA.
\newblock {\em science}, 334(6053):204--208, 2011.

\bibitem{russo2021physics}
J.~Russo, F.~Leoni, F.~Martelli, and F.~Sciortino.
\newblock The physics of Empty Liquids: from Patchy particles to Water.
\newblock {\em Reports on Progress in Physics}, 2021.

\bibitem{duguet2011design}
E.~Duguet, A.~D{\'e}sert, A.~Perro, and S.~Ravaine.
\newblock Design and elaboration of colloidal molecules: an overview.
\newblock {\em Chemical Society Reviews}, 40(2):941--960, 2011.

\bibitem{swinkels2021revealing}
P.~Swinkels, S.~Stuij, Z.~Gong, H.~Jonas, N.~Ruffino, B.~v.~d. Linden,
  P.~Bolhuis, S.~Sacanna, S.~Woutersen, and P.~Schall.
\newblock Revealing pseudorotation and ring-opening reactions in colloidal
  organic molecules.
\newblock {\em Nature communications}, 12(1):2810, 2021.

\bibitem{chakraborty2022self}
I.~Chakraborty, D.~J. Pearce, R.~W. Verweij, S.~C. Matysik, L.~Giomi, and D.~J.
  Kraft.
\newblock Self-assembly dynamics of reconfigurable colloidal molecules.
\newblock {\em ACS nano}, 16(2):2471--2480, 2022.

\bibitem{khalaf2023transfer}
R.~Khalaf, A.~Viamonte, E.~Ducrot, R.~M{\'e}rindol, and S.~Ravaine.
\newblock Transfer of multi-DNA patches by colloidal stamping.
\newblock {\em Nanoscale}, 2023.

\bibitem{Rothemund2006}
P.~W.~K. Rothemund.
\newblock {Folding DNA to create nanoscale shapes and patterns}.
\newblock {\em Nature}, 440(7082):297--302, 2006.

\bibitem{ngo2006tetrastack}
T.~Ngo, C.~Liddell, M.~Ghebrebrhan, and J.~Joannopoulos.
\newblock Tetrastack: Colloidal diamond-inspired structure with omnidirectional
  photonic band gap for low refractive index contrast.
\newblock {\em Applied physics letters}, 88(24):241920, 2006.

\bibitem{ducrot2018pyrochlore}
{\'E}.~Ducrot, J.~Gales, G.-R. Yi, and D.~J. Pine.
\newblock Pyrochlore lattice, self-assembly and photonic band gap
  optimizations.
\newblock {\em Optics Express}, 26(23):30052--30060, 2018.

\bibitem{romano2012patterning}
F.~Romano and F.~Sciortino.
\newblock Patterning symmetry in the rational design of colloidal crystals.
\newblock {\em Nature communications}, 3:975, 2012.

\bibitem{russo2022sat}
J.~Russo, F.~Romano, L.~Kroc, F.~Sciortino, L.~Rovigatti, and P.~{\v{S}}ulc.
\newblock SAT-assembly: A new approach for designing self-assembling systems.
\newblock {\em Journal of Physics: Condensed Matter}, 34(35):354002, 2022.

\bibitem{een2005minisat}
N.~Een.
\newblock MiniSat: A SAT solver with conflict-clause minimization.
\newblock In {\em Proc. SAT-05: 8th Int. Conf. on Theory and Applications of
  Satisfiability Testing}, pages 502--518, 2005.

\bibitem{poppleton2023oxdna}
E.~Poppleton, M.~Matthies, D.~Mandal, F.~Romano, P.~{\v{S}}ulc, and
  L.~Rovigatti.
\newblock oxDNA: coarse-grained simulations of nucleic acids made simple.
\newblock {\em Journal of Open Source Software}, 8(81):4693, 2023.

\bibitem{snodin2015introducing}
B.~E. Snodin, F.~Randisi, M.~Mosayebi, P.~{\v{S}}ulc, J.~S. Schreck, F.~Romano,
  T.~E. Ouldridge, R.~Tsukanov, E.~Nir, A.~A. Louis, et~al.
\newblock Introducing improved structural properties and salt dependence into a
  coarse-grained model of DNA.
\newblock {\em The Journal of chemical physics}, 142(23):06B613\_1, 2015.

\bibitem{sulc2012sequence}
P.~\v{S}ulc, F.~Romano, T.~E. Ouldridge, L.~Rovigatti, J.~P.~K. Doye, and A.~A.
  Louis.
\newblock Sequence-dependent thermodynamics of a coarse-grained {DNA} model.
\newblock {\em Journal of Chemical Physics}, 137(13):5101, 2012.

\bibitem{ouldridge2011structural}
T.~E. Ouldridge, A.~A. Louis, and J.~P. Doye.
\newblock Structural, mechanical, and thermodynamic properties of a
  coarse-grained {DNA} model.
\newblock {\em The Journal of chemical physics}, 134(8):02B627, 2011.

\bibitem{bohlin2022design}
J.~Bohlin, M.~Matthies, E.~Poppleton, J.~Procyk, A.~Mallya, H.~Yan, and
  P.~{\v{S}}ulc.
\newblock Design and simulation of DNA, RNA and hybrid protein--nucleic acid
  nanostructures with oxView.
\newblock {\em Nature protocols}, 17(8):1762--1788, 2022.

\bibitem{zhang2022spatially}
J.~Zhang, Y.~Xu, Y.~Huang, M.~Sun, S.~Liu, S.~Wan, H.~Chen, C.~Yang, Y.~Yang,
  and Y.~Song.
\newblock Spatially patterned neutralizing icosahedral DNA nanocage for
  efficient SARS-CoV-2 blocking.
\newblock {\em Journal of the American Chemical Society}, 144(29):13146--13153,
  2022.

\bibitem{tian2015prescribed}
Y.~Tian, T.~Wang, W.~Liu, H.~L. Xin, H.~Li, Y.~Ke, W.~M. Shih, and O.~Gang.
\newblock Prescribed nanoparticle cluster architectures and low-dimensional
  arrays built using octahedral DNA origami frames.
\newblock {\em Nature nanotechnology}, 10(7):637--644, 2015.

\bibitem{lin2013purification}
C.~Lin, S.~D. Perrault, M.~Kwak, F.~Graf, and W.~M. Shih.
\newblock Purification of DNA-origami nanostructures by rate-zonal
  centrifugation.
\newblock {\em Nucleic acids research}, 41(2):e40--e40, 2013.

\bibitem{wang2021dna}
Y.~Wang, L.~Dai, Z.~Ding, M.~Ji, J.~Liu, H.~Xing, X.~Liu, Y.~Ke, C.~Fan,
  P.~Wang, et~al.
\newblock DNA origami single crystals with Wulff shapes.
\newblock {\em Nature Communications}, 12(1):3011, 2021.

\bibitem{lee2023entropy}
S.~Lee, T.~Vo, and S.~C. Glotzer.
\newblock Entropy compartmentalization stabilizes open host--guest colloidal
  clathrates.
\newblock {\em Nature Chemistry}, pages 1--8, 2023.

\end{thebibliography}


\begin{thebibliography}{10}

\bibitem{romano2020designing}
F.~Romano, J.~Russo, L.~Kroc, and P.~{\v{S}}ulc.
\newblock Designing patchy interactions to self-assemble arbitrary structures.
\newblock {\em Physical Review Letters}, 125(11):118003, 2020.

\bibitem{russo2022sat}
J.~Russo, F.~Romano, L.~Kroc, F.~Sciortino, L.~Rovigatti, and P.~{\v{S}}ulc.
\newblock SAT-assembly: A new approach for designing self-assembling systems.
\newblock {\em Journal of Physics: Condensed Matter}, 34(35):354002, 2022.

\bibitem{een2005minisat}
N.~Een.
\newblock MiniSat: A SAT solver with conflict-clause minimization.
\newblock In {\em Proc. SAT-05: 8th Int. Conf. on Theory and Applications of
  Satisfiability Testing}, pages 502--518, 2005.

\bibitem{vsulc2012sequence}
P.~{\v{S}}ulc, F.~Romano, T.~E. Ouldridge, L.~Rovigatti, J.~P. Doye, and A.~A.
  Louis.
\newblock Sequence-dependent thermodynamics of a coarse-grained DNA model.
\newblock {\em The Journal of chemical physics}, 137(13):135101, 2012.

\bibitem{ouldridge2011structural}
T.~E. Ouldridge, A.~A. Louis, and J.~P. Doye.
\newblock Structural, mechanical, and thermodynamic properties of a
  coarse-grained {DNA} model.
\newblock {\em The Journal of chemical physics}, 134(8):02B627, 2011.

\bibitem{snodin2015introducing}
B.~E. Snodin, F.~Randisi, M.~Mosayebi, P.~{\v{S}}ulc, J.~S. Schreck, F.~Romano,
  T.~E. Ouldridge, R.~Tsukanov, E.~Nir, A.~A. Louis, et~al.
\newblock Introducing improved structural properties and salt dependence into a
  coarse-grained model of DNA.
\newblock {\em The Journal of chemical physics}, 142(23):06B613\_1, 2015.

\bibitem{russo2009reversible}
J.~Russo, P.~Tartaglia, and F.~Sciortino.
\newblock Reversible gels of patchy particles: role of the valence.
\newblock {\em The Journal of chemical physics}, 131(1):014504, 2009.

\bibitem{sulc2012sequence}
P.~\v{S}ulc, F.~Romano, T.~E. Ouldridge, L.~Rovigatti, J.~P.~K. Doye, and A.~A.
  Louis.
\newblock Sequence-dependent thermodynamics of a coarse-grained {DNA} model.
\newblock {\em Journal of Chemical Physics}, 137(13):5101, 2012.

\bibitem{rovigatti2015comparison}
L.~Rovigatti, P.~{\v{S}}ulc, I.~Z. Reguly, and F.~Romano.
\newblock A comparison between parallelization approaches in molecular dynamics
  simulations on GPUs.
\newblock {\em Journal of computational chemistry}, 36(1):1--8, 2015.

\bibitem{poppleton2023oxdna}
E.~Poppleton, M.~Matthies, D.~Mandal, F.~Romano, P.~{\v{S}}ulc, and
  L.~Rovigatti.
\newblock oxDNA: coarse-grained simulations of nucleic acids made simple.
\newblock {\em Journal of Open Source Software}, 8(81):4693, 2023.

\bibitem{poppleton2020design}
E.~Poppleton, J.~Bohlin, M.~Matthies, S.~Sharma, F.~Zhang, and P.~{\v{S}}ulc.
\newblock Design, optimization and analysis of large DNA and RNA nanostructures
  through interactive visualization, editing and molecular simulation.
\newblock {\em Nucleic acids research}, 48(12):e72--e72, 2020.

\bibitem{bohlin2022design}
J.~Bohlin, M.~Matthies, E.~Poppleton, J.~Procyk, A.~Mallya, H.~Yan, and
  P.~{\v{S}}ulc.
\newblock Design and simulation of DNA, RNA and hybrid protein--nucleic acid
  nanostructures with oxView.
\newblock {\em Nature protocols}, 17(8):1762--1788, 2022.

\bibitem{tian2015prescribed}
Y.~Tian, T.~Wang, W.~Liu, H.~L. Xin, H.~Li, Y.~Ke, W.~M. Shih, and O.~Gang.
\newblock Prescribed nanoparticle cluster architectures and low-dimensional
  arrays built using octahedral DNA origami frames.
\newblock {\em Nature nanotechnology}, 10(7):637--644, 2015.

\bibitem{zhang2022spatially}
J.~Zhang, Y.~Xu, Y.~Huang, M.~Sun, S.~Liu, S.~Wan, H.~Chen, C.~Yang, Y.~Yang,
  and Y.~Song.
\newblock Spatially patterned neutralizing icosahedral DNA nanocage for
  efficient SARS-CoV-2 blocking.
\newblock {\em Journal of the American Chemical Society}, 144(29):13146--13153,
  2022.

\bibitem{zadeh2011nupack}
J.~N. Zadeh, C.~D. Steenberg, J.~S. Bois, B.~R. Wolfe, M.~B. Pierce, A.~R.
  Khan, R.~M. Dirks, and N.~A. Pierce.
\newblock NUPACK: Analysis and design of nucleic acid systems.
\newblock {\em Journal of computational chemistry}, 32(1):170--173, 2011.

\bibitem{lin2013purification}
C.~Lin, S.~D. Perrault, M.~Kwak, F.~Graf, and W.~M. Shih.
\newblock Purification of DNA-origami nanostructures by rate-zonal
  centrifugation.
\newblock {\em Nucleic acids research}, 41(2):e40--e40, 2013.

\bibitem{wang2021dna}
Y.~Wang, L.~Dai, Z.~Ding, M.~Ji, J.~Liu, H.~Xing, X.~Liu, Y.~Ke, C.~Fan,
  P.~Wang, et~al.
\newblock DNA origami single crystals with Wulff shapes.
\newblock {\em Nature Communications}, 12(1):3011, 2021.

\bibitem{lewis2020single}
D.~J. Lewis, L.~Z. Zornberg, D.~J. Carter, and R.~J. Macfarlane.
\newblock Single-crystal Winterbottom constructions of nanoparticle
  superlattices.
\newblock {\em Nature materials}, 19(7):719--724, 2020.

\bibitem{tian2020ordered}
Y.~Tian, J.~R. Lhermitte, L.~Bai, T.~Vo, H.~L. Xin, H.~Li, R.~Li, M.~Fukuto,
  K.~G. Yager, J.~S. Kahn, et~al.
\newblock Ordered three-dimensional nanomaterials using DNA-prescribed and
  valence-controlled material voxels.
\newblock {\em Nature materials}, 19(7):789--796, 2020.

\bibitem{yager2014periodic}
K.~G. Yager, Y.~Zhang, F.~Lu, and O.~Gang.
\newblock Periodic lattices of arbitrary nano-objects: modeling and
  applications for self-assembled systems.
\newblock {\em Journal of Applied Crystallography}, 47(1):118--129, 2014.

\bibitem{wang2021designed}
S.-T. Wang, B.~Minevich, J.~Liu, H.~Zhang, D.~Nykypanchuk, J.~Byrnes, W.~Liu,
  L.~Bershadsky, Q.~Liu, T.~Wang, et~al.
\newblock Designed and biologically active protein lattices.
\newblock {\em Nature communications}, 12(1):3702, 2021.

\bibitem{santalucia2004thermodynamics}
J.~SantaLucia~Jr and D.~Hicks.
\newblock The thermodynamics of DNA structural motifs.
\newblock {\em Annu. Rev. Biophys. Biomol. Struct.}, 33:415--440, 2004.

\end{thebibliography}

\makeatletter\@input{xx.tex}\makeatother
   
\end{document}





\title{Supplementary Material}
\author{}


\setcounter{figure}{0}
 \makeatletter 
 \renewcommand{\thefigure}{S\@arabic\c@figure}
 \setcounter{equation}{0}
 \renewcommand{\theequation}{S\@arabic\c@equation}
 \setcounter{table}{0}
 \renewcommand{\thetable}{S\@arabic\c@table}
  \setcounter{section}{0}
 \renewcommand{\thesection}{S\@arabic\c@section}
\setcounter{secnumdepth}{3} 



\maketitle

 \setcounter{figure}{0}
 \makeatletter 
 \renewcommand{\thefigure}{S\@arabic\c@figure}
 \setcounter{equation}{0}
 \renewcommand{\theequation}{S\@arabic\c@equation}
 \setcounter{table}{0}
 \renewcommand{\thetable}{S\@arabic\c@table}
  \setcounter{section}{0}
 \renewcommand{\thesection}{S\@arabic\c@section}
  \renewcommand{\thesubsection}{S\@arabic\c@section.\@arabic\c@subsection}
   \renewcommand{\thesubsubsection}{S\@arabic\c@section.\@arabic\c@subsection.\@arabic\c@subsubsection}
\setcounter{secnumdepth}{4} 




\tableofcontents
\newpage
\section{\label{sec:computational}Computational design framework}

\subsection{SAT-assembly design of pyrochlore lattice}
We use the SAT-assembly framework to design nanostructures that self-assemble into the target lattice. We use the design framework introduced in Ref.~\cite{romano2020designing,russo2022sat} that formulates the inverse design problem as a Boolean Satisfiability problem (SAT) in terms of binary variables and logic clauses. SAT problems can be efficiently solved using available tools such as MiniSAT \cite{een2005minisat}. 

For completeness, we list here all the variables and clauses that were used to find the solution that forms pyrochlore lattice. The color interaction is given by binary variables $x^{\rm int}_{c_i,c_j}$ which are 1 if color $c_i$ is compatible with color $c_j$ and 0 otherwise. The patch coloring for each PP species is described by binary variables $x^{\rm pcol}_{s,p,c}$ which are 1 if patch $p$ of species $s$ has color $c$ and 0 otherwise. 
	The arrangement of the particle species in the lattice is described by $x^{L}_{l,s,o}$ which is 1 if the position $l$ is occupied by a PP of species $s$ in the specific orientation $o$. The variable $x^A_{l,k,c}$ is 1 if slot $k$ of lattice position $l$ is occupied by a patch with color $c$ and 0 otherwise. The clauses and variables are defined for all possible combinations of colors $c \in [1,N_c]$, patches $p \in [1,N_p]$, slots $k \in [1,N_p]$,  PP species $s \in [1,N_s]$, orientations $o \in [1,N_o]$, and lattice positions $l \in [1,L]$. For each particle orientation $o$, we assign a mapping $\phi_o$. The mapping $\phi_o(k) = p$ for a given orientation $o$ means that PP's patch $p$ overlaps with slot $k$ in a given lattice position, for example $\phi_1 = (1,2,3,4,5,6)\rightarrow(2,3,1,5,6,4)$. For a particle with 6 patches, there are $N_o = 12$ such orientations.  
 
As an input for the SAT solver, the problem has to be formulated in terms of clauses $C_j$, each of them containing variables $x_i$ connected by OR clauses. The final SAT problem corresponds to all respective clauses $C_j$ connected by AND clauses. To conform with this input format, the Boolean clauses introduced in Table 1 in the main text can be all reformulated as detailed below (we use symbols "$\land$", "$\lor$", and "$\neg$" to denote AND, OR, and NOT operations respectively):
%
\begin{enumerate}
\item  Each color $c_i$ can only bind to one other color $c_j$ (we exclude self-complementary colors):
\begin{equation}
 \label{eq_exactly_one_color}
\forall  c_i < c_j < c_k \in [1,N_c]: C^{\rm int}_{c_i,c_j,c_k} = \neg x^{\rm int}_{c_i,c_j} \lor \neg x^{\rm int}_{c_i,c_k} .
\end{equation}

\item Each patch $p$ of each PP species $s$ is assigned exactly one color:
\begin{equation} 
\label{eq_exactly_one_patch}
\forall s \in [1,N_s], p \in [1,N_p], c_l < c_k \in [1,N_c]: C^{\rm pcol}_{s,p,c_k,c_l} = \neg x^{pcol}_{s,p,c_k} \lor \neg x^{\rm pcol}_{s,p,c_l}.
\end{equation}

\item Each lattice position $l$ is only assigned exactly one PP species with exactly one assigned orientation:
\begin{equation} 
\label{eq_exactly_one_pos}
 \forall l \in [1,L], s_i < s_j \in [1,N_s], o_i < o_j \in [1,N_o]:  C^L_{l,s_i,o_i,s_j,o_j} = \neg x^L_{l,s_i,o_i} \lor \neg x^L_{l,s_j,o_j} .
\end{equation}

\item For all pairs of slots $k_i$ and $k_j$ that are in contact in neighboring lattice positions $l_i, l_j$ (as given by the unit cell topology listed in Table~\ref{tab:topo}), the patches that occupy them need to have complementary colors. 
\begin{equation*}
\forall c_i \leq c_j \in [1,N_c]: C^{\rm lint}_{l_i,k_i,l_j,k_j,c_i,c_j} = \left( x^A_{l_i,k_i,c_i} \land  x^A_{l_j,k_j,c_j} \right) \implies x^C_{c_i,c_j},
\end{equation*}
which can be equivalently rewritten as 
\begin{equation}
\label{eq_slots}
C^{\rm lint}_{l_i,k_i,l_j,k_j,c_i,c_j} = \neg x^A_{l_i,k_i,c_i} \lor \neg  x^A_{l_j,k_j,c_j} \lor x^C_{c_i,c_j},
\end{equation}

\item The slot of lattice position $l$ is colored with the same color as the patch of the PP species occupying it:
\begin{equation*}
\forall l \in [1,L], k \in [1,N_p], o \in [1,N_o], s \in [1,N_s], c \in [1,N_c]: C^{\rm LS}_{l,s,o,c,k} =   x^L_{l,s,o} \implies \left( x^A_{l, k, c} \iff x^{\rm pcol}_{s, \phi_o(k), c} \right) , 
\end{equation*}
 which can be equivalently rewritten as 
\begin{equation}
 C^{\rm LS}_{l,s,o,c,k} = \left( \neg  x^L_{l,s,o} \lor \neg x^A_{l, k, c} \lor x^{\rm pcol}_{s, \phi_r(p), c} \right) \land  \left(  \neg  x^L_{l,s,o} \lor  x^A_{l, k, c} \lor \neg x^{\rm pcol}_{s, \phi_o(k), c} \right) 
 \label{eq_latice_equiv}
\end{equation}

\item All $N_s$ PP species have to be used at least once in the assembled lattice: 
 \begin{equation}
 \forall s \in [1,N_s]: C^{\rm all\,spec.}_{s} =  \bigvee_{\forall l \in [1,L], o \in [1,N_o] } x^L_{l,s,o} ,
\end{equation}
where for each $s$, the conditions are connected by OR clause over all lattice positions and orientations.
\item Each color $c$ of $N_c$ total number of colors is assigned to at least one patch of one of the PP species: 
\begin{equation}
 \forall c \in [1,N_c]: C^{\rm all\,cols.}_{c} =  \bigvee_{\forall s \in [1,N_s], p \in [1,N_p] } x^{\rm pcol}_{s,p,c}
\end{equation}

\item Finally, we introduced an additional set of clauses that ensure that any pair of particles (of the same or different species) cannot bind by more than one bond at a time:
$\forall s_i, s_j \in [1,N_s], c^1_i,c^2_i,c^1_j,c^2_j \in [1,N_c]:$
\begin{equation}
  C^{\rm no\,two}_{s_i,s_j,p^i_1,p^i_2,p^j_1,p^j_2,c^1_i,c^2_i,c^1_j,c^2_j} = \neg \left(  x^{\rm pcol}_{s_i,p^i_1,c^1_i} \land x^{\rm pcol}_{s_i,p^i_2,c^2_i} \land x^{\rm pcol}_{s_j,p^j_1,c^1_j}  \land x^{\rm pcol}_{s_j,p^j_2,c^2_j} \land x^{\rm int}_{c^1_i,c^1_j}\land x^{\rm int}_{c^2_i,c^2_j} \right),
 \label{eq_no_more_than_two}
\end{equation}
where $p^i_1,p^i_2,p^j_1,p^j_2$ are all possible pairs of patches on PP of type $s_i$ and $s_j$ respectively for which there is a possible orientation so that they can both bind if they have compatible colors. 

\item We furthermore require that in the identified solution, no particle species has a patch that would be able to bind to any other patch on the same particle. This additional constraint is introduced due to constraints with DNA origami preparation, as each DNA origami with specific staple strands corresponding to patches is prepared separately, and we aim to avoid formation of dimers or unused staple strands binding to formed origami:
\begin{equation}
    \forall s \in [1,N_s],p_i,p_j \in [1,N_p,], c_k,c_l \in [1,N_c]: C^{\rm no\, self\, binding}_{s,p_i,p_j,c_k,c_l}: \neg x^{\rm pcol}_{s,p_i,c_k} \lor \neg x^{\rm pcol}_{s,p_j,c_l}  \lor \neg x^{\rm int}_{c_k,c_l}
    \label{eq_nosamespecies}
\end{equation}

\end{enumerate}

\begin{center}
\begin{table}
\centering
    \begin{tabular}{|l|llllll|}
\hline
PP species & \multicolumn{6}{|c|}{Patch Coloring}\\
\hline
\mbox{I: }  &(A,1) & (B,2) & (C,3) & (D,4)  & (E,5)  & (F,6) \\
\mbox{II: }  &(A,7) & (B,8) & (C,9) & (D,10) & (E,11) & (F,12) \\
\mbox{III: }  &(A,13) & (B,14) & (C,15) & (D,16) & (E,17) & (F,18) \\
\mbox{IV: }  &(A,19) & (B,20) & (C,21) & (D,22) & (E,23) & (F,24) \\

\hline
\multicolumn{7}{|c|}{Color interactions}\\
\hline
\multicolumn{7}{|l|}{
(1,15),
(2,8),
(12,17),
(13,20),
(16,23),
(3,21),
(4,18),
(5,11),
(6,24),
(7,19),
(9,14),
(10,22)} \\
\hline
\end{tabular}
\caption{Pyrochlore crystal lattice patchy particle design with 4 patchy particle species and 24 colors. The six patches on each species are labeled A-F respectively. Patch Coloring column shows colors (labeled 1-24) assigned to each patch and Color interactions show which colors are compatible. Compatible colors correspond to complementary DNA overhang strands in the DNA origami designs.}
\label{tab_ts}
\end{table}
\end{center}

The smallest solution in terms of number of different particle species $N_{\rm s}$ that satisfies all the clauses listed above, as identified with MiniSAT solver \cite{een2005minisat}, requires use of four distinct particle species. We chose species with maximum number of colors ($N_{\rm c} = 6 N_{\rm s}$), which are specified in Table \ref{tab_ts}.

 \begin{table}
 \small
 \centering
 \begin{tabular}{c|c|c|c}
 \hline
 Position $l_i$ & Slot $s_i$ & Position $l_j$ & Slot $s_j$\\ \hline
  0 &  0 &  7  &  0 \\
   0 &  1 &  4  &  3 \\
   0 &  2 &  1  &  5  \\
   0 &  3 &  9  &  1   \\
   0 &  4 &  6  &  4    \\
   0 &  5 &  3  &  2  \\
   10 &  1 &  12  &  1  \\
   1 &  0 &  12  &  4   \\
   10 &  3 &  11  &  1  \\
   1 &  1 &  13  &  3  \\
   11 &  5 &  14  &  2 \\
   1 &  2 &  2  &  5  \\
   12 &  5 &  13  &  5 \\
   13 &  0 &  15  &  0 \\
   1 &  3 &  7  &  1  \\
   14 &  4 &  15  &  4 \\
   1 &  4 &  4  &  4 \\
   2 &  0 &  14  &  0 \\
   2 &  1 &  11  &  3 \\
   2 &  2 &  3  &  5 \\
   2 &  3 &  12  &  3 \\
   2 &  4 &  13  &  4 \\
   3 &  0 &  9  &  0 \\
   3 &  1 &  6  &  3 \\
   3 &  3 &  14  &  1 \\
   3 &  4 &  11  &  4 \\
   4 &  0 &  5  &  4 \\
   4 &  1 &  15  &  3 \\
   4 &  2 &  14  &  5 \\
   4 &  5 &  7  &  2 \\
   5 &  0 &  12  &  0 \\
   5 &  1 &  6  &  1 \\
   5 &  2 &  10  &  2 \\
   5 &  3 &  14  &  3 \\
   5 &  5 &  15  &  5 \\
   6 &  0 &  10  &  0 \\
   6 &  2 &  12  &  2 \\
   6 &  5 &  9  &  2  \\
   7 &  3 &  8  &  1  \\
   7 &  4 &  10  &  4  \\
   7 &  5 &  11  &  2  \\
   8 &  0 &  11  &  0 \\
   8 &  2 &  10  &  5 \\
   8 &  3 &  13  &  1 \\
   8 &  4 &  9  &  4 \\
   8 &  5 &  15  &  2 \\
   9 &  3 &  15  &  1 \\
   9 &  5 &  13  &  2 \\  
\hline
\end{tabular}
 \label{tab:topo}
 \caption{Topology of the unit cell of a pyrochlore}
\end{table}

\subsection{Patchy particle simulations}
The patchy particle simulations were used to test solutions obtained from the SAT-solvers algorithm and probe their assembly into a target lattice.
In patchy-particle simulations of pyrochlore lattice assembly, each particle is represented by a sphere covered by 6 patches at distance $d_{\rm p} = 0.5$ distance units (d.u.) from the center of the sphere. To mimic the icosahedral DNA wireframe origami design, the positions of the patches correspond to the selected six vertices of an icosahedron. These positions, defined in terms of the orthonormal base associated with the patchy particle, are
\begin{equation}
 \mathbf{p}_1 = a \left( 0,1,\xi \right) ,\, \mathbf{p}_2 = a \left( 0,-1,\xi \right)  , \, \mathbf{p}_3 = a \left( \xi,0,1 \right),\,
 \mathbf{p}_4 = a \left(  0, -1, -\xi \right),\,  \mathbf{p}_5 = a \left( 0,1,-\xi \right), \,\mathbf{p}_6 = a \left( -\xi, 0, -1 \right)  ,
\end{equation}
where $\xi = (1 + \sqrt{5})/2$ and $a = d_{\rm p}/ \sqrt{1 + \xi^2 }$.

To model the octahedral wireframe DNA origami designs, we position the patches on the surface of the particle in cubic arrangement:
\begin{equation}
 \mathbf{p}_1 =  d_{\rm p} \left( 0,0,1 \right) ,\, \mathbf{p}_2 = d_{\rm p} \left( 1,0,0 \right)  , \, \mathbf{p}_3 = d_{\rm p} \left( 0,1,0 \right),\,
  \mathbf{p}_4 =  d_{\rm p} \left( 0,0,-1 \right) ,\, \mathbf{p}_5 = d_{\rm p} \left( -1,0,0 \right)  , \, \mathbf{p}_6 = d_{\rm p} \left( 0,-1,0 \right),\,
\end{equation}

The position of patch $i$ in the simulation box coordinate system is given by 
\begin{equation}
 \mathbf{r}_{p_i} = \mathbf{r}_{\rm cm} + {\mathrm{p}_i}_x \mathbf{e}_1 + {\mathrm{p}_i}_y \mathbf{e}_2  + {\mathrm{p}_i}_z \mathbf{e}_3 
\end{equation}
where $ \mathbf{r}_{\rm cm}$ is the position of the center of mass of the patchy particle, and $\mathbf{e}_{1,2,3}$ are the x, y, and z orthonormal base vectors associated with the patchy particle's orientation.

The interaction potential between a pair of patches on two distinct particles is 
\begin{equation}
    \label{eq_patch}
    V_{\rm patch}(r_p) =  \begin{cases} 
                          -1.001 \delta_{ij} \exp{\left[- \left( \frac{r_p}{\alpha} \right)^{10}\right]} - C & \text{if $r_p \leq r_{\rm pmax}  $} \\
                          0 & \text{otherwise}
                        \end{cases}
\end{equation}
where $\delta_{ij}$ is 1 if patch colors $i$ and $j$ can bind and 0 otherwise, $r_p$ is the distances between a pair of patches. 
The constant $C$ is set so that for $V_{\rm patch}(r_{pmax}) = 0$, $r_{\rm pmax} = 0.18\, \rm{d.u.}$ (distance units). The parameter $\alpha = 0.12$ d.u.~sets the patch width, where patches do not strongly interact if they are separated by distances larger than $\alpha = 0.12$. We approximately set the patch width based on oxDNA model simulations \cite{vsulc2012sequence,ouldridge2011structural,snodin2015introducing} of the DNA origami monomers (Sec.~\ref{sec:oxdd} and Fig.~\ref{fig:supmonomer}), where we measure the distribution of the distance between the last and the first bases of the single-stranded overhang. In order to capture the fact that single-stranded overhang can reach longer distances than its typical end-to-end distance, we approximately take the patch width $\alpha$ to be the mean plus standard deviation of the overhang length values measured in oxDNA molecular dynamics simulations at $25^{\circ}$C (ran for $10^9$ steps with time step $10$ fs). To convert this length from oxDNA to distance units of the patchy particle simulations, we divide it by the diameter of the wireframe origami, which we define as twice the mean distance from the center of mass of the DNA origami to the first base of the single-stranded overhang, as measured in the oxDNA simulations.

 The patchy particles further interact through excluded volume interactions ensuring that two particles do not overlap:
\begin{equation}
    V_{\rm exc}(r,\epsilon,\sigma,r^{\star}) = \begin{cases}
	V_{\rm LJ}(r, \epsilon, \sigma) & \text{if $r < r^{\star} $},\\
	\epsilon V_{\rm smooth} (r, b, r^c) & \text{if $r^{\star} < r < r^c$},\\
	0 & \text{otherwise}.
	\end{cases} 
\end{equation}
where $r$ is the distance between the centers of the patchy particles, and $\sigma$ is set to $2R = 0.8$, twice the desired radius 
of the patchy particle. The choice of a radius ($R = 0.4$ distance units) smaller than $d_{\rm p}$ has been done to mimic wireframe DNA origami, which are flexible and not spherical. In particular, we wanted the patchy model to be able to capture the fact that two faces of the icosahedral DNA origami, each with three patches, can be aligned in a way that can form up to three bonds with a second origami if patch coloring allows for it. Such states have been identified to lead to misassembly, and we use the SAT framework to avoid them.  The repulsive potential is a piecewise function, consisting of Lennard-Jones potential function
\begin{equation} 
V_{\rm LJ}(r,  \sigma) = 8 \left[ \left(\frac{\sigma}{r}\right) ^{12} - \left(\frac{\sigma}{r}\right) ^{6} \right].
\end{equation}
that is truncated using a quadratic smoothening function
\begin{equation}
V_{\rm smooth} (x, b, x^c) = b(x^c - x)^2, 
\end{equation}
with $b$ and $x_c$ are set so that the potential is a differentiable function that is equal to $0$ after a specified cutoff distance $r^c = 0.8$. 

The patchy particle systems are simulated using rigid-body Molecular Dynamics with an Andersen-like thermostat~\cite{russo2009reversible}. During the simulation, each patch was only able to be bound to one other patch at the time, and if the binding energy between a pair of patches, as given by Eq.~\eqref{eq_patch}, is smaller than 0, none of the patches can bind to any other patch until their pair interaction potential is again 0.

\begin{figure}
\centering
	\includegraphics[width=1.0\textwidth]{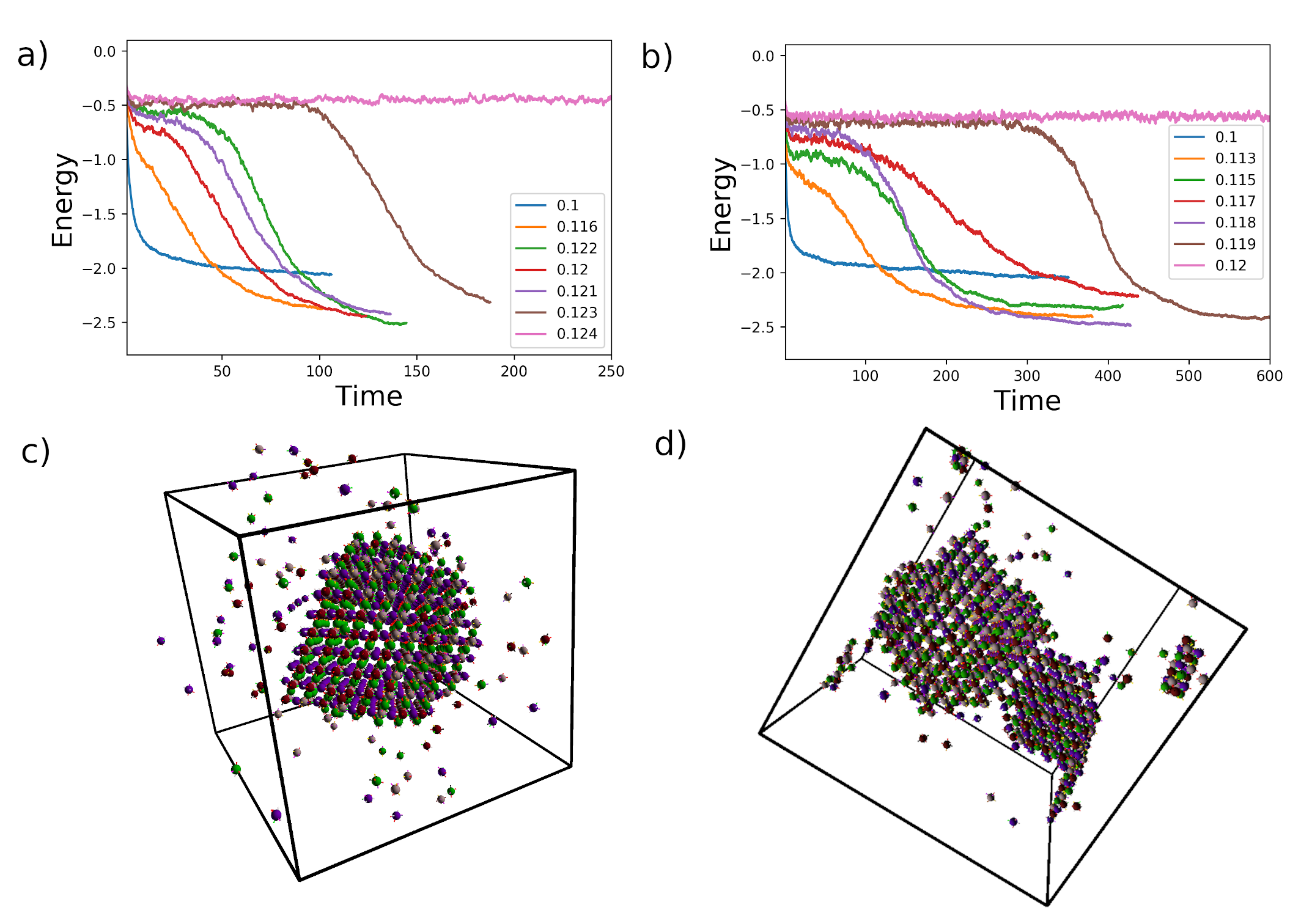}
	\caption{Energy per particle (in simulation energy units $\varepsilon$) as a function of simulation time units for different simulation temperatures (in units of $k_{\rm B}/\varepsilon$) for icosahedral patchy particle (a) and octahedral patchy particle (b). We show a snapshot of the last state of the simulation at temperature $T = 0.122\, k_{\rm B}/\varepsilon$ for icosahedral patchy particles (c) and at $T = 0.118\, k_{\rm B}/\varepsilon $ octahedral patchy particles (d). }
    \label{fig:sup_sim}
 \centering
\end{figure}

To verify designs obtained from SAT-assembly, we ran for each designed systems the patchy particle simulations (consisting of 2048 particles, 512 of each species) at a range of temperatures (from $0.1$ to $0.125$ in simulation units) at number density $0.1$ to identify an optimal temperature where the system nucleates and grows into a pyrochlore lattice (Fig.~\ref{fig:sup_sim}).  We found that at given density, the icosahedral patchy particle design nucleates into a pyrochlore lattice in a range of tested temperatures ($0.116$ to $0.123$). At lower temperatures, it forms quenched glassy state, and at higher temperatures remains in gas phase. For the octahedral patchy particle design, we found that the system successfully nucleated at temperatures $0.118$ and $0.119$. For larger temperatures that we considered (Fig.~\ref{fig:sup_sim}b), the systems did not successfully nucleate a crystal, and it misassembled at lower temperatures.

\begin{figure}
\centering
	\includegraphics[width=1.0\textwidth]{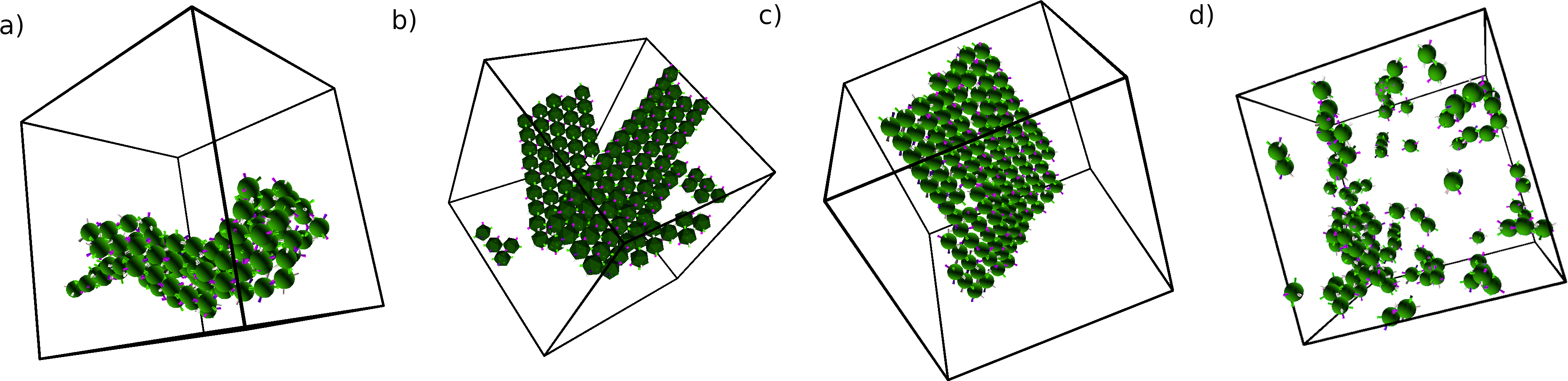}
	\caption{Snapshots from equilibrated final states of patchy particle simulations with a single particle species. At low temperature (a), the particles assemble into disordered quenched state. At intermediate temperature range, the particles assemble into alternative free-energy minima of a sheets where pairs of interacting particles interact through two patches (shown in (c) for spherical patchy simulation and in (b) for rigid body with icosahedral shape simulation). At high temperature, the system remains in gas state (d). }
    \label{fig:sup_miss}
 \centering
\end{figure}

Besides the model outlines above, we have further implemented an additional model to represent icosahedral origami designs: a rigid-body icosahedron model. The interactions between patches remain the same, given by Eq.~\eqref{eq_patch}. We however replace the excluded volume interaction by a new potential
\begin{equation}
   V_{\rm rigid\, b.}(\mathbf{r}_i,\mathbf{r}_j, \mathbf{\Omega}_i, \mathbf{\Omega}_j) = \begin{cases}
	\infty & \text{if icosahedrons $i$ and $j$ overlap},\\
	0 & \text{otherwise},
	\end{cases} 
\end{equation}
which we calculate for icosahedral particles $i$ and $j$ positioned at $\mathbf{r}_i$ ,$\mathbf{r}_j$ respectively, with their orientation given by $\mathbf{\Omega}_i$ and  $\mathbf{\Omega}_j$. Each particle is represented in this rigid body model as an icosahedron with distance from center of mass to its vertex equal to $R = 0.5$ distance units. Given the non-continuous nature of this potential, we use Monte Carlo simulations (with translations and rotations) to simulate these systems. Any move that would result in two icosahedral-shaped particles to overlap is always rejected, as it would result in infinite energy. We confirmed that the kinetically trapped state present for 1-species solution in the spherical system is also present in the rigid body icosahedron simulation. Similarly, we verified with the rigid body simulation that the solutions that are designed to avoid the state where one particle can bind to another one by two bonds at the same time eventually assemble into a pyrochlore lattice. 
The typical configurations identified in single-species systems that we studied are shown in Fig.~\ref{fig:sup_miss}.

Finally, we have also explored the role of number of colors when the number of particle species is fixed. The SAT clauses listed above have a two-species solution (which was also shown to form pyrochlore \cite{romano2020designing}) if we leave out the conditions listed in Eqs.~\ref{eq_nosamespecies}. For the two-species case, we have also tried solutions with number of colors equal to 6, 8, 10, and 12 respectively.
We  observed that solutions with $N_c = 10$ and $N_c = 8$ also assembled well into a pyrochlore, but took longer to nucleate than the $N_c = 12$ case. We also carried out simulations for a solution with $N_c = 6$ at a range of temperatures ranging from $0.12$ to $0.138\, k_{\rm B}/\epsilon$, but we only observed either gas state or glassy state formations and we did not observe successful pyrochlore nucleation in the simulation running time (up to $3 \times 10^9$ MD steps, corresponding to about 3 weeks of CPU running time). Hence, we focused only on the solution with maximum possible number of colors ($6 \times$ the number of species).

\subsection{OxDNA simulations}
\label{sec:oxdd}

\begin{figure}
\centering
	\includegraphics[width=0.7\textwidth]{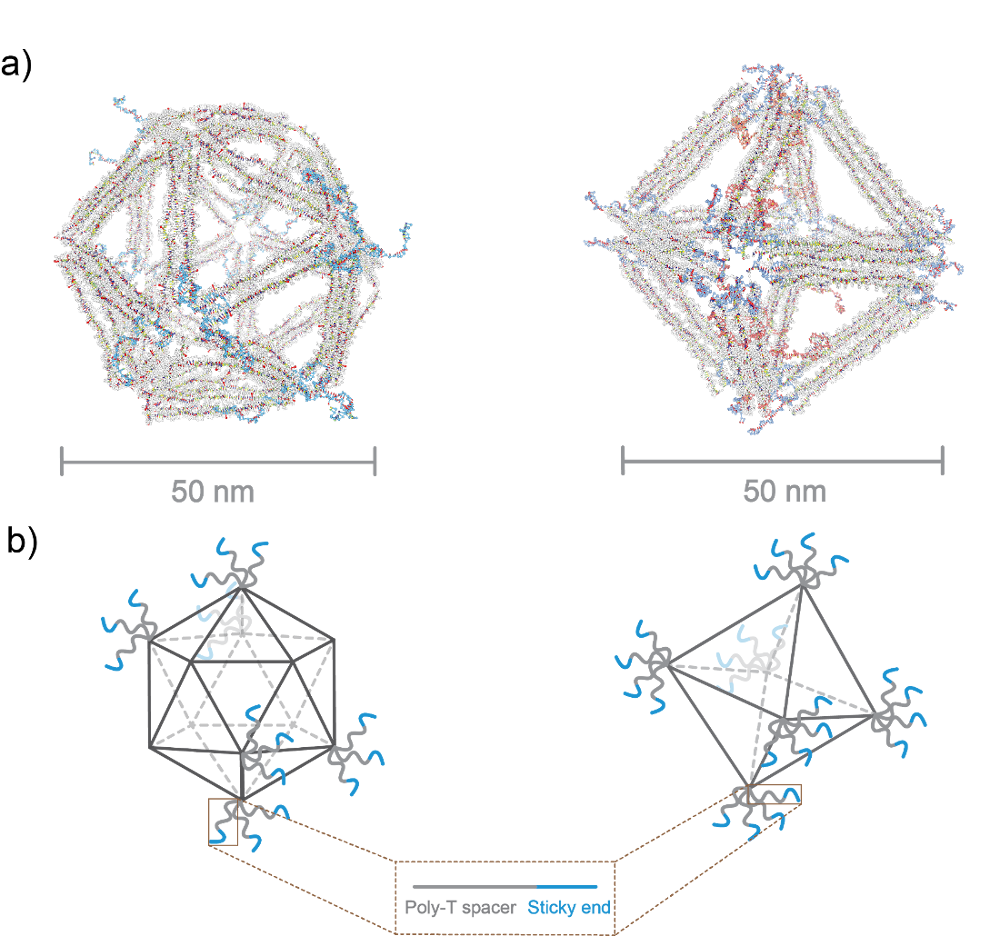}
	\caption{DNA origami design for the self-assembly of lattices. a) Origami design visualized in oxView \cite{bohlin2022design} with blue strands representing the handle strands. b) Schematics showing the origami design. For each handle strand, poly-T spacers are placed before the sticky end sequence for binding, allowing flexibility for the recognition and binding of the sticky end pairs in lattice annealing. }
    \label{fig:supmonomer}
 \centering
\end{figure}

\begin{figure}
\centering
	\includegraphics[width=1.0\textwidth]{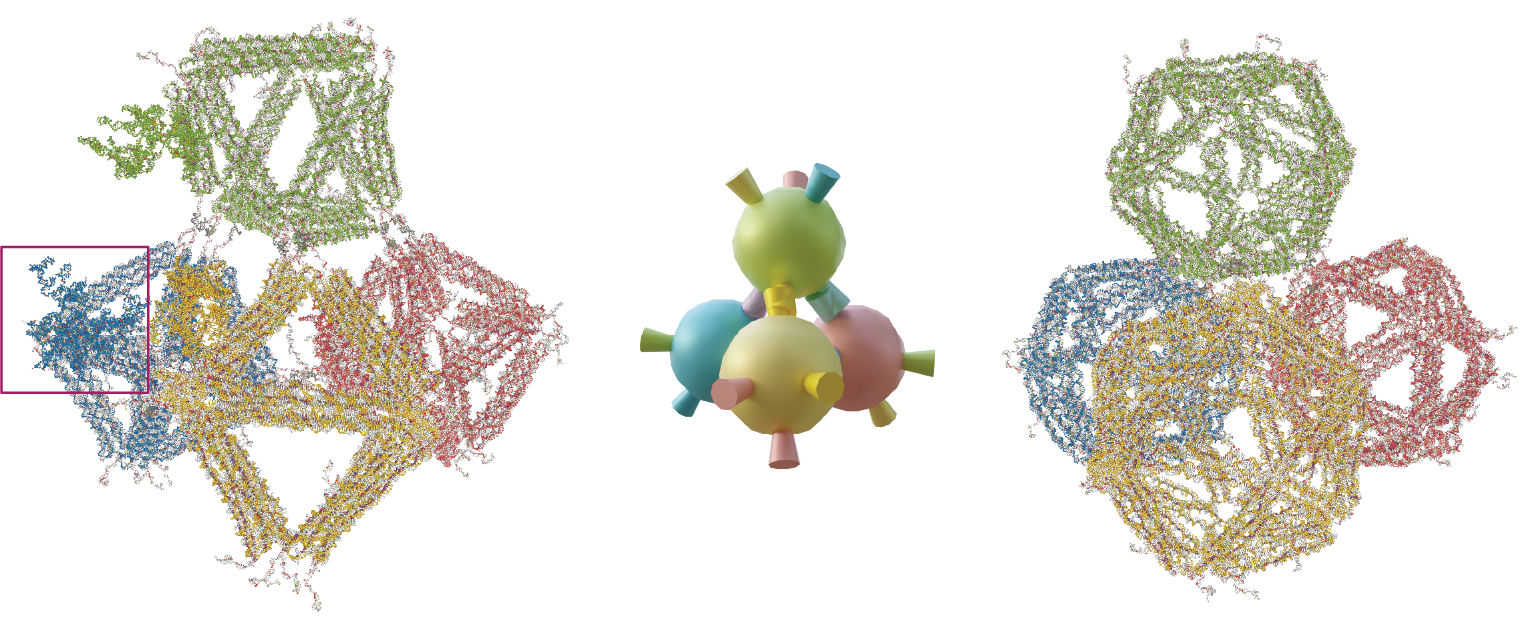}
	\caption{Tetramer in the topology of tetrahedron assembled with octahedron and icosahedron DNA origami, visualized in oxView. Tetrahedral assemblies stack on each other with the designed pathway forms further the pyrochlore lattice. The existence of scaffold loop in the octahedral DNA origami (one of them is highlighted in red square) could affect inter-particle connection due to the spatial proximity while icosahedral origami-based lattice design does not have such concern.}
    \label{fig:suptetramer}
 \centering
\end{figure}

The nucleotide-level simulations of DNA nanostructures have been carried out using the oxDNA2 parametrization of the coarse-grained model of DNA \cite{snodin2015introducing,ouldridge2011structural,sulc2012sequence}. All simulations were performed at 1 M salt concentration, with time-step of 15 fs and temperature $25^\circ$C unless stated otherwise. We used Andersen-like thermostat \cite{russo2009reversible} for molecular dynamics simulations. The simulations were performed using GPU implementation \cite{rovigatti2015comparison,poppleton2023oxdna} of the simulation code. The simulations of monomers and fourmers (Figs. \ref{fig:supmonomer} and \ref{fig:suptetramer}) were setup using oxView design tool \cite{poppleton2020design,bohlin2022design}, which was also used to interactively set the lengths of the single-stranded overhangs connecting the respective origamis. The simulations of the 2x2x2 unit cell cluster was setup by using the assembled part of the crystal from patchy particle simulation, where each DNA origami monomer were docked onto the position of the patchy particle and spring potential forces were applied to connect complementary DNA overhangs that represent the patches. The system was relaxed using Monte Carlo simulations and oxDNA relaxation potential, following the protocol from \cite{bohlin2022design}, before the production run of the crystal cluster was run.

\clearpage

\section{Experimental methods and materials}
\label{secc:expmethods}
\subsection{DNA origami and sequence design}

Octahedral DNA origami (ODO), originally designed and reported by Tian el al \cite{tian2015prescribed} is adapted and used here. Icosahedral DNA origami (IDO) was first introduced by Zhang et al \cite{zhang2022spatially} and further adapted and used here. For both structures, each edge is composed of at least 4 helices to ensure the structural rigidity. The handles for directional bonding are placed on the vertices of the polyhedral shape and are properly distributed for even connection. 

The sequence of handles are carefully designed so that the melting temperature difference between each pair of sticky ends are minimized and also the binding between each sequence pair is orthogonal to each other to prevent cross-talk. NUPACK\cite{zadeh2011nupack} is used for the sequence design, with customized scripts.

\subsection{DNA origami annealing and purification}


Octahedral DNA origami are folded by mixing 20 nM M13mp18 scaffold, 200 nM of each designed staples, including the one designed to bind gold nanoparticles, and specifically 400 nM of each of the handle (sticky end strands) making the connection in 1 X TAE $\rm Mg^{2+}$ buffer (40 mM Tris, 20 mM acetic acid, 1 mM EDTA, 12.5 mM $\rm Mg^{2+}$ pH 8.0). The mixed solution is then annealed by cooling from 75\textdegree C to 20 \textdegree C slowly to obtain the target DNA structure with the following program: from 75 \textdegree C to 65 \textdegree C at a rate of 0.1 \textdegree C per $0.5$ min, from 65 \textdegree C to 40 \textdegree C at a rate of $0.1$ \textdegree C per 4 min, from 40 \textdegree C to 20 \textdegree C at $0.1$ \textdegree C per 2 min, then held at 20 \textdegree C at the end of the cycle. Excess of the staple strands of the origami are purified through amicon ultrafiltration. Briefly, 100 kD amicon columns are passivated with fresh buffer and then annealed sample is loaded along with buffer to fill the column. The centrifugation is done with a rate between 2k rcf to 2.5k rcf followed by refilling the column with more buffer. Such procedure is done for overall 5 to 8 times, with higher loading amount taking more iterations to be purified. 

Icosahedral DNA origami is annealed with the same program with the octahedral origami but usually in a higher concentration (40 - 50 nM) for further purification. It is noted that the yield of correctly formed monomer icosahedral origami is lower than the octahedral one and requires more attention and care in purification to afford clean monomers for lattice assembly. Rate zonal centrifugation, originally reported by Lin et al \cite{lin2013purification}, is used to purify the monomers which guarantees higher yield compared to gel band excision and extraction. Briefly, in a 3.5 ml thick-wall ultracentrifuge tube, a glycerol gradient is created by adding \SI{400}{\micro\liter} of different percentage of glycerol solution in 1 x TAE $\rm Mg^{2+}$ buffer layer by layer. For overall 7 layers, the 45\% glycerol solution is added the first and therefore at the bottom layer and the 5\% glycerol solution is at the bottom with 5\% concentration decrement per layer in between. \SI{200}{\micro\liter} of annealed icosahedral DNA origamis are then applied on the top. Without waiting overnight for a quasi-continuous gradient, the tube is loaded onto the swinging-bucket rotor (Beckman SW 55 Ti) and spun at 50000 rpm for 1 hour at 4 \textdegree C. The centrifugation program is set to be "Slow" for acceleration and "No brake" for deceleration, allowing for slow and smooth rate transition for a better separation of the sample. Afterwards, fractions (\SI{200}{\micro\liter} per fraction) are taken out and placed in individual tubes which are further evaluated with agarose gel electrophresis (AGE). The fractions corresponding to monomers, ideally with no or limited contamination of multimers/malformed structure which appear above the monomer band in AGE, are collected and concentrated as well as buffer exchanged through amicon ultrafiltration. The buffer washing step in amicon ultrafiltration is done for 6 times with 2400 rcf to remove the leftover glycerol which could affect the diffusion of DNA origamis in solution if not completely filtered. 

For both DNA origami species, the concentration after purification is determined through Nanodrop based on the absorbance on 260 nm. Unused samples are stored at 4 \textdegree C refrigerator until further processing. 

\subsection{Fabrication of pyrochlore superlattices}

For each origami species there are 4 particle types with 6 corresponding sticky end sets to map the interaction matrix for the assembly of pyrochlore lattices. Purified origami monomers, namely 4 individual particle species, are mixed together with equal ratio (1:1:1:1) and the final monomer concentration is set to be 10 nM. The mixed samples are annealed with a customized protocol based on the melting temperature of the superlattices.

For the pyrochlore lattice assembled with octahedral DNA origami, the annealing protocol is as follows: 55 \textdegree C for 2 hours, 54 to 52 \textdegree C with 0.1 \textdegree C per 8 hours, 52 \textdegree C for 24 hours, and then incubated at RT for forever. 

For the pyrochlore lattice assembled with icosahedral DNA origami, the annealing protocol is as follows: 45 \textdegree C for 2 hours, 43 to 41 \textdegree C with 0.1 \textdegree C per 8 hours and then incubated at RT for forever. 

For the screening of the nucleation temperature for the pyrochlore lattice with icosahedral building blocks, the mixed solution is incubated at 45 \textdegree C for 2 hours followed by incubation at the designated temperatures (41.1 \textdegree C, 41.3 \textdegree C, 41.4 \textdegree C, 41.5 \textdegree C, 41.7 \textdegree C, 41.9 \textdegree C) and finally incubated at RT for forever. 

The thermal annealing of mixed solutions are done in thermocyclers above.
The mix-and-anneal strategy is done through a customized set-up in which PCR tubes containing the sample are fixed to shaker and the devices are placed in a thermal incubator. Particularly, unless noted, all the SEM images shown are annealed with thermocyclers. The annealing protocol set in thermal incubators is as follows: 45 \textdegree C for 2 hours, 43 to 41 \textdegree C with 0.2 \textdegree C per 24 hours and the system is then allowed to cool down to RT before being harvested for characterization. 

The fabricated lattices are stored in 4 \textdegree C after annealing. From our experience the assembled lattices are stable in fridge for more than 6 months - no visual differences are spotted with SEM examination after silica embedding for freshly annealed samples and fridge stored samples. Batch to batch difference exists and it is largely caused by pipeting and therefore stoichiometry accuracy, which leads to the change of average size of the lattices as visualized by SEM.  

\subsection{Measurement of the melting profile}

Melting profile is used to customize the annealing protocol for the superlattices. It is measured by monitoring the aggregate size for a range of temperature through Dynamic Light Scattering (DLS, Malvern Zetasizer Nano). Specifically, the temperature ramps down from 55 \textdegree C to 30 \textdegree C with a ramping rate of 0.2\textdegree C per 240 seconds. Disposable \SI{40}{\micro\liter} Cuvette is used after cleaning and kept capped during the measurement to reduce the evaporation of water. \SI{100}{\micro\liter} of sample, pre-mixed with purified all 4 particle species, is loaded so that the melting profile would be similar to the one associated to lattice annealing. 

\subsection{Silicification and SEM imaging}
\begin{figure}[h]
\centering
	\includegraphics[width=1.0\textwidth]{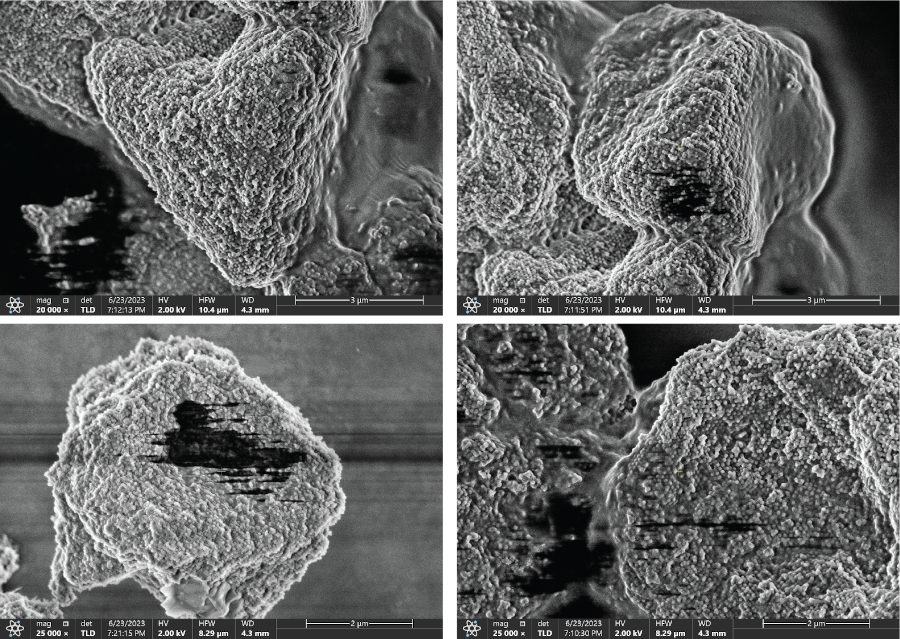}
	\caption{Typical examples showing the distorted/collapsed lattice imaged with SEM, with suboptimal silica coating being the major reason. }
    \label{fig:sembadsilica}
 \centering
\end{figure}

To preserve the structure of under EM conditions, a thin layer of silica was coated on the DNA origami superlattices with the method originally described by Wang et al\cite{wang2021dna} and further adapted by us. Briefly, the annealed sample is buffer exchanged through gently removing the supernatant and replace with fresh buffer of 1xTAE 12.5mM $\mathrm{Mg^{2+}}$. [$\mathrm{Mg^{2+}}$] reaches to the expected level after several times of pipetting, along with the amount of monomer and oligomer which would potentially affect the imaging quality of SEM. Samples are lightly centrifuged (1k rpm, 1min) to ensure the fabricated lattices are deposited at the bottom of the tube. TMAPS (50\% in Methanol, TCI), diluted with methanol, is then added to the sample with microsyringe and the solution mix is subjected to shaking (800rpm, 30min, 4\textdegree C) to facilitate the diffusion and therefore even binding of TMAPS to the DNA phosphate backbone. Subsequently, TEOS (Sigma-Alrich) is added to the solution after dilution with methanol and the same shaking protocol is executed for the sample. Notably, to embed silica onto large lattices elongated mixing (up to 2h) is needed to ensure the diffusion of silanes into the internal part of the lattice, as also indicated by Lewis et al \cite{lewis2020single}. A direct result of insufficient mixing is the collapse of lattice upon drying as shown in Fig.~\ref{fig:sembadsilica}. The molar ratio of nucleotide: TMAPS : TEOS is ~1:9:18 for the production of a thin layer of silica. The sample mixed with silanes is then incubated at RT for 12 hours undisturbed, wrapped within an aluminum foil. White cloudy precipitants are visible afterwards, indicating the successful formation of silica coated substances. Fresh ultrapure water is finally added to fill the tube to 2ml for the purpose of quenching the reaction, followed by light pipeting, centrifuging and supernatant removal to get rid of the excess silanes and silica particles. The washing process is done for at least 3 times with water, followed by one last washing step with isopropanol.

\SI{10}{\micro\liter} embedded sample is taken from the bottom of the tube and further dropped cast on a silicon wafer cleaned with ethanol. The droplet is then dried in room temperature to expose the silicated superlattices to the air. SEM images are obtained through Auiga (Zeiss) and Helios 5 UX (ThermoScientific). Accelerating voltages, currents and other imaging parameters are adjusted to minimize the charging effect. 

Cross-section of the silicated sample is created with Focused Ion Beam (FIB) installed on Helios 5 UX. A sacrificial layer of Platinum is deposited to the top of the sample to avoid unintended milling. For ion beam imaging, 30 kv and 7 pA is used to reduce the charging with charge neutralizer switched on while the current is increaed to 41 pA for the creation of cross-section to accelerate the process.  

\subsection{AuNP functionalization and incorporation}

10nm AuNP (Nanopartz) is functionalized with DNA through the salt aging method, adapted from Tian et al\cite{tian2020ordered}. Briefly, thiolated DNA is synthesized by IDT and reduced with TCEP (tris[2-carboxyethyl] phosphine) to be prepared for the conjugation onto AuNP. With a ratio of 1:100, excess TCEP is added to the solution of thiolated DNA and the mixture is shaked at room temperature for an hour with 800rpm. Desalting column (G-25, GE Healthcare) is used to remove the remaining TCEP along with the reduced side-product. The purified product is further added into the AuNP solution in a ratio of 1:300 and another hour is given to the mixture for initial DNA attachment. 
Phosphate buffer is then added until 10 mM concentration is reached. After 1 hour, 2 M Nacl solution is mixed with the solution from the previous step gradually to reach a final concentration of 300 mM over the course of 5 hours, sonication up to 30 seconds is allowed after the addition of NaCl. The solution is finally aged at room temperature for at least 18 hours. 
Excess DNA is removed from the functionalized AuNP through repeating the process of centrifugation at 15000 rcf and buffer exchange (1 x PBS, remove the supernatant and add fresh buffer). The concentration of AuNP is determined by measuring the absorbance at 520 nm. 

To incorporate the AuNP into the pyrochlore lattice, annealed sample is mixed with functionalized AuNP for 2 times excess. After gentle pipetting, the solution is annealed from 45 to 30 \textdegree C with a rate of 0.2 \textdegree C/hour.

\subsection{Grid preparation and imaging with TEM and STEM}

Carbon coated copper grids are glow discharged for 30 seconds and incubated at room temperature for 5 min before the following operation. To image the individual DNA origami building blocks for the assembly of the superlattice, \SI{5}{\micro\liter} solution of purified origami is dropcasted on the grid, followed by incubation for 5 min. Excess liquid is then blotted with a piece of filter paper and the grid is then negatively stained with \SI{5}{\micro\liter} solution of 2\% uranyl acetate for 90 s, afterwhich the grid is blotted thoroughly with filter paper and allowed to air dry for at least 20 min. To image the lattices, \SI{5}{\micro\liter} of silicated lattices, after thorough washing with water, is taken from the bottom of the tube and added to the glow discharged grid. Silicated lattice is used here to ensure the structural integrity. 20 min of incubation in the humid chamber is allowed for the lattice sample to sediment and the grid is blotted with filter paper afterwards. The grid is then allowed to air dry without further staining for EM imaging. 

TEM imaging is done on Talos L120C (TFS), operated at an accelerating voltage of 120 kV.

STEM images are taken using a Titan 300/80 (FEI) in high-angle annular dark field (HAADF) mode, operated at an accelerating voltage of 300 kV.

\subsection{SAXS measurements and modeling}
\label{secc:saxs}


\subsubsection{Representation of DNA origami for SAXS analysis}
\begin{figure}
\centering
	\includegraphics[width=0.8\textwidth]{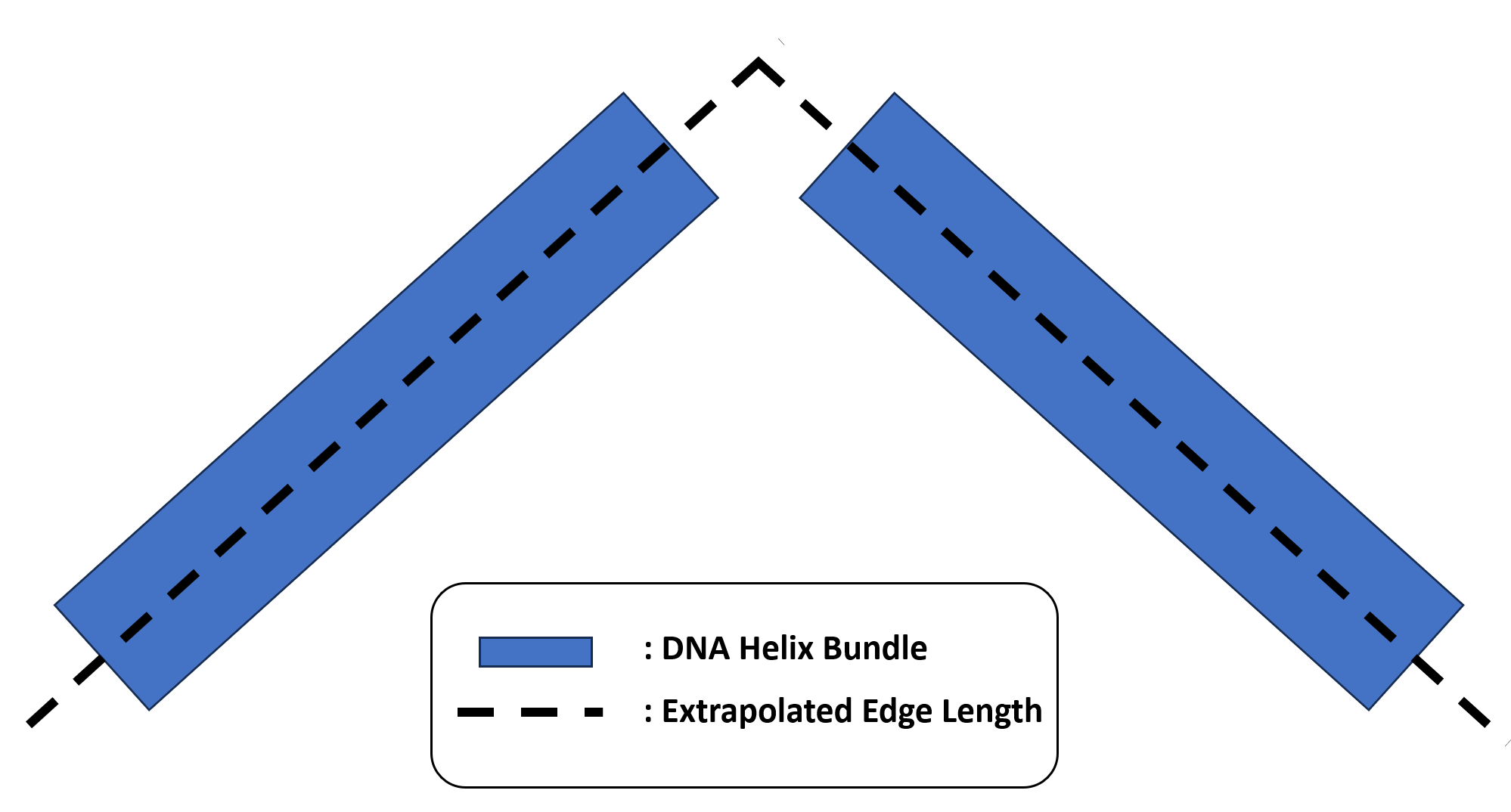}
	\caption{A 2D cartoon depicting how the junction between DNA bundles (blue rectangles) results in a truncated polyhedron, but that their spatial arrangement can be described mathematically via the extrapolated edge lengths (dashed lines)}
    \label{fig:daniel_fig1}
\end{figure}

For the SAXS analysis, we approximate the DNA origami wireframe structures as ideal polyhedrons. We extrapolate edges as lines from the center of cylindrical dsDNA bundles that will meet at the vertices of the ideal polyhedron, as shown in Fig.~\ref{fig:daniel_fig1}. This ideal polyhedron is used to model the spatial arrangements of the bundles for SAXS data analysis, in particular the form factor, of a given DNA wireframe. The analysis of the SAXS data in this work uses the parameters listed in Table \ref{tab:daniel_table1}.

\begin{table}[]
    \centering
    \resizebox{\textwidth}{!}{\begin{tabular}{|c|c|c|c|c|c|c|c|}
        \hline
        Frame Geometry & Bundle Length (nm) & Truncated Diagonal (nm) & Extrapolated Edge Length (nm) & Extrapolated Diagonal (nm) & Lattice Parameter (nm) & Nearest Neighbor (nm) & Inter-vertex Distance (nm) \\
        \hline
        Octahedron & 27.79 & 39.30 & 38.34 & 54.22 & 156.4 & 55.30 & 16.00 \\
        \hline
        Icosahedron & 21.42 & 40.74 & 27.86 & 53.0 & 159.1 & 56.26 & 15.52 \\
        \hline
    \end{tabular}}
    \caption{Parameters of the DNA origami wireframe structures and their extrapolated ideal polyhedral shapes.}
    \label{tab:daniel_table1}
\end{table}


The bundle length can be estimated by multiplying the number of base pair steps in the bundle by $0.34$ nm, the length of a base pair in the ideal B-form helix. From this bundle length, a truncated diagonal can be determined through geometric relations for a given frame geometry. The extrapolated edge length and diagonal are empirically determined from SAXS data through fitting the form factor of the DNA origami. After the unit cell is determined from SAXS data, the lattice parameter and nearest neighbor distances are used to estimate the inter-vertex distance between two bound vertices. The inter-vertex is the distance from the outer edge of one frame to another, which is the nearest neighbor distance subtracted by the truncated diagonal and is physically determined by the spacer and sticky end sequences of the binding patches on the frames. 

The parameters used to model the octahedron origami are in line with those used for prior SAXS analysis. The bundle length, 21.42 nm, used for the icosahedron origami comes from the design reported in the by Zhang et al \cite{zhang2022spatially}. There is a slight discrepancy between our estimate of the truncated diagonal, $40.74$ nm, and their reported truncated diagonal, $43$ nm, possibly due to the fact that their estimate might be based on negative TEM staining, which could have deformed the structures.

\subsubsection{SAXS Data Analysis}
The SAXS data analysis consists of three steps: 1) modelling the building block (DNA origami) used in the self-assembly, 2) arranging the building blocks into the expected unit cell, and 3) simulating the SAXS pattern from the model unit cell and fitting the pattern to the data. This process is described in greater detail in Section \ref{sec:saxs_modeling}. For DNA origami frames, the helix bundles that serve as the edges of the frames are represented as single cylinders. The bundle length and thickness are used as the cylinder height and diameter respectively. The cylinders are then rotated and translated such that the edge corresponding to the extrapolated polyhedron runs through the center of the cylinder, continuing until the frame is fully constructed. The icosahedron has 30 edges and therefore consists of 30 cylinders, whereas the octahedron has 12 edges and consists of 12 cylinders. For gold nanoparticles which are attached inside octahedral DNA origami, the known form factor of a sphere is used. Therefore, a form factor is fit to the data corresponding to systems with gold nanoparticles, such as the pyrochlore lattice made out of octahedral DNA origami with gold nanoparticles inside. When modelling the gold nanoparticle (AuNP), it is important to account for size polydispersity to better match the data as this can have a profound effect on the form factor of a sphere.

Once the building blocks are modelled, they need to be placed into the sites corresponding to the unit cell of the expected lattice. In this analysis, the sites corresponding to the pyrochlore lattice are determined through the Crystalmaker software (from CrystalMaker Software Ltd, Oxford, England (www.crystalmaker.com)), and exported as fractional coordinates that need to be scaled by the physical lattice parameters. Theoretical Bragg peaks corresponding to the pyrochlore are generated with the software CrystalDiffract.

The lattice parameters are extracted from the experimental data by converting the first peak position from q-space to real space and accounting for the peak indices, using the following formula: 

\begin{equation}
    q_{hkl} = \frac{2 \pi}{a} \sqrt{h^2 + k^2 + l^2}
\end{equation}

Where $q_{hkl}$ is the $q$ position of any given peak, $a$ is the lattice parameter of a cubic lattice, and $h$, $k$, and $l$, are the crystallographic indices for the peak. The first peak in the SAXS data used to determine the lattice parameter in the pyrochlore lattices corresponds to the [1 1 1] peak. From this formula, the octahedron origami pyrochlore lattice has a lattice parameter of $156.4$ nm and the icosahedron origami pyrochlore lattice has a lattice parameter of $159.1$ nm. The unit cell and building blocks are then appropriately scaled to fit the SAXS data. Fine tuning of other parameters, such as peak shape and Debye-Waller factor, a thermal disorder parameter, further improves the fitting of the model. 

\begin{figure}
\centering
	\includegraphics[width=0.6\textwidth]{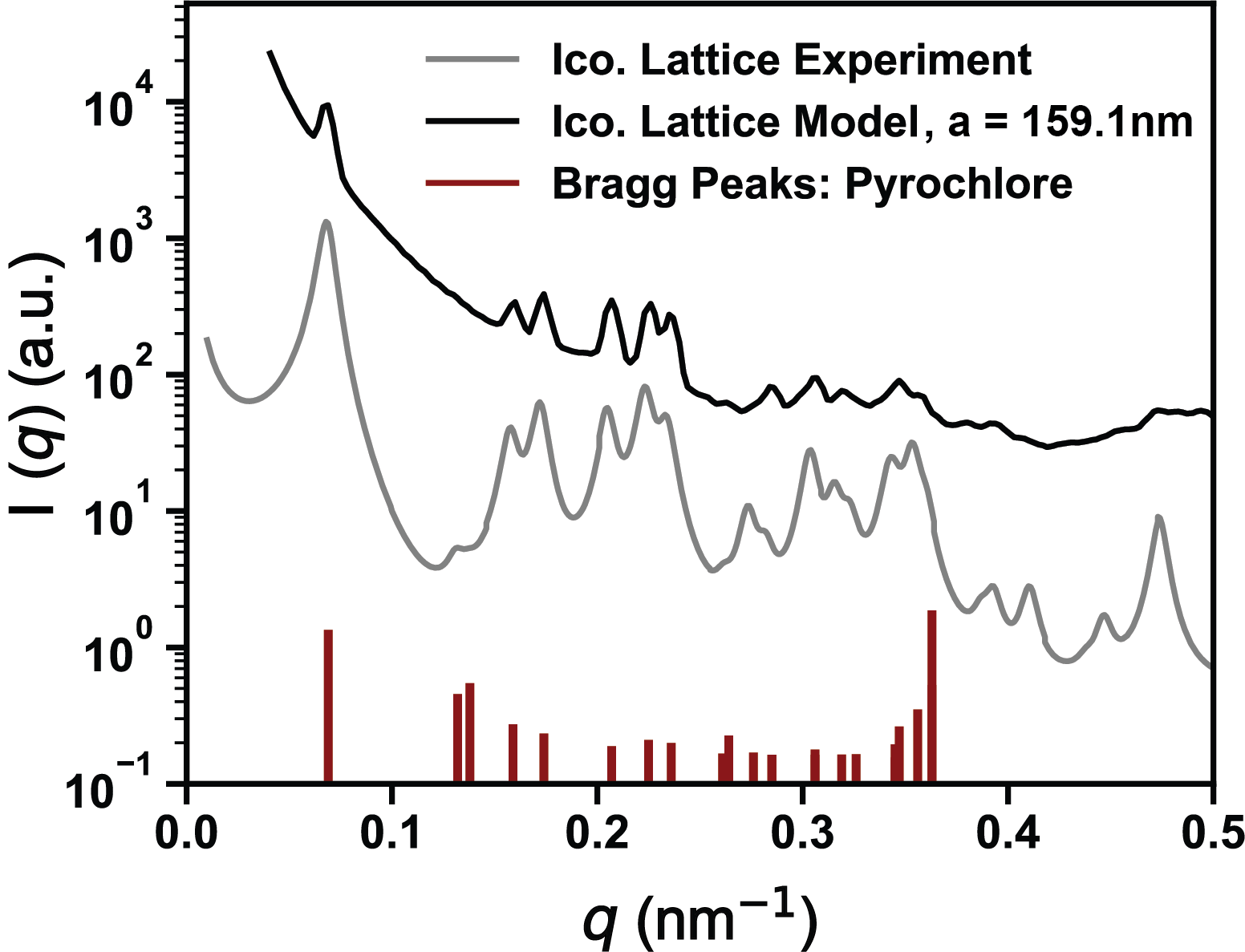}
	\caption{Scattered intensity, I(q), of the pyrochlore lattice constructed with icosahedral DNA origami, where the blue and black curves correspond to the experimental data and modelling respectively. The red lines correspond to the Bragg peak positions of a pyrochlore lattice.}
    \label{fig:daniel_fig2}
 \centering
\end{figure}

\begin{figure}
\centering
	\includegraphics[width=0.55\textwidth]{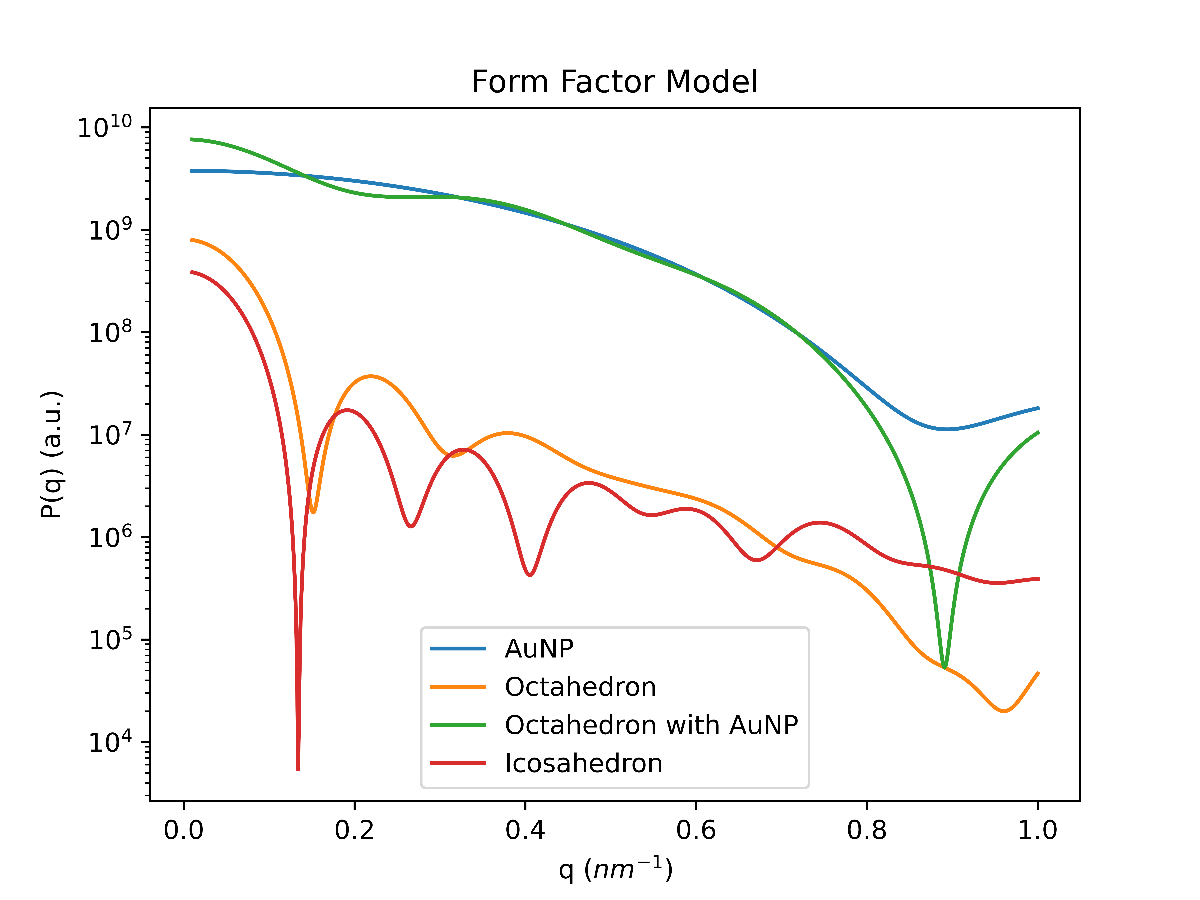}
	\caption{Modelled form factor, P(q), profiles for the different origami frames and gold nanoparticles.}
    \label{fig:daniel_fig4}
 \centering
\end{figure}

The icosahedron origami pyrochlore lattice data shows good agreement with the I(q) model, as shown in Figure \ref{fig:daniel_fig2}, and agrees with most of the model in terms of peak height ratios. The model departs in terms of relative peak intensity at the peaks around $0.28$ nm$^{-1}$ and diverges when approaching the high q space data, due to instrument limitations. The scattering data from the pyrochlore constructed with icosahedra origami is left as $I(q)$ because of the difficulty in deconvoluting the form factor of an empty icosahedral frame from the structure factor of the pyrochlore. The analytical form factors of the different objects are shown in Figure \ref{fig:daniel_fig4}. From the SAXS data, the icosahedron origami frames have an extrapolated diagonal of $53.0$ nm. The icosahedron pyrochlore lattice parameter is $159.1$ nm, and the nearest neighbor distance is $56.26$ nm. 

\begin{figure}
\centering
	\includegraphics[width=0.6\textwidth]{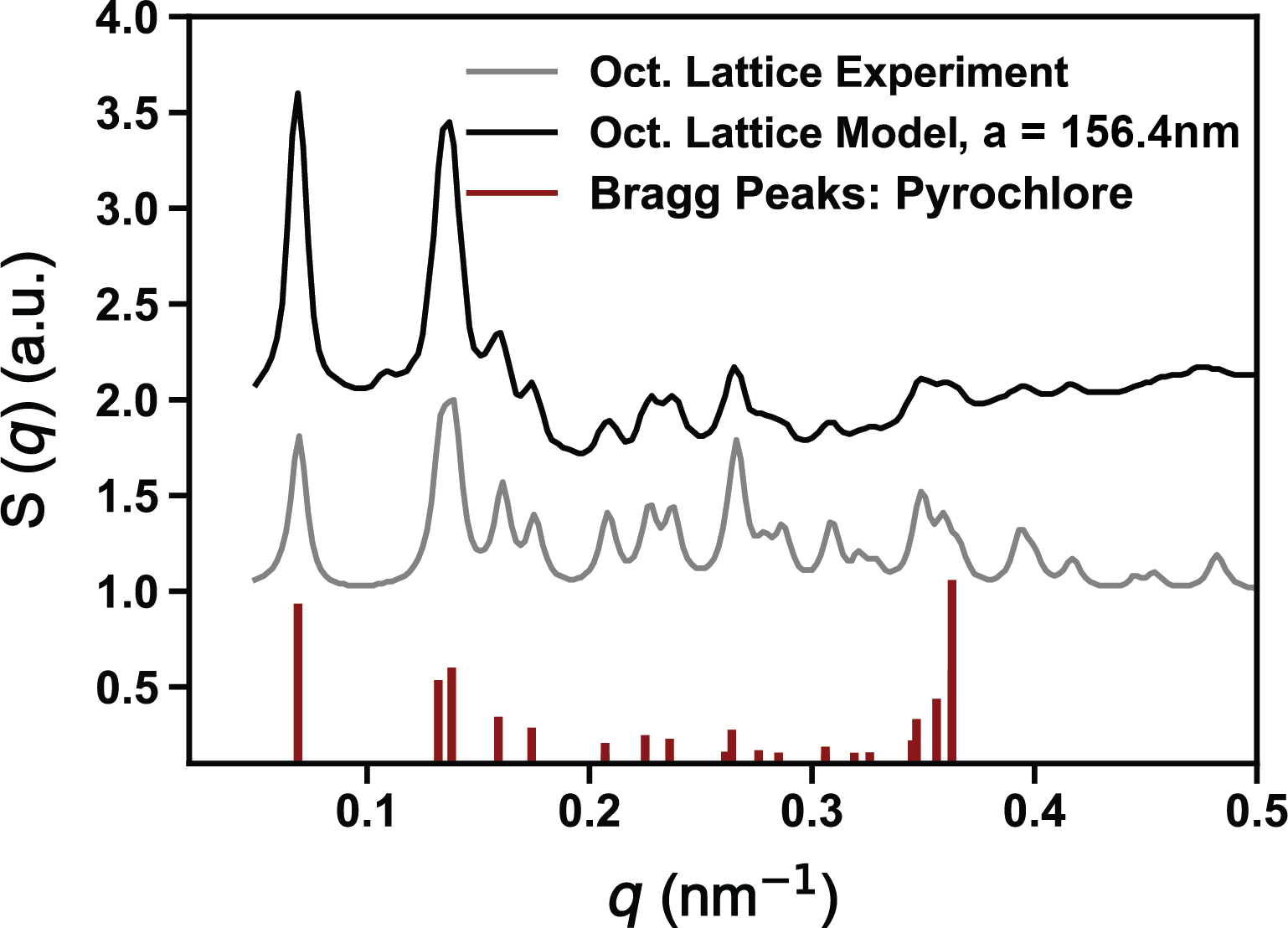}
	\caption{Structure factor, S(q), data from the pyrochlore of octahedral origami loaded with gold nanoparticles, where the blue and black curves correspond to the experimental data and modelling respectively. The red lines correspond to the Bragg peaks of the pyrochlore lattice. }
    \label{fig:daniel_fig3}
 \centering
\end{figure}

The structure factor data is in good agreement with the modelled structure factor for the pyrochlore of octahedron loaded with gold nanoparticles (AuNP), as shown in Figure \ref{fig:daniel_fig3}. The data is presented as structure factor data because the AuNPs provide a simple measurable form factor to remove from the data. The S(q) data show a small peak around $0.15$ nm$^{-1}$ that does not appear in either the model structure factor or the Bragg peaks from the pyrochlore, which can be due possible defects in the crystal. The data eventually diverges in the high q space, due to instrument limitations. The octahedron origami pyrochlore has a lattice parameter of $156.4$ nm, which results in a nearest neighbor distance of $55.30$ nm. 

The complexity of the form factor contribution to scattering intensity is dependent upon the complexity of the object. A polydisperse AuNP has low complexity and is therefore easier to deconvolute from the scattering intensity data of the octahedron pyrochlore. However, it is more difficult to deconvolute the icosahedron form factor’s contribution to the scattering intensity because of its more complex features. It is challenging to determine the capability of the instrument to measure the complex features of the icosahedral form factor. Therefore, the data from the icosahedral pyrochlore is displayed as I(q) rather than S(q).

\subsubsection{Small-Angle X-Ray Scattering (SAXS) Data Acquisition of DNA Origami Lattices}

Origami assemblies were measured in solution at the Complex Materials Scattering (CMS, 11-BM) beamline at the National Synchrotron Light Source II (NSLS-II), Brookhaven National Laboratory (BNL). The photon energy of the beam at CMS is $13.5$ keV with a beam size of $200$ $\mu$m $\times$ $200$ $\mu$m with an approximate flux of $1011$ photons/sec. The sample-to-detector distance was set at $5.05$ m. The detector used for the measurements was a Pilatus $1$M with a pixel size of $172$ $\mu$m $\times$ $172$ $\mu$m. Samples were loaded into borosilicate glass capillaries that were then sealed with glue or wax. The bottom of the capillaries were then probed at the bottom of the capillary where origami lattices had settled. Two-dimensional scattering patterns were captured on are detectors downstream of the sample. The two-dimensional scattering patterns were then converted into one-dimensional scattering intensity profiles ($I(q)$) via azimuthal integration, as a function of the scattering vector $q$, where  $q = \frac{4 \pi}{\lambda} \sin \left( \frac{\theta}{2}\right)$. $\lambda$ and $\theta$ are the wavelength of the incident x-rays and the full scattering angle, respectively. The resulting I(q) curves spanned from around 0.04 nm$^{–1}$ to 1 nm$^{–1}$ with a resolution of 0.002 nm$^{–1}$, in reciprocal space.

\subsubsection{SAXS modeling}
\label{sec:saxs_modeling}
SAXS modeling was primarily performed using ScatterSim software, a python software package for simulating the 1D curves associated with crystalline superlattices built from arbitrary nano-objects \cite{yager2014periodic}. 

\textbf{Modeling Polydisperse Gold Nanospheres:}
	The model for polydisperse AuNP was built using “PolydisperseNanoObject” in the ScatterSim package \cite{yager2014periodic}. This generates a specified object while varying a key parameter of the object. In this case, “PolydisperseNanoObject” was given a gold sphere generated by “SphereNanoObject”. We used ($r_{\rm AuNP}$) must be supplied and here, $r_{\rm AuNP}$ = $4.99$ nm. This radius was determined by analytically fitting the underlying form factor of intensity data gathered at the beamtime with spheres of different size. “PolydisperseNanoObject” requires an additional parameter, $\sigma_r$, which accounts for the standard deviation of the sphere. In this work, the $\sigma_r$ of the AuNP was found by a fit to be 0.483 nm. The scattering length density (SLD) of the aqueous environment was
$SLD_{\rm water} = 9.43 \times 10^{-6}$ \AA$^{-2}$
and the SLD of gold is
$SLD_{\rm gold} = 119.16 \times 10^{-6}$ \AA$^{-2}$. With these parameters, the form factor of AuNP can be calculated using:
\begin{equation}
    F_{\rm sphere} \left( q, R, \rho' \right) = \frac{4 \pi}{3} R^3 \rho \frac{sin (q R) - q R cos (q R) }{ (q R)^3}
\end{equation}
where $\rho'$ is the scattering contrast $SLD_{\rm gold} - SLD_{\rm water}$.
We then applying polydispersity dependent on radius $r$ with average value $\overline{r}$ with its standard deviation $\sigma_r$ 
\begin{equation}
\langle F(q) \rangle_r = \int P \left( h,\overline{r}, \sigma_r \right) F_{\rm sphere} (q,h,\rho') dh
\end{equation}
and 
\begin{equation}
\langle \left| F(q) \right|^2 \rangle_r = \int P(h) \left| F_{\rm sphere} (q,h,\rho')  \right|^2 dh 
\end{equation}
and where the Gaussian distribution is 
\begin{equation}
P(r,\overline{r},\sigma_r) = \frac{1}{\sqrt{2 \pi \sigma^2_r}}e^\frac{{- (r - \overline{r} )^2}}{2 \sigma_r^2} .
\end{equation}
In ScatterSim this is accomplished by randomly creating spheres with radii that are sampled from the corresponding probability distribution and averaging their resultant form factor amplitudes or intensities ($F(q)$ and $|F(q) |^2$ respectively).

\textbf{Modeling Octahedral DNA origami:}
The octahedron model was built using “OctahedronCylindersNanoObject in the ScatterSim package. This generates an octahedral DNA frame where each of six helix bundles (6HB) at the 12 edges of the octahedron frame is approximated as a cylinder with a specified radius ($r_{\rm 6HB}$) and height ($h_{\rm 6HB}$). The parameters used in this work were as follows: $r_{\rm 6HB} = 4.156$ nm, $h_{\rm 6HB} = 27.79 $ nm. Additionally, an extrapolated edge length ($L_{\rm octahedron}$ ) of the DNA octahedron must be supplied, since the true shape of the origami is a truncated octahedron. In this work, the edge length was $L_{\rm octahedron} = 38.34$ nm. The SLD of 6HB is approximated as that of double stranded DNA (dsDNA), $SLD_{\rm dsDNA} = 11 \times 10^{-6} $ \AA$^{-2}$. These parameters agree with both the design and known physical features of the octahedron design as well as previous work \cite{tian2020ordered,wang2021designed}. 
The octahedral DNA origami loaded with a polydisperse gold nanoparticle (Octa-AuNP) was treated as a non-overlapping “composite” object of the two different components. This is possible to model in ScatterSim as a “CompositeNanoObject”, where the interference between sub-components is accounted for. The SAXS modeling with these objects did not include distortions of the octahedron shapes and polydispersity of their constituent cylinder components. Adding these distortions introduces many additional parameters, increases model complexity, and increases calculation times. Additionally, the resulting model has negligible improvement in fit over the results shown in this work.

\textbf{Modeling Icosahedral DNA origami:}
	The icosahedral model was built using “CylinderNanoObject” in ScatterSim to generate each individual four helix bundle (4HB) at each edge of the icosahedron. Each cylinder was then rotated and translated to the positions corresponding to the edge positions of the icosahedron. All 30 cylinders were then combined through “CompositeNanoObject”. Each cylinder was set with a specified radius and height of $r_{\rm 4HB} = 2.5$ nm, $h_{\rm 4HB}$ = 21.42 nm. The extrapolated edge length of the icosahedron was set to $L_{\rm icosahedron} = 27.86$ nm. These parameters are aligned with the dimensions simulated via oxDNA and the previously reported design \cite{zhang2022spatially}. The SLD of the 4HB is approximated as SLD$_{\rm dsDNA}$.

\textbf{Modeling Crystal Lattices of DNA origami:}
	To model the pyrochlore lattices of DNA origami, the Octa-AuNP or icosahedron composites were used as simple objects for the construction of the pyrochlore unit cell using ScatterSim. Once the unit cell is properly organized, the scattering intensity profile can be calculated via: 
\begin{equation}
I(q) = c Z_0(q)G(q) + P(q)(1-\beta(q) G(q))
\end{equation}
 where c is an overarching scaling factor, P(q) is the form factor contribution of the composite object which contains all objects in the lattice to the signal intensity, and the lattice factor is:
 \begin{equation}
Z_0(q) = \frac{1}{q^2} \sum_{hkl}^{m_{hkl}} \left| \sum_{j=1}^N \left\langle F_j(q_{hkl}) \right\rangle_{\epsilon} e^{2 \pi i (x_j h + y_j k +z_j l)} \right|^2 L(q - q_{hkl} )    
 \end{equation}
 where $L(q - q_{hkl})$ is a peak shape function. Lattice structure and symmetry were taken into account through sampling over the correspondent Miller indices $q_{hkl}$. $G(q)$ is the Debye-Waller factor which accounts for thermal vibrations within the lattice and is defined as:
 \begin{equation}
 G(q) =  e^{-q^2 \sigma^2_{\rm Drms}},
 \end{equation}
 where $\sigma^2_{\rm Drms}$ is the rms displacement of lattice elements. Finally 
 \begin{equation}
\beta(q) = \frac{ \left| \langle F(q) \rangle_{\epsilon} \right|^2}{\left\langle |F(q)|^2 \right\rangle_{\epsilon}}
 \end{equation}
 which comes from any polydispersity in parameter $\epsilon$. When thermal vibrations increase, the Debye-Waller factor, $G(q)$, results in reduced ordered scattering from $Z_0(q)$ and nd increased diffuse scattering from unordered elements. In this work, the effect of $\beta (q)$ was only considered for the radius of the AuNP.
SAXS modeling of lattice scattering is then carried out as follows: First, given certain lattice components, we used morphological and scattering parameters as described by designs and informed by those obtained from single-particle form factor analysis. This includes parameters like size, shape, and SLD. Next, we built lattice components and calculated P(q) and $\beta (q)$. The lattice components were then arranged into the unit cell of a pyrochlore lattice, and lattice constants were extracted by fitting the primary peak of the experimental data with the constructed lattice model. Finally, $G(q)$ , peak shape, scaling factors, and background were adjusted to best match modeled scattering with experimental measurements. In this work, no direct non-linear least squares fitting of intensity profiles was conducted because convergence of such fitting is limited due to the contribution to scattering of diffuse scattering elements and the time required by the complex architecture of a collection of composite objects.
\clearpage

\section{Additional experimental graphs and figures}
\label{suppfigs}
\begin{figure}[h]
\centering
	\includegraphics[width=0.8\textwidth]{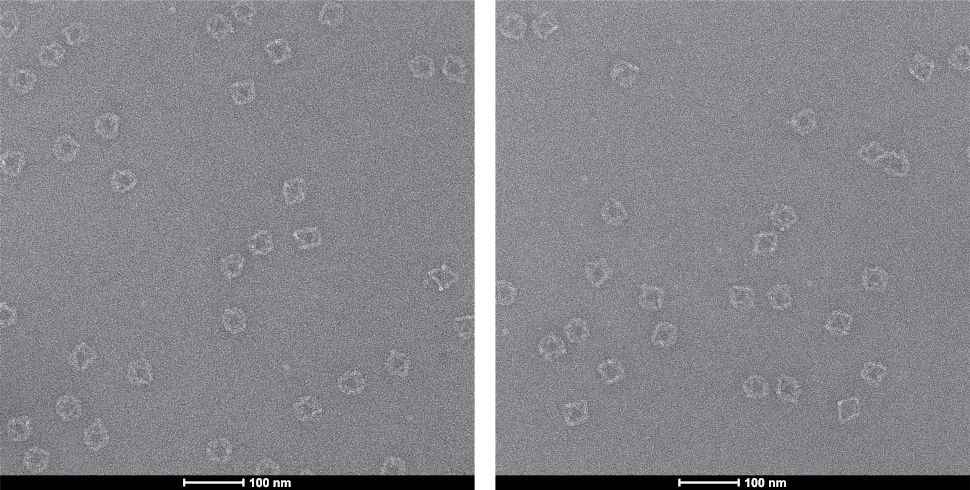}
	\caption{Negative stain TEM images of the octahedral DNA origami, purified with ultrafiltration.}
    \label{fig:supGelOctRZP}
 \centering
\end{figure}

\begin{figure}[h]
\centering
	\includegraphics[width=0.8\textwidth]{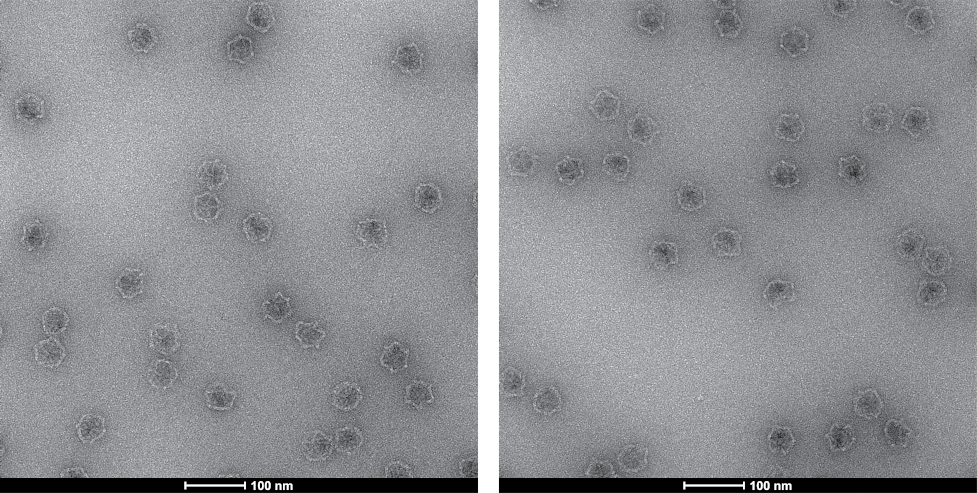}
	\caption{Negative stain TEM images of the icosahedral DNA origami, purified with rate zonal purification and therefore free from the malformed multimers that would potentially affect the lattice growth.}
    \label{fig:supGelIcoRZP}
 \centering
\end{figure}
\clearpage

\begin{figure}[h]
\centering
	\includegraphics[width=1.0\textwidth]{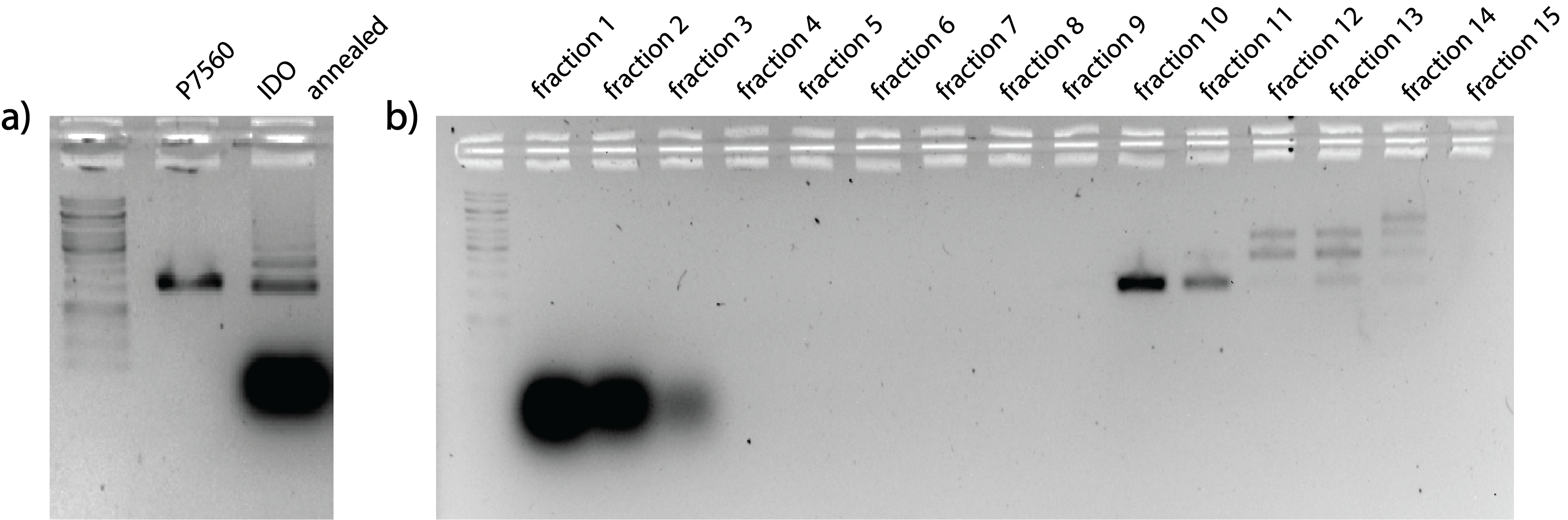}
	\caption{Agarose gel result on the synthesis and purification of IDO. a) Raw IDO from annealing shows a mixture of targeted structure with unwanted multi-mer/malformed structures above the targeted gel band. b) Purification of the annealed IDO with rate zonal centrifugation: gel result on all the fractions collected after the ultra-centrifugation. Batch to batch differences exist for the purification and in this case, only fraction 10 and 11 are collected for further assembly process.}
    \label{fig:supGelIcoGelRZP}
 \centering
\end{figure}
\clearpage

\begin{figure}[h]
\centering
	\includegraphics[width=0.7\textwidth]{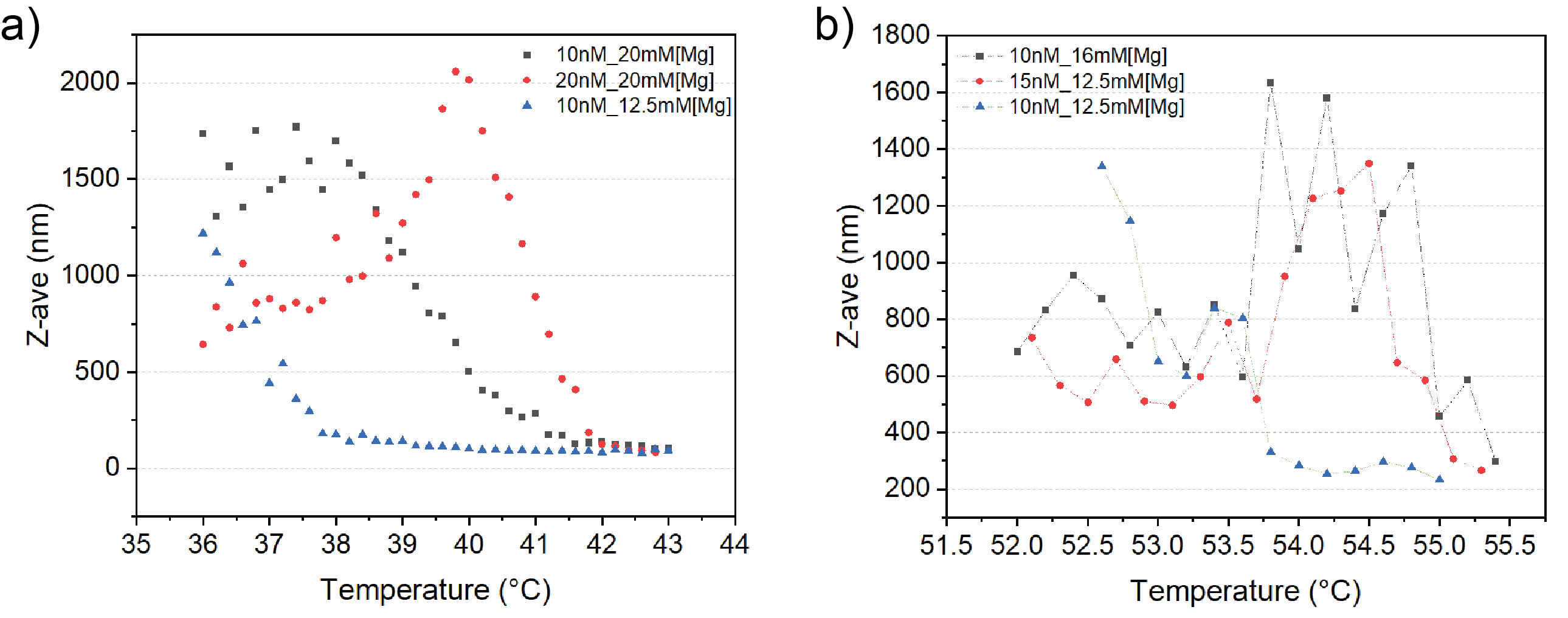}
	\caption{DLS spectrum of the self-assembly systems involved at different conditions. DLS measures the change of Z-average (intensity weighted mean hydrodynamic size of the ensemble) against temperature. In a typical experiment, the cuvette is equilibrated at the starting temperature for at least 5 min before the execution of program, which when initiated decreases the temperature at a rate of \SI{0.2}{\celsius} per 4 min. In the figure the spectra corresponding to a) icosahedron and b) octahedron building blocks are shown, with varied concentration of $\rm Mg^{2+}$ and the assembling unit themselves. Note in b) the data points are connected with dashed lines for better differentiation. A qualitative comparison shows that the concentration increase of either the units or $\rm Mg^{2+}$ increase the melting temperature and shrink the "annealing window" (the temperature starting from origamis just bind to each other to the point no further increasing of the assembly size is observed). It is noted that for the spectra of the lattice assembly with octahedron origami, the initial size is much higher than the one shown in panel a and also the expected monomer size. We reason that it is because of the strong binding strength of the handle to connect octahedron units, as the melting temperature of the lattice is close to \SI{54}{\celsius} and incubation of the system at \SI{55}{\celsius} for limited time does not dissociate all the connected units into dispersed monomers. In the meantime, the starting temperature can only be set at \SI{55}{\celsius} or lower as higher temperature will dissolve the origami itself. While we elongated the incubation time of the system at \SI{55}{\celsius} for up to 2 hours, we suspect still not all the building blocks will stay mono-dispersed for the assembly and may contribute to the final inferior quality of the pyrochlore lattice assembled  with the octahedron origami. }
    \label{fig:supdls}
 \centering
\end{figure}
\clearpage

\clearpage

\begin{figure}[h]
\centering
	\includegraphics[width=0.9\textwidth]{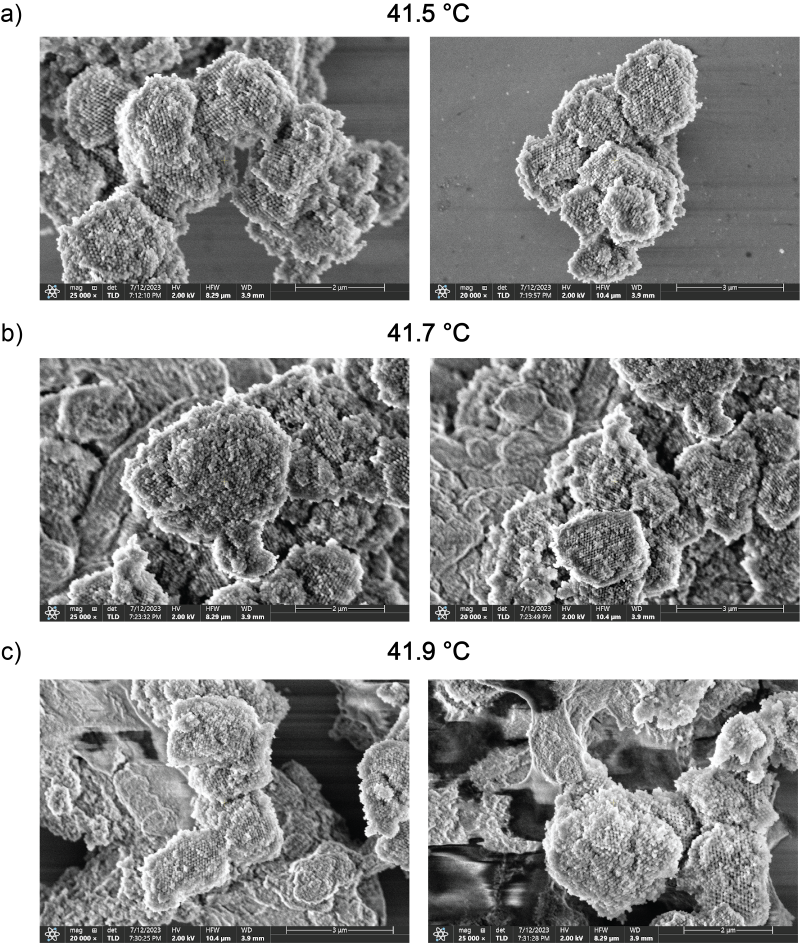}
	\caption{Isothermal annealing of the pyrochlore lattice with the icosahedral building blocks. The mixed samples are incubated at a higher temperature for the dissociation of the interactions and then allowed to be incubated at the specified temperature for a week. The temperature series in this figure is 41.5 \textdegree C,41.7 \textdegree C and 41.9 \textdegree C.}
    \label{fig:semSingleTempHigher}
 \centering
\end{figure}
\clearpage

\begin{figure}[h]
\centering
	\includegraphics[width=0.9\textwidth]{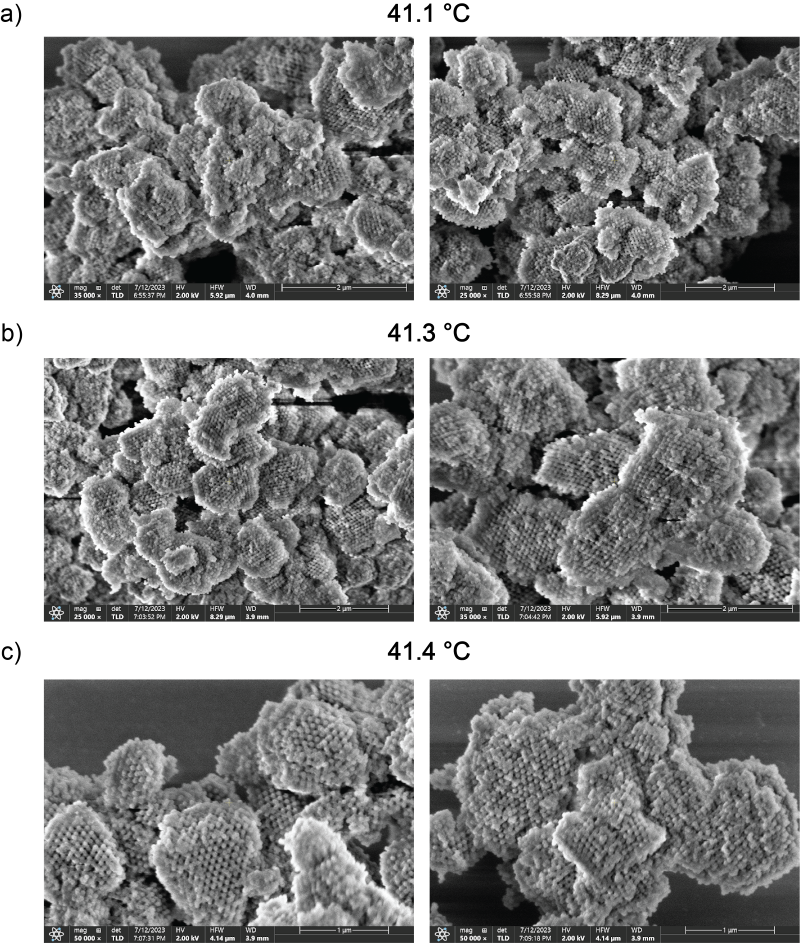}
	\caption{Isothermal annealing of the pyrochlore lattice with the icosahedral building blocks. The temperature series in this figure is 41.1 \textdegree C,41.3 \textdegree C and 41.4 \textdegree C.}
    \label{fig:semSingleTemplower}
 \centering
\end{figure}
\clearpage

\begin{figure}[h]
\centering
	\includegraphics[width=0.8\textwidth]{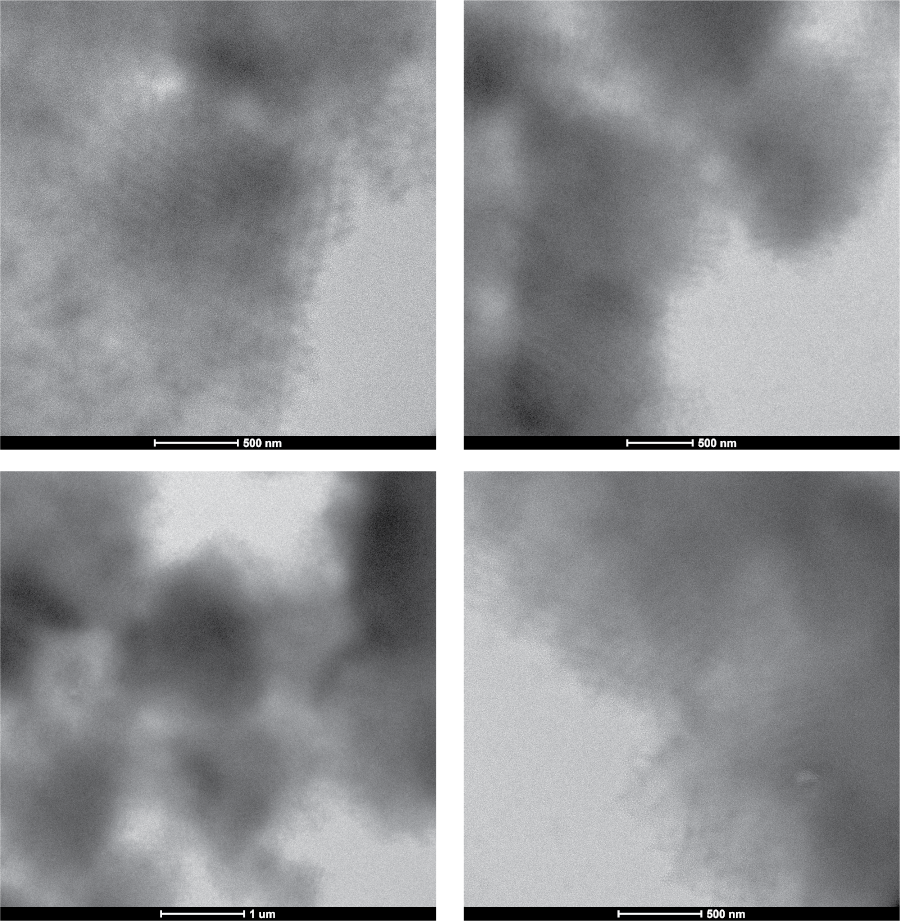}
	\caption{TEM imaging the silicated pyrochlore lattice assembled with icosahedron DNA origami. Despite the lack of negative staining, due to the elevated sample height and silica embedding, it is straightforward to identify and therefore examine the assembled lattices in low magnification mode. On the other hand, it becomes nearly impossible for the electron beam to penetrate through the superlattice, except for the small grains or the peripheral regions with reduced sample height. The images shown are taken based on such principle for the purpose of examining the result of the programmed assembly.    }
    \label{fig:temimages}
 \centering
\end{figure}
\clearpage

\begin{figure}[h]
\centering
	\includegraphics[width=0.8\textwidth]{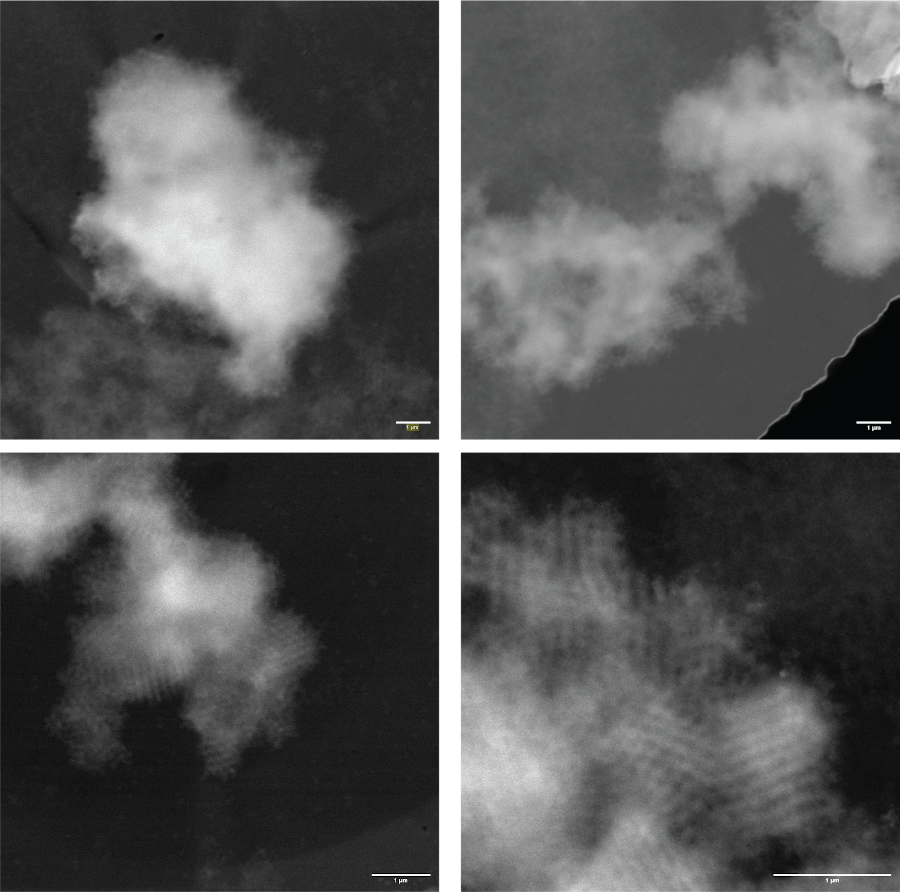}
	\caption{STEM imaging the silicated pyrochlore lattice assembled with icosahedron DNA origami in HAADF mode. While at a higher acceleration voltage and better contrast with STEM-HAADF, similar issues as mentioned with TEM imaging exist. Images are taken on a sample assembled in a less optimal condition in this case, in which small grains are more abundant to be imaged and examined. }
    \label{fig:stemtrials}
 \centering
\end{figure}
\clearpage

\begin{figure}[h]
\centering
	\includegraphics[width=1.0\textwidth]{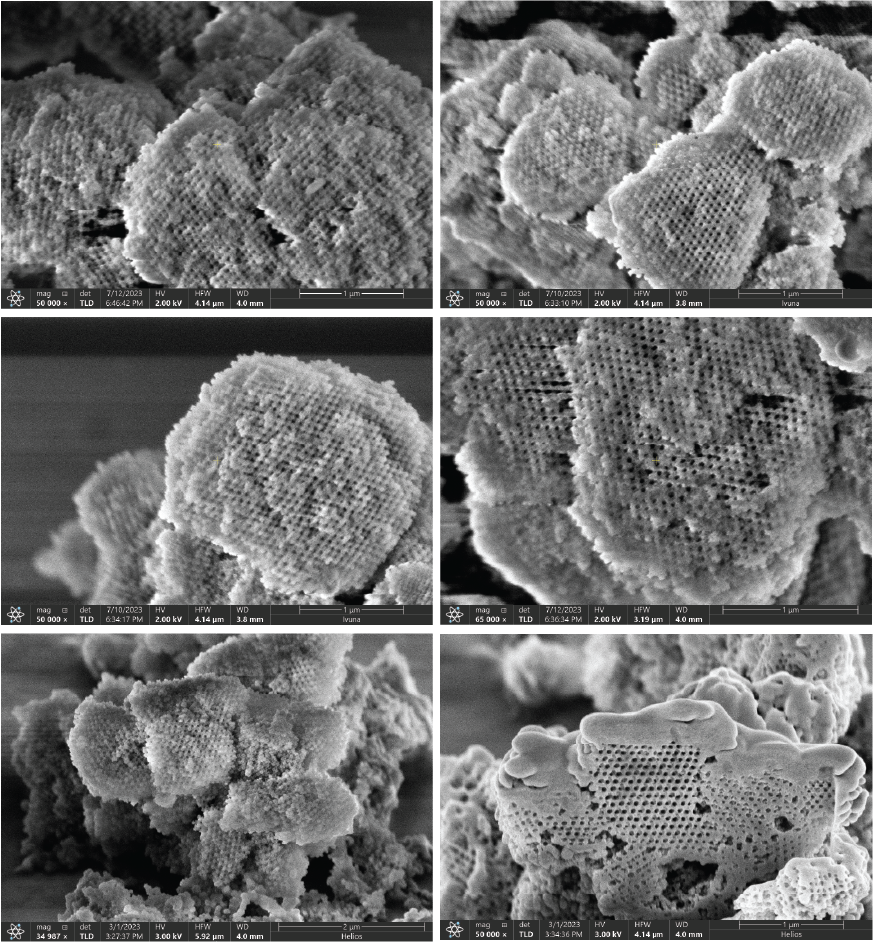}
	\caption{Additional figures for the ODO pyrochlore lattices with zoomed-in and out views and cross-sections of the lattices generated by FIB. The blur of the edges and boundary are likely caused by the charging effect and possibly slightly overcoating of the silica.}
    \label{fig:supoctlattice}
 \centering
\end{figure}
\clearpage


\begin{figure}[h]
\centering
	\includegraphics[width=1.0\textwidth]{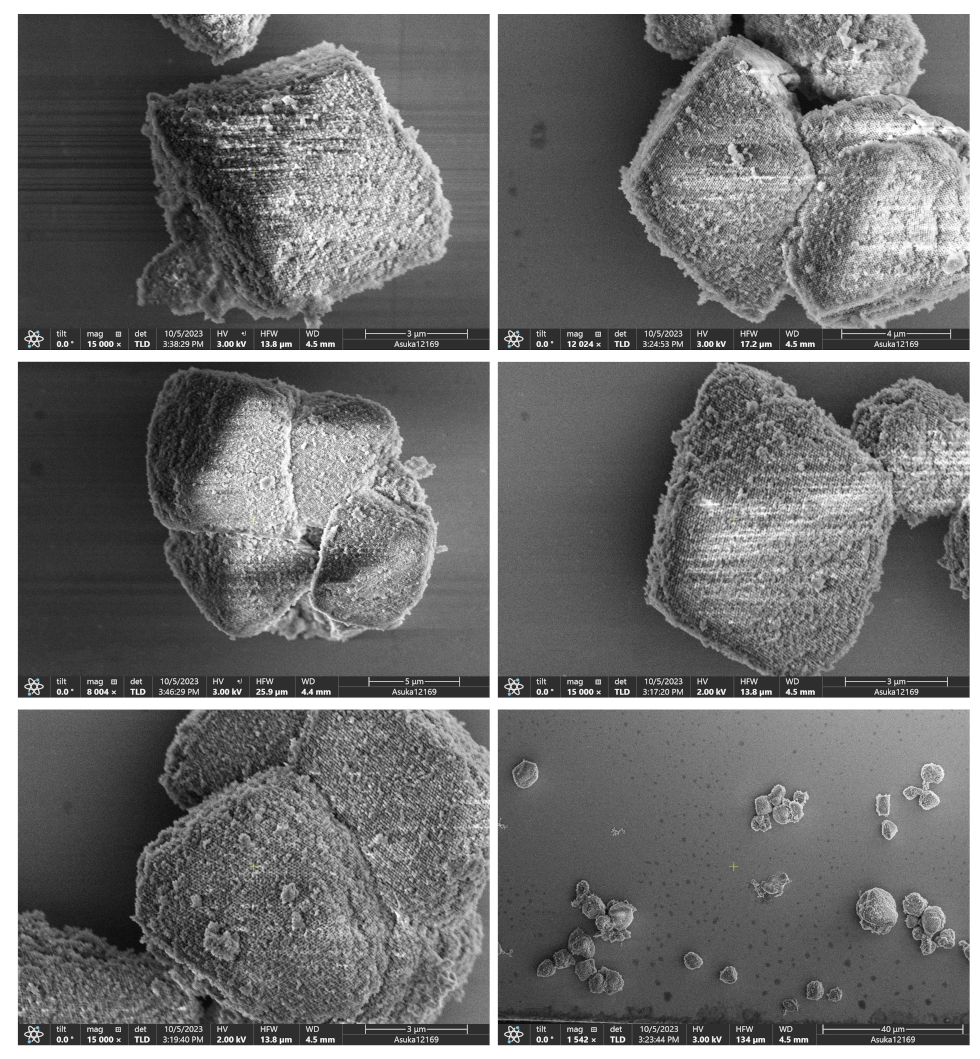}
	\caption{Additional figures for the IDO pyrochlore lattices annealed with thermocycler and embedded in silica with zoomed-in views and zoomed-out view (bottom right). Charging effect is observed as no metal was sputtered onto the surface.}
    \label{fig:supicolattice}
 \centering
\end{figure}
\clearpage

\begin{figure}[h]
\centering
	\includegraphics[width=1.0\textwidth]{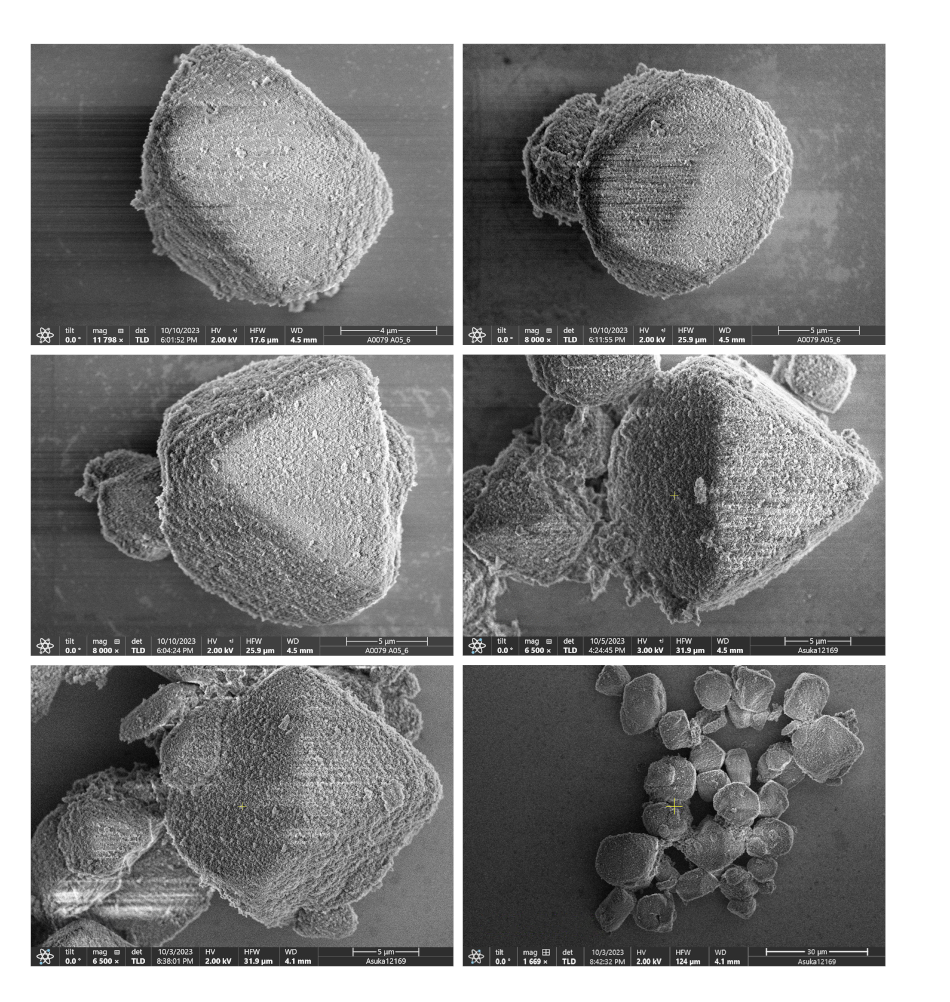}
	\caption{Additional figures for the IDO pyrochlore lattices annealed within a thermal incubator while actively mixed and further embedded in silica with zoomed-in views and zoomed-out view (bottom right). Charging effect is observed as no metal was sputtered onto the surface.}
    \label{fig:supicolattice}
 \centering
\end{figure}
\clearpage

\section{DNA Sequences}
\label{sec:sequences}
\subsection{Sequence design}

In order to find a set of 24 sequences (12 complementary pairs) for the design consisting of 4 particle species, each with 6 unique patches, we have first enumerated all possible duplex sequences of length 7. We have then selected a subset which had nearly the same binding-free energy according to the SantaLucia nearest-neighbor model \cite{santalucia2004thermodynamics}. We removed any sequences that would be predicted to form a stable secondary structure. We then considered all possible pairs of the overhang sequences (which were not complementary to each other) and used NUPACK \cite{zadeh2011nupack} to find the lowest free energy of binding that these non-complementary sequences can form. We have then selected a subset of 12 pairs of sequences so that the lowest free-energy of a duplex formed between two non-complementary sequences was as high as possible across the pairs considered.  The resulting handle sequences have been incorporated into a staple strand with poly-T overhangs. All these sequences have been placed at the 3' end of the staple strands. The sequences are listed as Handles" in the following tables, that list all the sequences that compose each DNA origami particle. The staple strands used in all the DNA origami species are the same except for the staples sequences that include the handles.

\subsection{Icosahedral DNA origami}
\subsubsection{Staple sequences}

\begin{spacing}{1.35}\centering
\begin{longtable}{@{} |p{.16\textwidth}|p{.84\textwidth}| @{}} 

\hline
\endhead
\hline
\endfoot
\endlastfoot

core\_strand\_5 &	\seqsplit{ATTAAACCTTTTTATTAGCCCAATAGGAACCC} \\ \hline
core\_strand\_10 &	\seqsplit{GAAGTGTTTCTGTCCAGATAAGTCCTGAACAA} \\ \hline
core\_strand\_17 &	\seqsplit{ATCTAAAGTTTTCTGTCGAGCACGTATAACGT} \\ \hline
core\_strand\_24 &	\seqsplit{GAAAGCCGGAAAGGAGAAGTACGGTGTCTGGA} \\ \hline
core\_strand\_31 &	\seqsplit{AGCCCCAATTGTAAACACTACGTGAACCATCA} \\ \hline
core\_strand\_39 &	\seqsplit{AGGCGGATCCCGGAATGTAGCGACAGAATCAA} \\ \hline
core\_strand\_50 &	\seqsplit{TGACAACAGGATTAGGCTCAGCAGCGAAAGAC} \\ \hline
core\_strand\_56 &	\seqsplit{AAGCGAACTAGAGAGTAAAAAAAAGGCTCCAA} \\ \hline
core\_strand\_63 &	\seqsplit{ATACGTAAGAGGCAAACTTTACCCTGACTATT} \\ \hline
core\_strand\_70 &	\seqsplit{GCATTAACTAGACTGGTACCAAAAACATTATG} \\ \hline
core\_strand\_76 &	\seqsplit{ATAACCTGAAAAGGTGTCAGGTCATTGCCTGA} \\ \hline
core\_strand\_83 &	\seqsplit{ATCAAAATCCCTCAGATGGTAATAAGTTTTAA} \\ \hline
core\_strand\_93 &	\seqsplit{CAGACCAGTGATATTCTGCTCATTCAGTGAAT} \\ \hline
core\_strand\_98 &	\seqsplit{GCAACACTAACCAAAACCGCGACCTGCTCCAT} \\ \hline
core\_strand\_105 &	\seqsplit{CCTTATGCGAGCGGAACATTATTACAGGTAGA} \\ \hline
core\_strand\_112 &	\seqsplit{GTTGTAAACGTTATTATAAGTGTCCTTAGTGC} \\ \hline
core\_strand\_118 &	\seqsplit{ACTCCAGCGGTGAGAAGCGATCGGTGCGGGCC} \\ \hline
core\_strand\_126 &	\seqsplit{AGAGATAAGAACCACCAAGAAAAGTAAGCAGA} \\ \hline
core\_strand\_136 &	\seqsplit{ACAATAACAAAGTCAGAAGATGATGAAACAAA} \\ \hline
core\_strand\_142 &	\seqsplit{ACAACTAAAGGATTTACAAAATTATTTGCACG} \\ \hline
core\_strand\_149 &	\seqsplit{GAGTGAATAATTTTCCGAACCTCAAATATCAA} \\ \hline
core\_strand\_156 &	\seqsplit{CGCATTTCAACCACCACTCGATAAAGACGGAG} \\ \hline
core\_strand\_162 &	\seqsplit{CCGACAGTGGGCACGAGACCGTAATGGGATAG} \\ \hline
core\_strand\_169 &	\seqsplit{ATTACGCAAAGGTGGCTTATTTATCCCAATCC} \\ \hline
core\_strand\_179 &	\seqsplit{TTTCATCTTCAAGATTTTACTAGAAAAAGCCT} \\ \hline
core\_strand\_184 &	\seqsplit{GAAAGCGTGGCTATTATCCGGCTTAGGTTGGG} \\ \hline
core\_strand\_191 &	\seqsplit{AACAACGCCCAGTAATCAATCGTCTGAAATGG} \\ \hline
core\_strand\_198 &	\seqsplit{TTTCTTTTCCTGAGTAGCCCCAGCAGGCGAAA} \\ \hline
core\_strand\_204 &	\seqsplit{TTAAATTTAACGCCATACTGCCCGCTTTCCAG} \\ \hline
core\_strand\_212 &	\seqsplit{CCCAATAGATTCTAAGTCAACCGATTGAGGGA} \\ \hline
core\_strand\_3 &	\seqsplit{TGTAGCATCGCCACCCTAACACTGCCGCCACCGAGAACAA} \\ \hline
core\_strand\_11 &	\seqsplit{AATAATCGGACGACAAATATCCCAGTAATTCTCCACCGAG} \\ \hline
core\_strand\_18 &	\seqsplit{AAAGGGATACCACACCCGTTAGAACACGCTGCGACGTTAG} \\ \hline
core\_strand\_25 &	\seqsplit{GATTTAGTAGAGCTTATCCATATATTTTTGCGAGGAAGGG} \\ \hline
core\_strand\_34 &	\seqsplit{GGAACCCTAAGAACGTAAGTTTTTCAAGAGTCTAAGCAAA} \\ \hline
core\_strand\_43 &	\seqsplit{GGTCATAGCATTAGCATTAGCGTCCCAGTAGCGTTGATAT} \\ \hline
core\_strand\_48 &	\seqsplit{AAGACTTTTATTCGGTAACGAGGGAGAGGCTGAGACTCCT} \\ \hline
core\_strand\_57 &	\seqsplit{ATTTCTTAAGAAAGGATTTAATTGTCAGCGGATCCAACAG} \\ \hline
core\_strand\_64 &	\seqsplit{CCGAAAGACTCAAATGAAGCAAAGTATTCATTCCAACCTA} \\ \hline
core\_strand\_71 &	\seqsplit{TAAAAATTCAAAGAATATACTTTTCAGAGGGGGTAATAGT} \\ \hline
core\_strand\_78 &	\seqsplit{AGCATGTCAGCTGATAAGCAAACAGATATTCATTTGGGGC} \\ \hline
core\_strand\_86 &	\seqsplit{CCCCTGCCTGAATTTAGTGCCTTGTGGAAAGCACCACCGG} \\ \hline
core\_strand\_91 &	\seqsplit{TTGAGATGCAAGAGTACCCTGACGTGAGGCAGGTCAGACG} \\ \hline
core\_strand\_99 &	\seqsplit{GAACTGACCGCGAAACGCCGGAACCAGCGATTACCAGACG} \\ \hline
core\_strand\_106 &	\seqsplit{GATACATACCAGTCAGTCAGTTGAGCGGAACAAAGAAACC} \\ \hline
core\_strand\_113 &	\seqsplit{CTAATCTACAGGAGAACAACCTTACTCGTATTAAATCCTT} \\ \hline
core\_strand\_121 &	\seqsplit{GATTAAGTGCCGGAAAATTACGCCAATGTGTAGGTAAAGA} \\ \hline
core\_strand\_129 &	\seqsplit{ACGGAATAATAATAAGCAAAGTTACCACCACCCTCAGAGC} \\ \hline
core\_strand\_134 &	\seqsplit{ACCTTTTTCCAAGTTAAAACAAAAGAGAATTAACTGAACA} \\ \hline
core\_strand\_143 &	\seqsplit{GATGAATAATCCTGATAAATAAAGGCAATTCAATAGATAA} \\ \hline
core\_strand\_150 &	\seqsplit{AATTGAGGAACAGTGCTCAATATCTATTAACAATCGTCGC} \\ \hline
core\_strand\_157 &	\seqsplit{TTCCTGTGGACCTCCTGGTACCGACATCGCCATTAAAAAT} \\ \hline
core\_strand\_164 &	\seqsplit{TTGAGGGGTTAAATGTGGTGTAGACCAGCTTTTAAGCAAC} \\ \hline
core\_strand\_172 &	\seqsplit{CCTTTACACGCTAACGAACGATTTTCCTGAATAGAAAATA} \\ \hline
core\_strand\_177 &	\seqsplit{ACGCTCAAAATACCGATCATATGCTGCGGGAGGTTTTGAA} \\ \hline
core\_strand\_185 &	\seqsplit{AAGAACGCAAGAGTCACTATATGTAGATTAAGGACAATAT} \\ \hline
core\_strand\_192 &	\seqsplit{GGGACATTACCGCCAGCATTGGCATATCCAGACAGAGGCA} \\ \hline
core\_strand\_199 &	\seqsplit{CAAAAGAAATTGCCCTTGATGGTGTTTGATTAGTAATAAC} \\ \hline
core\_strand\_207 &	\seqsplit{CGGGGAGATGTAAAGCCCTGTCGTAGCCGGAATTTTTTAA} \\ \hline
core\_strand\_215 &	\seqsplit{ACCGTCACCACAATCAAAATATTGAATAAGTTAGAAGGCT} \\ \hline
core\_strand\_6 &	\seqsplit{ATGTACCGTCAGAACCCACTCATCCTCAGAACTCCACAGAGTTTAGTAAGTTTCGT} \\ \hline
core\_strand\_12 &	\seqsplit{GAAAAATATAAACAACCAGTGAGGGTCCAGACGCTGTCTTAAGGTAAATCCTAATT} \\ \hline
core\_strand\_19 &	\seqsplit{GCTTTCCTCGCCGCGCTCTTTCCAGCGTAACCTTTAGACAGTAGCGGTTCAGAGCG} \\ \hline
core\_strand\_26 &	\seqsplit{AGTTTCATATTGCTGAGGCGAGAAGATGGCTTTTGACCATAGAGGTCAACAGTTGA} \\ \hline
core\_strand\_35 &	\seqsplit{CCCAAATCGGACTCCAAGATTGTACACTATTAAAAGGGAGGTTTGGAATGGGGTCG} \\ \hline
core\_strand\_42 &	\seqsplit{GTTTGCCTAGGCCGGATCGAGAGGACCATTACCCCCCTTACAAAATCAAGACTGTA} \\ \hline
core\_strand\_51 &	\seqsplit{AGCATCGGCGCTGAGGCCACGCATAACCGATATTCATGAGAAGTATTATAGCAACG} \\ \hline
core\_strand\_58 &	\seqsplit{AAGGAGCCACAACTAAAAGCAAACGTGAGAATAACAGCTTCAACAGTTTATCGGTT} \\ \hline
core\_strand\_65 &	\seqsplit{ATAGTCAGCTTTAAACCGAAGGCAGAATCCCCCTTCAAATGTCATAAACGGATTGC} \\ \hline
core\_strand\_72 &	\seqsplit{ACCCTGTATAGCAAAAAATCATACAGGCAAGGTTTAGAACAGTTTTGCGCGGGAGA} \\ \hline
core\_strand\_79 &	\seqsplit{GAGTCTGGAATTAATGTATTTTCAACCGTTCTAATCATATATCAATATAGAGAATC} \\ \hline
core\_strand\_85 &	\seqsplit{CGGGGTCACCGTTCCACCAGAGCCGCAGTCTCTATTTCGGAGCCAGAAAGTAACAG} \\ \hline
core\_strand\_94 &	\seqsplit{AAGGCTTGATCTTGACGCTGGCTGACCTTCATGTTTAATTCAGGAGGTAGAAACAC} \\ \hline
core\_strand\_100 &	\seqsplit{GTTACTTAAAAGTACACCTCGTTTATACCAAGCAACTTTGTTGACCCCGAGGCGCA} \\ \hline
core\_strand\_107 &	\seqsplit{AAGATTCAGACGTTGGGAACTGGCTCATTATAACGCCAAAATCATTTTGATTTAGG} \\ \hline
core\_strand\_114 &	\seqsplit{TGAATTGTGCCAGGGTCAGTGCCAAGCTTTCTTTTACGCTTTCGACAATGACAATG} \\ \hline
core\_strand\_122 &	\seqsplit{TCTTCGCTCCAGGCAACGGCACCGCTTCTGGTTGGGTAACGCCTGAGTAGCTGGCG} \\ \hline
core\_strand\_128 &	\seqsplit{TAGCCGAAAGCAAGAAAATTGAGTTAAGCCCACCCAAAAGCCTCAGAGCCAGAAGG} \\ \hline
core\_strand\_137 &	\seqsplit{CATCAAGACAAAATCGCTGATTGCTTTGAATATAATGGAATTAGACGGTTAATTAC} \\ \hline
core\_strand\_144 &	\seqsplit{TAAAACAGTGTTTGGAAGCCGTCATCAATATATACAGTAAAGATGATGAAATTGCG} \\ \hline
core\_strand\_151 &	\seqsplit{ACCCTCAACACGCTGATTCTGTAACCGCCTGCAAGGTTATGCGGTCAGTGGTCAGT} \\ \hline
core\_strand\_158 &	\seqsplit{GATCCCCGGGTTGGTGCATTTCTCCGAACTCTTGAAATTGCCCTAAAAGCTCGAAT} \\ \hline
core\_strand\_165 &	\seqsplit{GTCACGTTGAGCGAGTGCCATCTGCATCAACAACGACGACTCCTGTAGTGGGCGCA} \\ \hline
core\_strand\_171 &	\seqsplit{AAATAAGAAGCGTCTTGCAAACGTCTTACCAAGAGAGAATCAATTTTATTTGTTTA} \\ \hline
core\_strand\_180 &	\seqsplit{GTTTAGTACCGTGTGAAAATTTAATGGTTTGACAGTAGGGTCCCGACTGTTATACA} \\ \hline
core\_strand\_186 &	\seqsplit{TTATATAAATAGTGAAGTGGCACAACGCTGAGGAGAAAACGATAGCTTAAATGCTG} \\ \hline
core\_strand\_193 &	\seqsplit{ATTATTTACCATTGCAAATTTAGGACAATATTCTGGCCAAGCTGGTAAGATTCACC} \\ \hline
core\_strand\_200 &	\seqsplit{ATCCTGTTTCACCGCCAGACGGGCAACAGCTGTAGCCCGAAATACTTCGTTCCGAA} \\ \hline
core\_strand\_208 &	\seqsplit{TCGGGAAACTGGGGTGTCAGCTCAGCATAAAGGGCGGTTTAACATACGGCCAGCTG} \\ \hline
core\_strand\_214 &	\seqsplit{GGGAAGGTATAGAAAATCAGATATTATTTTGTCGACTTGACACCACGGACGGAAAT} \\ \hline
core\_strand\_4 &	\seqsplit{GCAAGCCGAAGTACCGGCCACCCTCAGAGCCA} \\ \hline
core\_strand\_13 &	\seqsplit{TAAAAGAGTTTATAATATGTTCAGCTAATGCA} \\ \hline
core\_strand\_20 &	\seqsplit{TAAATGAATTTTGTCGTTAATGCGCCGCTACA} \\ \hline
core\_strand\_27 &	\seqsplit{AAGAAAGCGCGAACGTATATAATGCTGTAGCT} \\ \hline
core\_strand\_36 &	\seqsplit{TATTTAAAAAACAGGAACGTCAAAGGGCGAAA} \\ \hline
core\_strand\_45 &	\seqsplit{AAGTATAGAAGTGCCGAACGTCACCAATGAAA} \\ \hline
core\_strand\_49 &	\seqsplit{CAAGAGAAACCATCGCCTTGCAGGGAGTTAAA} \\ \hline
core\_strand\_59 &	\seqsplit{GTCAGGATCAGACCGGAGGAATTGCGAATAAT} \\ \hline
core\_strand\_66 &	\seqsplit{AAACGAAATGCCACTAAGTTCAGAAAACGAGA} \\ \hline
core\_strand\_73 &	\seqsplit{AAAATGTTATCCAATATTAAGCAATAAAGCCT} \\ \hline
core\_strand\_80 &	\seqsplit{GCGAGCTGTTTAGCTACCGGAGAGGGTAGCTA} \\ \hline
core\_strand\_88 &	\seqsplit{AACCGCCTCACCGGAAGTAAGCGTCATACATG} \\ \hline
core\_strand\_92 &	\seqsplit{ATTGGCCTGCGCATAGAAGAACCGGATATTCA} \\ \hline
core\_strand\_101 &	\seqsplit{ACGATAAAATCATAACACGGAGATTTGTATCA} \\ \hline
core\_strand\_108 &	\seqsplit{ACCAGAAGGATTTTAAGAAGAAAAATCTACGT} \\ \hline
core\_strand\_115 &	\seqsplit{TGCCCGAAACGACGGCGGATGTTCTTCTAAGT} \\ \hline
core\_strand\_123 &	\seqsplit{TTCAAAAGCAGCTTTCAGCGCCATTCGCCATT} \\ \hline
core\_strand\_131 &	\seqsplit{CGCCACCACCCACAAGACAATGAAATAGCAAT} \\ \hline
core\_strand\_135 &	\seqsplit{CCCTGAACGGATTCGCCGCAGAGGCGAATTAT} \\ \hline
core\_strand\_145 &	\seqsplit{TACATTTGTAGATTAGTTATACTTCTGAATAA} \\ \hline
core\_strand\_152 &	\seqsplit{TATTAATTAACCTTGCGAGCCAGCAGCAAATG} \\ \hline
core\_strand\_159 &	\seqsplit{ACCGAACGACATAAATTAATGAGTAAACAGGG} \\ \hline
core\_strand\_166 &	\seqsplit{TCGTCGGTGCGGCCCTAACAACCCGTCGGATT} \\ \hline
core\_strand\_174 &	\seqsplit{CATACATAGTATGTTATCCAGAGCCTAATTTG} \\ \hline
core\_strand\_178 &	\seqsplit{GCCTTAAATCTGACCTTAAATAAGGCGTTAAA} \\ \hline
core\_strand\_187 &	\seqsplit{TTTTGAATAAGAATACTTTATCAAAATCATAG} \\ \hline
core\_strand\_194 &	\seqsplit{TTTTCGAGCAACATGTACAGGAAAAACGCTCA} \\ \hline
core\_strand\_201 &	\seqsplit{ATCACTTGCACCAGTGTGGCCCTGAGAGAGTT} \\ \hline
core\_strand\_209 &	\seqsplit{CCAATAGGTTGTTAAACCTAATGAGTGAGCTA} \\ \hline
core\_strand\_217 &	\seqsplit{TATCCGGTCAAGCAAATTCATATGGTTTACCA} \\ \hline

corner\_strand\_1 &	\seqsplit{TATAATTTTTAGTACCGACAATCCTTATCATTTTTTTCCAAGAACGGGT} \\ \hline
corner\_strand\_2 &	\seqsplit{GCGCCAAAGACATTCAGGGATAGCATTCATCG} \\ \hline
corner\_strand\_7 &	\seqsplit{GGGCGCGTACTATACTACAACGCC} \\ \hline
corner\_strand\_8 &	\seqsplit{CCACCCTCATTTTAGAAACCAATC} \\ \hline
corner\_strand\_14 &	\seqsplit{ATCGGCAAAATTTTATCAACAATATCACGCA} \\ \hline
corner\_strand\_15 &	\seqsplit{GAACGCGCCTGTTGGAGGCCGATT} \\ \hline
corner\_strand\_16 &	\seqsplit{TCACCTTTTTGTACTCAGGAGCAGCCCTCATATTTTTGTTAGCGTAACG} \\ \hline
corner\_strand\_22 &	\seqsplit{TTTTTGAGAGATTGCGAACGAGTA} \\ \hline
corner\_strand\_23 &	\seqsplit{TTCCACCCCCGATTTATTTTTGAGCTTGACGGG} \\ \hline
corner\_strand\_30 &	\seqsplit{AGACATTTTTGTCAAATCACCGTACCCCGGTTTTTTTGATAATCAGAAA} \\ \hline
corner\_strand\_33 &	\seqsplit{CAACATGTTTTATAGCACTAAATC} \\ \hline
corner\_strand\_37 &	\seqsplit{ACTCACATTAATTGGGCGATGGCCCGTTAATA} \\ \hline
corner\_strand\_38 &	\seqsplit{GGGTTTTTTTTTGCTCAGTACC} \\ \hline
corner\_strand\_41 &	\seqsplit{AAAAGTTTTTAAACGCAAAGAGCCATTTGGGATTTTTATTAGA} \\ \hline
corner\_strand\_44 &	\seqsplit{GCTTTTGATGATTTCGGCATTTTC} \\ \hline
corner\_strand\_47 &	\seqsplit{CCATCGATAGCATGGGATCGTCACCATTAGCG} \\ \hline
corner\_strand\_52 &	\seqsplit{AATAATTTTTATCCTCATTAAAACCTATTATTTTTTTCTGAAA} \\ \hline
corner\_strand\_53 &	\seqsplit{ATGACCATAAATTTTTGAGGACTA} \\ \hline
corner\_strand\_54 &	\seqsplit{GGCCGCTTTTGCTTTTCGAGGTGA} \\ \hline
corner\_strand\_55 &	\seqsplit{CCAATTTTTTACTGCGGAATCATCGCGTTTTATTTTTATTCGAGCTTCA} \\ \hline
corner\_strand\_61 &	\seqsplit{AATTTTTTCACGTTAAGAGGAAGC} \\ \hline
corner\_strand\_62 &	\seqsplit{CATGAGAAGTTTCCATTTTTTTAAACGGGTAAA} \\ \hline
corner\_strand\_67 &	\seqsplit{GACGGTCAATCTCAAAAATCAGGTAGAATAC} \\ \hline
corner\_strand\_68 &	\seqsplit{AAAGGGGGATGTCAACGCAAGGA} \\ \hline
corner\_strand\_69 &	\seqsplit{TTCTATTTTTCTAATAGTAGTA} \\ \hline
corner\_strand\_77 &	\seqsplit{AACCGTCTATCATATCGTAAAACT} \\ \hline
corner\_strand\_81 &	\seqsplit{CAGAGCATAAAGTCTACAAAGGCTAGCATCAA} \\ \hline
corner\_strand\_82 &	\seqsplit{GCCAGTTAGCGTTTGCTTTTTCATCTTTTCATA} \\ \hline
corner\_strand\_84 &	\seqsplit{AAACCGAGGAATACAGGAGTGTACGCCGCCA} \\ \hline
corner\_strand\_87 &	\seqsplit{GCTACAGAGGCTACAGTTAATGC} \\ \hline
corner\_strand\_89 &	\seqsplit{ACTAATTTTTAACACTCATCTAAAGAGGACAGTTTTTATGAACGGTGTA} \\ \hline
corner\_strand\_90 &	\seqsplit{TGCCCGTATAATACGTAACAAAGCACAAACA} \\ \hline
corner\_strand\_95 &	\seqsplit{TAATAAAACGAATGTAAATTGGGC} \\ \hline
corner\_strand\_96 &	\seqsplit{TTACCCAAATCATATAAGGGAACC} \\ \hline
corner\_strand\_97 &	\seqsplit{AAGTTTTTTTTGAGTAACATTAGGAATTACGATTTTTGGCATAGTAAGA} \\ \hline
corner\_strand\_102 &	\seqsplit{AGCCTTTATTTTTTGTGTCGAAATTAGCGAG} \\ \hline
corner\_strand\_103 &	\seqsplit{TCGCCTGATAAATCAACTAATGCA} \\ \hline
corner\_strand\_104 &	\seqsplit{CCGCCTTTTTGCCAGCATTGATCAACTTTAATTTTTTCATTGTGAATTA} \\ \hline
corner\_strand\_110 &	\seqsplit{CTCCGTGGGAACTTAACCCCGCTT} \\ \hline
corner\_strand\_111 &	\seqsplit{GCAATGCCAGGGTTTTTTTTTCCCAGTCACGAC} \\ \hline
corner\_strand\_116 &	\seqsplit{AATACCACATTTATGCGCACGACTATTTTAA} \\ \hline
corner\_strand\_117 &	\seqsplit{AATTCTTTTTGCGTCTGGCCTAGTATCGGCCTTTTTTCAGGAAGATCGC} \\ \hline
corner\_strand\_119 &	\seqsplit{AGGCTTTTTTTTTGCAAAAGACCTCATATATTTTTTTTTAAAT} \\ \hline
corner\_strand\_120 &	\seqsplit{GGTTGTGAATTCTTGCTGCAAGGC} \\ \hline
corner\_strand\_124 &	\seqsplit{GATGAACGGTATCTGTTGGGAAGGAGGCCGG} \\ \hline
corner\_strand\_125 &	\seqsplit{TTGAGTTTTTCGCTAATATCAG} \\ \hline
corner\_strand\_130 &	\seqsplit{CCAGTTACAAAATACGCAATAATA} \\ \hline
corner\_strand\_133 &	\seqsplit{AGCTATCTTACCTCCTGAGCAAAAGAGGGTAA} \\ \hline
corner\_strand\_139 &	\seqsplit{AAAAATCTAAAGTTCATTTGAATT} \\ \hline
corner\_strand\_140 &	\seqsplit{TCATTTCAATTATGTTTAACGTCA} \\ \hline
corner\_strand\_141 &	\seqsplit{ATAAATTTTTACAGAGGTGAGCTAAAATATCTTTTTTTTAGGAGCACTA} \\ \hline
corner\_strand\_147 &	\seqsplit{TGGAAGGGTTAGTCAGTTGAAAGG} \\ \hline
corner\_strand\_148 &	\seqsplit{GCGCAACAGTACATAATTTTTATCAATATATGT} \\ \hline
corner\_strand\_154 &	\seqsplit{CATTAATGAATTGTCATAGCTGT} \\ \hline
corner\_strand\_155 &	\seqsplit{GGCCTTTTTTTGAATCGGCTGA} \\ \hline
corner\_strand\_163 &	\seqsplit{CAGGCTGCGCAATCATCTGCCAGT} \\ \hline
corner\_strand\_167 &	\seqsplit{CTTAAGCTACGTTAAACGGCGGATTATATAGG} \\ \hline
corner\_strand\_168 &	\seqsplit{CCCTCTTTTTAGAACCGCCACAACTGGCATGATTTTTTTAAGACTCCTT} \\ \hline
corner\_strand\_170 &	\seqsplit{TATTCATTAAATTAAACAGCCATAAACATAT} \\ \hline
corner\_strand\_173 &	\seqsplit{ATTTAACAATTTGAAAATAGCAG} \\ \hline
corner\_strand\_176 &	\seqsplit{ACGTCAAAAATTCCGGAATCATAAAGTTGCT} \\ \hline
corner\_strand\_181 &	\seqsplit{TGGAAATACCTATGTATAAAGCCA} \\ \hline
corner\_strand\_182 &	\seqsplit{TAAGAATAAACATATCGCAAGACA} \\ \hline
corner\_strand\_183 &	\seqsplit{TCAAATTTTTCTATCGGCCTTCAGAGATAGAATTTTTCCCTTCTGACCT} \\ \hline
corner\_strand\_188 &	\seqsplit{TCGTAATCATGTACCTTTTTAACCGTCTTTA} \\ \hline
corner\_strand\_189 &	\seqsplit{GTCTGAGAGACTTCAGTAATAAAA} \\ \hline
corner\_strand\_190 &	\seqsplit{GGCGTTTTTTTTTAGCGAACCCTTAATTGAGATTTTTATCGCCATATTT} \\ \hline
corner\_strand\_195 &	\seqsplit{TACGAGCATGTTCATTTTGACGCTAAGAGAA} \\ \hline
corner\_strand\_196 &	\seqsplit{AGGTGCCGTAATCCCTTATAAAT} \\ \hline
corner\_strand\_197 &	\seqsplit{CACACGCGTATTGGGCTTTTTGCCAGGGTGGTT} \\ \hline
corner\_strand\_202 &	\seqsplit{AGTCACACGACTCCACGCTGGTTTGAAGAAC} \\ \hline
corner\_strand\_203 &	\seqsplit{TTTTGTTTTTTTAAAATTCGCA} \\ \hline
corner\_strand\_205 &	\seqsplit{ATGCGTTTTTCGAACTGATAGTTATCCGCTCATTTTTCAATTC} \\ \hline
corner\_strand\_206 &	\seqsplit{GCAGCAAGCGGTTCGGCCAACGCG} \\ \hline
corner\_strand\_210 &	\seqsplit{TCGTAACCGTGTTGCGTTGCGCTCCAAAAAT} \\ \hline
corner\_strand\_211 &	\seqsplit{TAGGATTTTTATCATTACCGCG} \\ \hline
corner\_strand\_213 &	\seqsplit{AATTCTTACCATAAAGGGCGACATAACGCGA} \\ \hline
corner\_strand\_216 &	\seqsplit{GCGCGTTTTCATGGTGAATTATC} \\ \hline

\caption{Staple sequence for the icosahedral DNA origami} 

\label{tab:Icostaple}
\end{longtable}
\end{spacing}

\clearpage

\subsubsection{Handle sequences}

\begin{spacing}{1.35}\centering

\begin{longtable}{@{} |p{.18\textwidth}|  p{.82\textwidth} | @{}} 
\hline
\endhead
\hline
\endfoot
\endlastfoot

Handle\_A0\_1 &	\seqsplit{TTCCCAATTCTTTTGAAAATCTCCACCTTTATTTTTTTTTTTTTTTTGTGTGTC} \\ \hline
Handle\_A0\_2 &	\seqsplit{ATTGCTTTTTTCCTTTTGATATAGATACATTTTTTTTCGCAAATGGTCATTTTTTTTTTTTTTTTGTGTGTC} \\ \hline
Handle\_A0\_3 &	\seqsplit{ATCAAAAAGATTCTAAATCGGTTGATAGCGTTTTTTTTTTTTTTTTTGTGTGTC} \\ \hline
Handle\_A1\_1 &	\seqsplit{TTTGCTTTTTTAAACAACTTTGATACCGATAGTTTTTTTGCGCCGACAATTTTTTTTTTTTTTTTACATCGC} \\ \hline
Handle\_A1\_2 &	\seqsplit{TATCAGCTTGCTTGGTTGCTTTGAATGGGATTTTTTTTTTTTTTTTTACATCGC} \\ \hline
Handle\_A1\_3 &	\seqsplit{CACCAGTACAATGCACCGTAATCAAGGTGTATTTTTTTTTTTTTTTTACATCGC} \\ \hline
Handle\_A2\_1 &	\seqsplit{TAGGGTTTTTCGCTGGCAAGTGGAACGGTACGTTTTTCCAGAATCCTGATTTTTTTTTTTTTTTTATGTCGC} \\ \hline
Handle\_A2\_2 &	\seqsplit{GGAGCTAAACATAATATGCAACTACGGGCGCTTTTTTTTTTTTTTTTATGTCGC} \\ \hline
Handle\_A2\_3 &	\seqsplit{AATTATTTTTACCGTTGTAGCGATAGGGTTGATTTTTGTGTTGTTTTTTTTTTTTTTTTATGTCGC} \\ \hline
Handle\_A3\_1 &	\seqsplit{ATTTTTTTTTGCACCCAGCTAAACATAAAAACTTTTTAGGGAATTTTTTTTTTTTTTTCTTTCGGT} \\ \hline
Handle\_A3\_2 &	\seqsplit{ATGCAAATCCATCATCACCTTGCTCTTAGAATTTTTTTTTTTTTTTCTTTCGGT} \\ \hline
Handle\_A3\_3 &	\seqsplit{TCCTTTTTTTGAAAACATAGCTTTTTCAAATATTTTTTATTTTAGTTAATTTTTTTTTTTTTTTCTTTCGGT} \\ \hline
Handle\_A4\_1 &	\seqsplit{TCCCGCCAAAATAACCTACCATATGAAGTATTTTTTTTTTTTTTTTTGGAGCTA} \\ \hline
Handle\_A4\_2 &	\seqsplit{TGGCAAATCAATGGTGCTTGTTACGCAGAAGTTTTTTTTTTTTTTTTGGAGCTA} \\ \hline
Handle\_A4\_3 &	\seqsplit{TAGACTTTTTTTTACAAACAACGCCCTGGAGTTTTTTGACTCTATGATATTTTTTTTTTTTTTTTGGAGCTA} \\ \hline
Handle\_A5\_1 &	\seqsplit{TAGATTTTCAGTCTAACGGAACAATTATCATTTTTTTTTTTTTTTTTAGAGGCA} \\ \hline
Handle\_A5\_2 &	\seqsplit{CAGAACGAGTATGAAGCCCTTTTTACCAGAGTTTTTTTTTTTTTTTTAGAGGCA} \\ \hline
Handle\_A5\_3 &	\seqsplit{CATATTTTTTTCCTGATTATCCAGTACCTTTTTTTTTACATCGGGAGAATTTTTTTTTTTTTTTTAGAGGCA} \\ \hline
	
Handle\_B0\_1 &	\seqsplit{TTCCCAATTCTTTTGAAAATCTCCACCTTTATTTTTTTTTTTTTTTTAGCTTGC} \\ \hline
Handle\_B0\_2 &	\seqsplit{ATTGCTTTTTTCCTTTTGATATAGATACATTTTTTTTCGCAAATGGTCATTTTTTTTTTTTTTTTAGCTTGC} \\ \hline
Handle\_B0\_3 &	\seqsplit{ATCAAAAAGATTCTAAATCGGTTGATAGCGTTTTTTTTTTTTTTTTTAGCTTGC} \\ \hline
Handle\_B1\_1 &	\seqsplit{TTTGCTTTTTTAAACAACTTTGATACCGATAGTTTTTTTGCGCCGACAATTTTTTTTTTTTTTTGCGATGTA} \\ \hline
Handle\_B1\_2 &	\seqsplit{TATCAGCTTGCTTGGTTGCTTTGAATGGGATTTTTTTTTTTTTTTTGCGATGTA} \\ \hline
Handle\_B1\_3 &	\seqsplit{CACCAGTACAATGCACCGTAATCAAGGTGTATTTTTTTTTTTTTTTGCGATGTA} \\ \hline
Handle\_B2\_1 &	\seqsplit{TAGGGTTTTTCGCTGGCAAGTGGAACGGTACGTTTTTCCAGAATCCTGATTTTTTTTTTTTTTTCGGATTGT} \\ \hline
Handle\_B2\_2 &	\seqsplit{GGAGCTAAACATAATATGCAACTACGGGCGCTTTTTTTTTTTTTTTCGGATTGT} \\ \hline
Handle\_B2\_3 &	\seqsplit{AATTATTTTTACCGTTGTAGCGATAGGGTTGATTTTTGTGTTGTTTTTTTTTTTTTTTCGGATTGT} \\ \hline
Handle\_B3\_1 &	\seqsplit{ATTTTTTTTTGCACCCAGCTAAACATAAAAACTTTTTAGGGAATTTTTTTTTTTTTTTCGGAACAT} \\ \hline
Handle\_B3\_2 &	\seqsplit{ATGCAAATCCATCATCACCTTGCTCTTAGAATTTTTTTTTTTTTTTCGGAACAT} \\ \hline
Handle\_B3\_3 &	\seqsplit{TCCTTTTTTTGAAAACATAGCTTTTTCAAATATTTTTTATTTTAGTTAATTTTTTTTTTTTTTTCGGAACAT} \\ \hline
Handle\_B4\_1 &	\seqsplit{TCCCGCCAAAATAACCTACCATATGAAGTATTTTTTTTTTTTTTTTTAGCTCCA} \\ \hline
Handle\_B4\_2 &	\seqsplit{TGGCAAATCAATGGTGCTTGTTACGCAGAAGTTTTTTTTTTTTTTTTAGCTCCA} \\ \hline
Handle\_B4\_3 &	\seqsplit{TAGACTTTTTTTTACAAACAACGCCCTGGAGTTTTTTGACTCTATGATATTTTTTTTTTTTTTTTAGCTCCA} \\ \hline
Handle\_B5\_1 &	\seqsplit{TAGATTTTCAGTCTAACGGAACAATTATCATTTTTTTTTTTTTTTTACTTGACG} \\ \hline
Handle\_B5\_2 &	\seqsplit{CAGAACGAGTATGAAGCCCTTTTTACCAGAGTTTTTTTTTTTTTTTACTTGACG} \\ \hline
Handle\_B5\_3 &	\seqsplit{CATATTTTTTTCCTGATTATCCAGTACCTTTTTTTTTACATCGGGAGAATTTTTTTTTTTTTTTACTTGACG} \\ \hline
	
Handle\_C0\_1 &	\seqsplit{TTCCCAATTCTTTTGAAAATCTCCACCTTTATTTTTTTTTTTTTTTCTCGTTGT} \\ \hline
Handle\_C0\_2 &	\seqsplit{ATTGCTTTTTTCCTTTTGATATAGATACATTTTTTTTCGCAAATGGTCATTTTTTTTTTTTTTTCTCGTTGT} \\ \hline
Handle\_C0\_3 &	\seqsplit{ATCAAAAAGATTCTAAATCGGTTGATAGCGTTTTTTTTTTTTTTTTCTCGTTGT} \\ \hline
Handle\_C1\_1 &	\seqsplit{TTTGCTTTTTTAAACAACTTTGATACCGATAGTTTTTTTGCGCCGACAATTTTTTTTTTTTTTTACAATCCG} \\ \hline
Handle\_C1\_2 &	\seqsplit{TATCAGCTTGCTTGGTTGCTTTGAATGGGATTTTTTTTTTTTTTTTACAATCCG} \\ \hline
Handle\_C1\_3 &	\seqsplit{CACCAGTACAATGCACCGTAATCAAGGTGTATTTTTTTTTTTTTTTACAATCCG} \\ \hline
Handle\_C2\_1 &	\seqsplit{TAGGGTTTTTCGCTGGCAAGTGGAACGGTACGTTTTTCCAGAATCCTGATTTTTTTTTTTTTTTGACACACA} \\ \hline
Handle\_C2\_2 &	\seqsplit{GGAGCTAAACATAATATGCAACTACGGGCGCTTTTTTTTTTTTTTTGACACACA} \\ \hline
Handle\_C2\_3 &	\seqsplit{AATTATTTTTACCGTTGTAGCGATAGGGTTGATTTTTGTGTTGTTTTTTTTTTTTTTTGACACACA} \\ \hline
Handle\_C3\_1 &	\seqsplit{ATTTTTTTTTGCACCCAGCTAAACATAAAAACTTTTTAGGGAATTTTTTTTTTTTTTTCCGCTTTA} \\ \hline
Handle\_C3\_2 &	\seqsplit{ATGCAAATCCATCATCACCTTGCTCTTAGAATTTTTTTTTTTTTTTCCGCTTTA} \\ \hline
Handle\_C3\_3 &	\seqsplit{TCCTTTTTTTGAAAACATAGCTTTTTCAAATATTTTTTATTTTAGTTAATTTTTTTTTTTTTTTCCGCTTTA} \\ \hline
Handle\_C4\_1 &	\seqsplit{TCCCGCCAAAATAACCTACCATATGAAGTATTTTTTTTTTTTTTTTCGTCAAGT} \\ \hline
Handle\_C4\_2 &	\seqsplit{TGGCAAATCAATGGTGCTTGTTACGCAGAAGTTTTTTTTTTTTTTTCGTCAAGT} \\ \hline
Handle\_C4\_3 &	\seqsplit{TAGACTTTTTTTTACAAACAACGCCCTGGAGTTTTTTGACTCTATGATATTTTTTTTTTTTTTTCGTCAAGT} \\ \hline
Handle\_C5\_1 &	\seqsplit{TAGATTTTCAGTCTAACGGAACAATTATCATTTTTTTTTTTTTTTTACCGAAAG} \\ \hline
Handle\_C5\_2 &	\seqsplit{CAGAACGAGTATGAAGCCCTTTTTACCAGAGTTTTTTTTTTTTTTTACCGAAAG} \\ \hline
Handle\_C5\_3 &	\seqsplit{CATATTTTTTTCCTGATTATCCAGTACCTTTTTTTTTACATCGGGAGAATTTTTTTTTTTTTTTACCGAAAG} \\ \hline
	
Handle\_D0\_1 &	\seqsplit{TTCCCAATTCTTTTGAAAATCTCCACCTTTATTTTTTTTTTTTTTTGCAAGCTA} \\ \hline
Handle\_D0\_2 &	\seqsplit{ATTGCTTTTTTCCTTTTGATATAGATACATTTTTTTTCGCAAATGGTCATTTTTTTTTTTTTTTGCAAGCTA} \\ \hline
Handle\_D0\_3 &	\seqsplit{ATCAAAAAGATTCTAAATCGGTTGATAGCGTTTTTTTTTTTTTTTTGCAAGCTA} \\ \hline
Handle\_D1\_1 &	\seqsplit{TTTGCTTTTTTAAACAACTTTGATACCGATAGTTTTTTTGCGCCGACAATTTTTTTTTTTTTTTACAACGAG} \\ \hline
Handle\_D1\_2 &	\seqsplit{TATCAGCTTGCTTGGTTGCTTTGAATGGGATTTTTTTTTTTTTTTTACAACGAG} \\ \hline
Handle\_D1\_3 &	\seqsplit{CACCAGTACAATGCACCGTAATCAAGGTGTATTTTTTTTTTTTTTTACAACGAG} \\ \hline
Handle\_D2\_1 &	\seqsplit{TAGGGTTTTTCGCTGGCAAGTGGAACGGTACGTTTTTCCAGAATCCTGATTTTTTTTTTTTTTTGCGACATA} \\ \hline
Handle\_D2\_2 &	\seqsplit{GGAGCTAAACATAATATGCAACTACGGGCGCTTTTTTTTTTTTTTTGCGACATA} \\ \hline
Handle\_D2\_3 &	\seqsplit{AATTATTTTTACCGTTGTAGCGATAGGGTTGATTTTTGTGTTGTTTTTTTTTTTTTTTGCGACATA} \\ \hline
Handle\_D3\_1 &	\seqsplit{ATTTTTTTTTGCACCCAGCTAAACATAAAAACTTTTTAGGGAATTTTTTTTTTTTTTTATGTTCCG} \\ \hline
Handle\_D3\_2 &	\seqsplit{ATGCAAATCCATCATCACCTTGCTCTTAGAATTTTTTTTTTTTTTTATGTTCCG} \\ \hline
Handle\_D3\_3 &	\seqsplit{TCCTTTTTTTGAAAACATAGCTTTTTCAAATATTTTTTATTTTAGTTAATTTTTTTTTTTTTTTATGTTCCG} \\ \hline
Handle\_D4\_1 &	\seqsplit{TCCCGCCAAAATAACCTACCATATGAAGTATTTTTTTTTTTTTTTTTAAAGCGG} \\ \hline
Handle\_D4\_2 &	\seqsplit{TGGCAAATCAATGGTGCTTGTTACGCAGAAGTTTTTTTTTTTTTTTTAAAGCGG} \\ \hline
Handle\_D4\_3 &	\seqsplit{TAGACTTTTTTTTACAAACAACGCCCTGGAGTTTTTTGACTCTATGATATTTTTTTTTTTTTTTTAAAGCGG} \\ \hline
Handle\_D5\_1 &	\seqsplit{TAGATTTTCAGTCTAACGGAACAATTATCATTTTTTTTTTTTTTTTTGCCTCTA} \\ \hline
Handle\_D5\_2 &	\seqsplit{CAGAACGAGTATGAAGCCCTTTTTACCAGAGTTTTTTTTTTTTTTTTGCCTCTA} \\ \hline
Handle\_D5\_3 &	\seqsplit{CATATTTTTTTCCTGATTATCCAGTACCTTTTTTTTTACATCGGGAGAATTTTTTTTTTTTTTTTGCCTCTA} \\ \hline

\caption{Handle sequence for the icosahedral DNA origami} 

\label{tab:Icohandle}
\end{longtable}
\end{spacing}

\clearpage

\subsection{Octahedral DNA origami}

\subsubsection{Staple sequences}

\begin{spacing}{1.35}\centering
\begin{longtable}{@{} |p{.18\textwidth}|  p{.82\textwidth} | @{}} 
\hline
\endhead
\hline
\endfoot
\endlastfoot
core\_octa-1 &	\seqsplit{TCAAAGCGAACCAGACCGTTTTATATAGTC} \\ \hline
core\_octa-2 &	\seqsplit{GCTTTGAGGACTAAAGAGCAACGGGGAGTT} \\ \hline
core\_octa-3 &	\seqsplit{GTAAATCGTCGCTATTGAATAACTCAAGAA} \\ \hline
core\_octa-4 &	\seqsplit{AAGCCTTAAATCAAGACTTGCGGAGCAAAT} \\ \hline
core\_octa-5 &	\seqsplit{ATTTTAAGAACTGGCTTGAATTATCAGTGA} \\ \hline
core\_octa-6 &	\seqsplit{GTTAAAATTCGCATTATAAACGTAAACTAG} \\ \hline
core\_octa-7 &	\seqsplit{AGCACCATTACCATTACAGCAAATGACGGA} \\ \hline
core\_octa-8 &	\seqsplit{ATTGCGTAGATTTTCAAAACAGATTGTTTG} \\ \hline
core\_octa-9 &	\seqsplit{TAACCTGTTTAGCTATTTTCGCATTCATTC} \\ \hline
core\_octa-10 &	\seqsplit{GTCAGAGGGTAATTGAGAACACCAAAATAG} \\ \hline
core\_octa-11 &	\seqsplit{CTCCAGCCAGCTTTCCCCTCAGGACGTTGG} \\ \hline
core\_octa-12 &	\seqsplit{GTCCACTATTAAAGAACCAGTTTTGGTTCC} \\ \hline
core\_octa-13 &	\seqsplit{TAAAGGTGGCAACATAGTAGAAAATAATAA} \\ \hline
core\_octa-14 &	\seqsplit{GATAAGTCCTGAACAACTGTTTAAAGAGAA} \\ \hline
core\_octa-15 &	\seqsplit{GGTAATAGTAAAATGTAAGTTTTACACTAT} \\ \hline
core\_octa-16 &	\seqsplit{TCAGAACCGCCACCCTCTCAGAGTATTAGC} \\ \hline
core\_octa-17 &	\seqsplit{AAGGGAACCGAACTGAGCAGACGGTATCAT} \\ \hline
core\_octa-18 &	\seqsplit{GTAAAGATTCAAAAGGCCTGAGTTGACCCT} \\ \hline
core\_octa-19 &	\seqsplit{AGGCGTTAAATAAGAAGACCGTGTCGCAAG} \\ \hline
core\_octa-20 &	\seqsplit{CAGGTCGACTCTAGAGCAAGCTTCAAGGCG} \\ \hline
core\_octa-21 &	\seqsplit{CAGAGCCACCACCCTCTCAGAACTCGAGAG} \\ \hline
core\_octa-22 &	\seqsplit{TTCACGTTGAAAATCTTGCGAATGGGATTT} \\ \hline
core\_octa-23 &	\seqsplit{AAGTTTTAACGGGGTCGGAGTGTAGAATGG} \\ \hline
core\_octa-24 &	\seqsplit{TTGCGTATTGGGCGCCCGCGGGGTGCGCTC} \\ \hline
core\_octa-25 &	\seqsplit{GTCACCAGAGCCATGGTGAATTATCACCAATCAGAAAAGCCT} \\ \hline
core\_octa-26 &	\seqsplit{GGACAGAGTTACTTTGTCGAAATCCGCGTGTATCACCGTACG} \\ \hline
core\_octa-27 &	\seqsplit{CAACATGATTTACGAGCATGGAATAAGTAAGACGACAATAAA} \\ \hline
core\_octa-28 &	\seqsplit{AACCAGACGCTACGTTAATAAAACGAACATACCACATTCAGG} \\ \hline
core\_octa-29 &	\seqsplit{TGACCTACTAGAAAAAGCCCCAGGCAAAGCAATTTCATCTTC} \\ \hline
core\_octa-30 &	\seqsplit{TGCCGGAAGGGGACTCGTAACCGTGCATTATATTTTAGTTCT} \\ \hline
core\_octa-31 &	\seqsplit{AGAACCCCAAATCACCATCTGCGGAATCGAATAAAAATTTTT} \\ \hline
core\_octa-32 &	\seqsplit{GCTCCATTGTGTACCGTAACACTGAGTTAGTTAGCGTAACCT} \\ \hline
core\_octa-33 &	\seqsplit{AGTACCGAATAGGAACCCAAACGGTGTAACCTCAGGAGGTTT} \\ \hline
core\_octa-34 &	\seqsplit{CAGTTTGAATGTTTAGTATCATATGCGTAGAATCGCCATAGC} \\ \hline
core\_octa-35 &	\seqsplit{AAGATTGTTTTTTAACCAAGAAACCATCGACCCAAAAACAGG} \\ \hline
core\_octa-36 &	\seqsplit{TCAGAGCGCCACCACATAATCAAAATCAGAACGAGTAGTATG} \\ \hline
core\_octa-37 &	\seqsplit{GATGGTTGGGAAGAAAAATCCACCAGAAATAATTGGGCTTGA} \\ \hline
core\_octa-38 &	\seqsplit{CTCCTTAACGTAGAAACCAATCAATAATTCATCGAGAACAGA} \\ \hline
core\_octa-39 &	\seqsplit{AGACACCTTACGCAGAACTGGCATGATTTTCTGTCCAGACAA} \\ \hline
core\_octa-40 &	\seqsplit{GCCAGCTAGGCGATAGCTTAGATTAAGACCTTTTTAACCTGT} \\ \hline
core\_octa-41 &	\seqsplit{CCGACTTATTAGGAACGCCATCAAAAATGAGTAACAACCCCA} \\ \hline
core\_octa-42 &	\seqsplit{GTCCAATAGCGAGAACCAGACGACGATATTCAACGCAAGGGA} \\ \hline
core\_octa-43 &	\seqsplit{CCAAAATACAATATGATATTCAACCGTTAGGCTATCAGGTAA} \\ \hline
core\_octa-44 &	\seqsplit{AACAGTACTTGAAAACATATGAGACGGGTCTTTTTTAATGGA} \\ \hline
core\_octa-45 &	\seqsplit{TTTCACCGCATTAAAGTCGGGAAACCTGATTTGAATTACCCA} \\ \hline
core\_octa-46 &	\seqsplit{GAGAATAGAGCCTTACCGTCTATCAAATGGAGCGGAATTAGA} \\ \hline
core\_octa-47 &	\seqsplit{ATAATTAAATTTAAAAAACTTTTTCAAACTTTTAACAACGCC} \\ \hline
core\_octa-48 &	\seqsplit{GCACCCAGCGTTTTTTATCCGGTATTCTAGGCGAATTATTCA} \\ \hline
core\_octa-49 &	\seqsplit{GGAAGCGCCCACAAACAGTTAATGCCCCGACTCCTCAAGATA} \\ \hline
core\_octa-50 &	\seqsplit{GTTTGCCTATTCACAGGCAGGTCAGACGCCACCACACCACCC} \\ \hline
core\_octa-51 &	\seqsplit{CGCGAGCTTAGTTTTTCCCAATTCTGCGCAAGTGTAAAGCCT} \\ \hline
core\_octa-52 &	\seqsplit{AGAAGCAACCAAGCCAAAAGAATACACTAATGCCAAAACTCC} \\ \hline
core\_octa-53 &	\seqsplit{ATTAAGTATAAAGCGGCAAGGCAAAGAAACTAATAGGGTACC} \\ \hline
core\_octa-54 &	\seqsplit{CAGTGCCTACATGGGAATTTACCGTTCCACAAGTAAGCAGAT} \\ \hline
core\_octa-55 &	\seqsplit{ATAAGGCGCCAAAAGTTGAGATTTAGGATAACGGACCAGTCA} \\ \hline
core\_octa-56 &	\seqsplit{TGCTAAACAGATGAAGAAACCACCAGAATTTAAAAAAAGGCT} \\ \hline
core\_octa-57 &	\seqsplit{CAGCCTTGGTTTTGTATTAAGAGGCTGACTGCCTATATCAGA} \\ \hline
core\_octa-58 &	\seqsplit{CGGAATAATTCAACCCAGCGCCAAAGACTTATTTTAACGCAA} \\ \hline
core\_octa-59 &	\seqsplit{CGCCTGAATTACCCTAATCTTGACAAGACAGACCATGAAAGA} \\ \hline
core\_octa-60 &	\seqsplit{ACGCGAGGCTACAACAGTACCTTTTACAAATCGCGCAGAGAA} \\ \hline
core\_octa-61 &	\seqsplit{CAGCGAACATTAAAAGAGAGTACCTTTACTGAATATAATGAA} \\ \hline
core\_octa-62 &	\seqsplit{GGACGTTTAATTTCGACGAGAAACACCACCACTAATGCAGAT} \\ \hline
core\_octa-63 &	\seqsplit{AAAGCGCCAAAGTTTATCTTACCGAAGCCCAATAATGAGTAA} \\ \hline
core\_octa-64 &	\seqsplit{GAGCTCGTTGTAAACGCCAGGGTTTTCCAAAGCAATAAAGCC} \\ \hline
core\_octa-65 &	\seqsplit{AATTATTGTTTTCATGCCTTTAGCGTCAGATAGCACGGAAAC} \\ \hline
core\_octa-66 &	\seqsplit{AAGTTTCAGACAGCCGGGATCGTCACCCTTCTGTAGCTCAAC} \\ \hline
core\_octa-67 &	\seqsplit{ACAAAGAAATTTAGGTAGGGCTTAATTGTATACAACGGAATC} \\ \hline
core\_octa-68 &	\seqsplit{AACAAAAATAACTAGGTCTGAGAGACTACGCTGAGTTTCCCT} \\ \hline
core\_octa-69 &	\seqsplit{CATAACCTAAATCAACAGTTCAGAAAACGTCATAAGGATAGC} \\ \hline
core\_octa-70 &	\seqsplit{CACGACGAATTCGTGTGGCATCAATTCTTTAGCAAAATTACG} \\ \hline
core\_octa-71 &	\seqsplit{CCTACCAACAGTAATTTTATCCTGAATCAAACAGCCATATGA} \\ \hline
core\_octa-72 &	\seqsplit{GATTATAAAGAAACGCCAGTTACAAAATTTACCAACGTCAGA} \\ \hline
core\_octa-73 &	\seqsplit{AGTAGATTGAAAAGAATCATGGTCATAGCCGGAAGCATAAGT} \\ \hline
core\_octa-74 &	\seqsplit{TAGAATCCATAAATCATTTAACAATTTCTCCCGGCTTAGGTT} \\ \hline
core\_octa-75 &	\seqsplit{AAAGGCCAAATATGTTAGAGCTTAATTGATTGCTCCATGAGG} \\ \hline
core\_octa-76 &	\seqsplit{CCAAAAGGAAAGGACAACAGTTTCAGCGAATCATCATATTCC} \\ \hline
core\_octa-77 &	\seqsplit{GAAATCGATAACCGGATACCGATAGTTGTATCAGCTCCAACG} \\ \hline
core\_octa-78 &	\seqsplit{TGAATATTATCAAAATAATGGAAGGGTTAATATTTATCCCAA} \\ \hline
core\_octa-79 &	\seqsplit{GAGGAAGCAGGATTCGGGTAAAATACGTAAAACACCCCCCAG} \\ \hline
core\_octa-80 &	\seqsplit{GGTTGATTTTCCAGCAGACAGCCCTCATTCGTCACGGGATAG} \\ \hline
core\_octa-81 &	\seqsplit{CAAGCCCCCACCCTTAGCCCGGAATAGGACGATCTAAAGTTT} \\ \hline
core\_octa-82 &	\seqsplit{TGTAGATATTACGCGGCGATCGGTGCGGGCGCCATCTTCTGG} \\ \hline
core\_octa-83 &	\seqsplit{CATCCTATTCAGCTAAAAGGTAAAGTAAAAAGCAAGCCGTTT} \\ \hline
core\_octa-84 &	\seqsplit{CAGCTCATATAAGCGTACCCCGGTTGATGTGTCGGATTCTCC} \\ \hline
core\_octa-85 &	\seqsplit{CATGTCACAAACGGCATTAAATGTGAGCAATTCGCGTTAAAT} \\ \hline
core\_octa-86 &	\seqsplit{AGCGTCACGTATAAGAATTGAGTTAAGCCCTTTTTAAGAAAG} \\ \hline
core\_octa-87 &	\seqsplit{TATAAAGCATCGTAACCAAGTACCGCACCGGCTGTAATATCC} \\ \hline
core\_octa-88 &	\seqsplit{ATAGCCCGCGAAAATAATTGTATCGGTTCGCCGACAATGAGT} \\ \hline
core\_octa-89 &	\seqsplit{AGACAGTTCATATAGGAGAAGCCTTTATAACATTGCCTGAGA} \\ \hline
core\_octa-90 &	\seqsplit{AACAGGTCCCGAAATTGCATCAAAAAGATCTTTGATCATCAG} \\ \hline
core\_octa-91 &	\seqsplit{ACTGCCCTTGCCCCGTTGCAGCAAGCGGCAACAGCTTTTTCT} \\ \hline
core\_octa-92 &	\seqsplit{TCAAAGGGAGATAGCCCTTATAAATCAAGACAACAACCATCG} \\ \hline
core\_octa-93 &	\seqsplit{GTAATACGCAAACATGAGAGATCTACAACTAGCTGAGGCCGG} \\ \hline
core\_octa-94 &	\seqsplit{GAGATAACATTAGAAGAATAACATAAAAAGGAAGGATTAGGA} \\ \hline
core\_octa-95 &	\seqsplit{CAGATATTACCTGAATACCAAGTTACAATCGGGAGCTATTTT} \\ \hline
core\_octa-96 &	\seqsplit{CATATAACTAATGAACACAACATACGAGCTGTTTCTTTGGGG} \\ \hline
core\_octa-97 &	\seqsplit{ATGTTTTGCTTTTGATCGGAACGAGGGTACTTTTTCTTTTGATAAGAGGTCATT} \\ \hline
core\_octa-98 &	\seqsplit{CTTCGCTGGGCGCAGACGACAGTATCGGGGCACCGTCGCCATTCAGGCTGCGCA} \\ \hline
core\_octa-99 &	\seqsplit{GATATTCTAAATTGAGCCGGAACGAGGCCCAACTTGGCGCATAGGCTGGCTGAC} \\ \hline
core\_octa-100 &	\seqsplit{TGTCGTCATAAGTACAGAACCGCCACCCATTTTCACAGTACAAACTACAACGCC} \\ \hline
core\_octa-101 &	\seqsplit{CGATTATAAGCGGAGACTTCAAATATCGCGGAAGCCTACGAAGGCACCAACCTA} \\ \hline
core\_octa-102 &	\seqsplit{AACATGTACGCGAGTGGTTTGAAATACCTAAACACATTCTTACCAGTATAAAGC} \\ \hline
core\_octa-103 &	\seqsplit{GTCTGGATTTTGCGTTTTAAATGCAATGGTGAGAAATAAATTAATGCCGGAGAG} \\ \hline
core\_octa-104 &	\seqsplit{GCCTTGAATCTTTTCCGGAACCGCCTCCCAGAGCCCAGAGCCGCCGCCAGCATT} \\ \hline
core\_octa-105 &	\seqsplit{CGCTGGTGCTTTCCTGAATCGGCCAACGAGGGTGGTGATTGCCCTTCACCGCCT} \\ \hline
core\_octa-106 &	\seqsplit{ACATAACTTGCCCTAACTTTAATCATTGCATTATAACAACATTATTACAGGTAG} \\ \hline
core\_octa-107 &	\seqsplit{TTATTTTTACCGACAATGCAGAACGCGCGAAAAATCTTTCCTTATCATTCCAAG} \\ \hline
core\_octa-108 &	\seqsplit{TTTCAATAGAAGGCAGCGAACCTCCCGATTAGTTGAAACAATAACGGATTCGCC} \\ \hline
core\_octa-109 &	\seqsplit{GGGCGACCCCAAAAGTATGTTAGCAAACTAAAAGAGTCACAATCAATAGAAAAT} \\ \hline
core\_octa-110 &	\seqsplit{ATGACCACTCGTTTGGCTTTTGCAAAAGTTAGACTATATTCATTGAATCCCCCT} \\ \hline
core\_octa-111 &	\seqsplit{TCCAAATCTTCTGAATTATTTGCACGTAGGTTTAACGCTAACGAGCGTCTTTCC} \\ \hline
core\_octa-112 &	\seqsplit{GGGTTATTTAATTACAATATATGTGAGTAATTAATAAGAGTCAATAGTGAATTT} \\ \hline

\caption{Staple sequence for the octahedral DNA origami }

\label{tab:Octstaple}
\end{longtable}
\end{spacing}

\clearpage

\subsubsection{Functionalization sequence}

\begin{spacing}{1.35}\centering
\begin{longtable}{@{} |p{.18\textwidth}|  p{.82\textwidth} | @{}}

\hline
\endhead
\hline
\endfoot
\endlastfoot

Oct\_AuNPhandles\_1 &	\seqsplit{ATCCATCACTTCATACTCTACGTTGTTGTTGTTGTTGTTGGGGTGCCAGTTGAGACCATTAGATACAATTTTCACTGTGTGAAATTGTTATCC} \\ \hline
Oct\_AuNPhandles\_2 &	\seqsplit{ATCCATCACTTCATACTCTACGTTGTTGTTGTTGTTGTTTCAGAGCTGGGTAAACGACGGCCAGTGCGATCCCCGTAGTAGCATTAACATCCA} \\ \hline
Oct\_AuNPhandles\_3 &	\seqsplit{ATCCATCACTTCATACTCTACGTTGTTGTTGTTGTTGTTTTAGCGGTACAGAGCGGGAGAATTAACTGCGCTAATTTCGGAACCTATTATTCT} \\ \hline
Oct\_AuNPhandles\_4 &	\seqsplit{ATCCATCACTTCATACTCTACGTTGTTGTTGTTGTTGTTTGATTATCAACTTTACAACTAAAGGAATCCAAAAAGTTTGAGTAACATTATCAT} \\ \hline
Oct\_AuNPhandles\_5 &	\seqsplit{ATCCATCACTTCATACTCTACGTTGTTGTTGTTGTTGTTGTAGCGCCATTAAATTGGGAATTAGAGCGCAAGGCGCACCGTAATCAGTAGCGA} \\ \hline
Oct\_AuNPhandles\_6 &	\seqsplit{ATCCATCACTTCATACTCTACGTTGTTGTTGTTGTTGTTAGCCGAAAGTCTCTCTTTTGATGATACAAGTGCCTTAAGAGCAAGAAACAATGA} \\ \hline
Oct\_AuNPhandles\_7 &	\seqsplit{ATCCATCACTTCATACTCTACGTTGTTGTTGTTGTTGTTGTGGGAAATCATATAAATATTTAAATTGAATTTTTGTCTGGCCTTCCTGTAGCC} \\ \hline
Oct\_AuNPhandles\_8 &	\seqsplit{ATCCATCACTTCATACTCTACGTTGTTGTTGTTGTTGTTCCCACGCGCAAAATGGTTGAGTGTTGTTCGTGGACTTGCTTTCGAGGTGAATTT} \\ \hline
Oct\_AuNPstaples\_1 &	\seqsplit{GGGGTGCCAGTTGAGACCATTAGATACAATTTTCACTGTGTGAAATTGTTATCC} \\ \hline
Oct\_AuNPstaples\_2 &	\seqsplit{TCAGAGCTGGGTAAACGACGGCCAGTGCGATCCCCGTAGTAGCATTAACATCCA} \\ \hline
Oct\_AuNPstaples\_3 &	\seqsplit{TTAGCGGTACAGAGCGGGAGAATTAACTGCGCTAATTTCGGAACCTATTATTCT} \\ \hline
Oct\_AuNPstaples\_4 &	\seqsplit{TGATTATCAACTTTACAACTAAAGGAATCCAAAAAGTTTGAGTAACATTATCAT} \\ \hline
Oct\_AuNPstaples\_5 &	\seqsplit{GTAGCGCCATTAAATTGGGAATTAGAGCGCAAGGCGCACCGTAATCAGTAGCGA} \\ \hline
Oct\_AuNPstaples\_6 &	\seqsplit{AGCCGAAAGTCTCTCTTTTGATGATACAAGTGCCTTAAGAGCAAGAAACAATGA} \\ \hline
Oct\_AuNPstaples\_7 &	\seqsplit{GTGGGAAATCATATAAATATTTAAATTGAATTTTTGTCTGGCCTTCCTGTAGCC} \\ \hline
Oct\_AuNPstaples\_8 &	\seqsplit{CCCACGCGCAAAATGGTTGAGTGTTGTTCGTGGACTTGCTTTCGAGGTGAATTT} \\ \hline
AuNP\_conjugation &
TATGAAGTGATGGATGAT-SH \\ \hline

\caption{Functionalization sequence for the octahedral DNA origami. The one named with "AuNPhandle" is used to capture DNA functionalized AuNP while the one named with "AnNPstaple" is just the staple without the handle for doing so. The one named as AuNP\_conjugation is used to be conjugated to AuNPs.} 

\label{tab:OctAunp}
\end{longtable}
\end{spacing}

\clearpage

\subsubsection{Handle sequence }
\begin{spacing}{1.35}\centering
\begin{longtable}{@{} |p{.18\textwidth}|  p{.82\textwidth} | @{}} 
\hline
\endhead
\hline
\endfoot
\endlastfoot

Handle\_A0\_0 &	\seqsplit{GCTCACAATTCCGTGAGCTAACTCACTGGAAGTAATGGTCAATTTTTTTTTTTTTTTTTTTTTTCCCTGCTCC} \\ \hline
Handle\_A0\_1 &	\seqsplit{GGCCCTGAGAGAAGCAGGCGAAAATCATTGCGTAGAGGCGGTTTTTTTTTTTTTTTTTTTTTTTCCCTGCTCC} \\ \hline
Handle\_A0\_2 &	\seqsplit{CTTAAACAGCTTATATATTCGGTCGCTTGATGGGGAACAAGATTTTTTTTTTTTTTTTTTTTTTCCCTGCTCC} \\ \hline
Handle\_A0\_3 &	\seqsplit{TTTGCGGATGGCCAACTAAAGTACGGGCTTGCAGCTACAGAGTTTTTTTTTTTTTTTTTTTTTTCCCTGCTCC} \\ \hline
Handle\_A1\_0 &	\seqsplit{ATCAAAATCATATATGTAAATGCTGAACAAACACTTGCTTCTTTTTTTTTTTTTTTTTTTTTTTGACATGCGC} \\ \hline
Handle\_A1\_1 &	\seqsplit{CAACGCTCAACAGCAGAGGCATTTTCAATCCAATGATAAATATTTTTTTTTTTTTTTTTTTTTTGACATGCGC} \\ \hline
Handle\_A1\_2 &	\seqsplit{AACGGGTATTAAGGAATCATTACCGCCAGTAATTCAACAATATTTTTTTTTTTTTTTTTTTTTTGACATGCGC} \\ \hline
Handle\_A1\_3 &	\seqsplit{TGATTGCTTTGAGCAAAAGAAGATGAAATAGCAGAGGTTTTGTTTTTTTTTTTTTTTTTTTTTTGACATGCGC} \\ \hline
Handle\_A2\_0 &	\seqsplit{ACTGTTGGGAAGCAGCTGGCGAAAGGATAGGTCAAGATCGCATTTTTTTTTTTTTTTTTTTTTTCGCCAAGTC} \\ \hline
Handle\_A2\_1 &	\seqsplit{ATAAATCATACATAAATCGGTTGTACTGTGCTGGCATGCCTGTTTTTTTTTTTTTTTTTTTTTTCGCCAAGTC} \\ \hline
Handle\_A2\_2 &	\seqsplit{GGTAGCTATTTTAGAGAATCGATGAAAACATTAAATGTGTAGTTTTTTTTTTTTTTTTTTTTTTCGCCAAGTC} \\ \hline
Handle\_A2\_3 &	\seqsplit{AGCTTTCATCAACGGATTGACCGTAAAATCGTATAATATTTTTTTTTTTTTTTTTTTTTTTTTTCGCCAAGTC} \\ \hline
Handle\_A3\_0 &	\seqsplit{AATAGCAATAGCACCAGAAGGAAACCTAAAGCCACTGGTAATTTTTTTTTTTTTTTTTTTTTTTGCGTGTCAC} \\ \hline
Handle\_A3\_1 &	\seqsplit{GACAGGAGGTTGAAACAAATAAATCCGCCCCCTCCGCCACCCTTTTTTTTTTTTTTTTTTTTTTGCGTGTCAC} \\ \hline
Handle\_A3\_2 &	\seqsplit{CAGAATCAAGTTTCGGCATTTTCGGTTAAATATATCACCAGTTTTTTTTTTTTTTTTTTTTTTTGCGTGTCAC} \\ \hline
Handle\_A3\_3 &	\seqsplit{TCATATGGTTTACGATTGAGGGAGGGAAACGCAATACATACATTTTTTTTTTTTTTTTTTTTTTGCGTGTCAC} \\ \hline
Handle\_A4\_0 &	\seqsplit{CAAATGCTTTAAAAAATCAGGTCTTTAAGAGCAGCCAGAGGGTTTTTTTTTTTTTTTTTTTTTTGCGGCATAC} \\ \hline
Handle\_A4\_1 &	\seqsplit{AAACGAAAGAGGGCGAAACAAAGTACTGACTATATTCGAGCTTTTTTTTTTTTTTTTTTTTTTTGCGGCATAC} \\ \hline
Handle\_A4\_2 &	\seqsplit{CTTCATCAAGAGAAATCAACGTAACAGAGATTTGTCAATCATTTTTTTTTTTTTTTTTTTTTTTGCGGCATAC} \\ \hline
Handle\_A4\_3 &	\seqsplit{AAAGATTCATCAGGAATTACGAGGCATGCTCATCCTTATGCGTTTTTTTTTTTTTTTTTTTTTTGCGGCATAC} \\ \hline
Handle\_A5\_0 &	\seqsplit{AGAGCCTAATTTGATTTTTTGTTTAAATCCTGAAATAAAGAATTTTTTTTTTTTTTTTTTTTTTGGCTGCAAC} \\ \hline
Handle\_A5\_1 &	\seqsplit{TTTGCGGAACAATGGCAATTCATCAATCTGTATAATAATTTTTTTTTTTTTTTTTTTTTTTTTTGGCTGCAAC} \\ \hline
Handle\_A5\_2 &	\seqsplit{GAAACATGAAAGCTCAGTACCAGGCGAAAAATGCTGAACAAATTTTTTTTTTTTTTTTTTTTTTGGCTGCAAC} \\ \hline
Handle\_A5\_3 &	\seqsplit{TGTAGCATTCCAACGTTAGTAAATGAAGTGCCGCGCCACCCTTTTTTTTTTTTTTTTTTTTTTTGGCTGCAAC} \\ \hline
Handle\_B0\_0 &	\seqsplit{GCTCACAATTCCGTGAGCTAACTCACTGGAAGTAATGGTCAATTTTTTTTTTTTTTTTTTTTTTCGCTGTACC} \\ \hline
Handle\_B0\_1 &	\seqsplit{GGCCCTGAGAGAAGCAGGCGAAAATCATTGCGTAGAGGCGGTTTTTTTTTTTTTTTTTTTTTTTCGCTGTACC} \\ \hline
Handle\_B0\_2 &	\seqsplit{CTTAAACAGCTTATATATTCGGTCGCTTGATGGGGAACAAGATTTTTTTTTTTTTTTTTTTTTTCGCTGTACC} \\ \hline
Handle\_B0\_3 &	\seqsplit{TTTGCGGATGGCCAACTAAAGTACGGGCTTGCAGCTACAGAGTTTTTTTTTTTTTTTTTTTTTTCGCTGTACC} \\ \hline
Handle\_B1\_0 &	\seqsplit{ATCAAAATCATATATGTAAATGCTGAACAAACACTTGCTTCTTTTTTTTTTTTTTTTTTTTTTTGCGCATGTC} \\ \hline
Handle\_B1\_1 &	\seqsplit{CAACGCTCAACAGCAGAGGCATTTTCAATCCAATGATAAATATTTTTTTTTTTTTTTTTTTTTTGCGCATGTC} \\ \hline
Handle\_B1\_2 &	\seqsplit{AACGGGTATTAAGGAATCATTACCGCCAGTAATTCAACAATATTTTTTTTTTTTTTTTTTTTTTGCGCATGTC} \\ \hline
Handle\_B1\_3 &	\seqsplit{TGATTGCTTTGAGCAAAAGAAGATGAAATAGCAGAGGTTTTGTTTTTTTTTTTTTTTTTTTTTTGCGCATGTC} \\ \hline
Handle\_B2\_0 &	\seqsplit{ACTGTTGGGAAGCAGCTGGCGAAAGGATAGGTCAAGATCGCATTTTTTTTTTTTTTTTTTTTTTCGCTCGAAC} \\ \hline
Handle\_B2\_1 &	\seqsplit{ATAAATCATACATAAATCGGTTGTACTGTGCTGGCATGCCTGTTTTTTTTTTTTTTTTTTTTTTCGCTCGAAC} \\ \hline
Handle\_B2\_2 &	\seqsplit{GGTAGCTATTTTAGAGAATCGATGAAAACATTAAATGTGTAGTTTTTTTTTTTTTTTTTTTTTTCGCTCGAAC} \\ \hline
Handle\_B2\_3 &	\seqsplit{AGCTTTCATCAACGGATTGACCGTAAAATCGTATAATATTTTTTTTTTTTTTTTTTTTTTTTTTCGCTCGAAC} \\ \hline
Handle\_B3\_0 &	\seqsplit{AATAGCAATAGCACCAGAAGGAAACCTAAAGCCACTGGTAATTTTTTTTTTTTTTTTTTTTTTTGCTGACGAC} \\ \hline
Handle\_B3\_1 &	\seqsplit{GACAGGAGGTTGAAACAAATAAATCCGCCCCCTCCGCCACCCTTTTTTTTTTTTTTTTTTTTTTGCTGACGAC} \\ \hline
Handle\_B3\_2 &	\seqsplit{CAGAATCAAGTTTCGGCATTTTCGGTTAAATATATCACCAGTTTTTTTTTTTTTTTTTTTTTTTGCTGACGAC} \\ \hline
Handle\_B3\_3 &	\seqsplit{TCATATGGTTTACGATTGAGGGAGGGAAACGCAATACATACATTTTTTTTTTTTTTTTTTTTTTGCTGACGAC} \\ \hline
Handle\_B4\_0 &	\seqsplit{CAAATGCTTTAAAAAATCAGGTCTTTAAGAGCAGCCAGAGGGTTTTTTTTTTTTTTTTTTTTTTGTATGCCGC} \\ \hline
Handle\_B4\_1 &	\seqsplit{AAACGAAAGAGGGCGAAACAAAGTACTGACTATATTCGAGCTTTTTTTTTTTTTTTTTTTTTTTGTATGCCGC} \\ \hline
Handle\_B4\_2 &	\seqsplit{CTTCATCAAGAGAAATCAACGTAACAGAGATTTGTCAATCATTTTTTTTTTTTTTTTTTTTTTTGTATGCCGC} \\ \hline
Handle\_B4\_3 &	\seqsplit{AAAGATTCATCAGGAATTACGAGGCATGCTCATCCTTATGCGTTTTTTTTTTTTTTTTTTTTTTGTATGCCGC} \\ \hline
Handle\_B5\_0 &	\seqsplit{AGAGCCTAATTTGATTTTTTGTTTAAATCCTGAAATAAAGAATTTTTTTTTTTTTTTTTTTTTTCCAGAGCCC} \\ \hline
Handle\_B5\_1 &	\seqsplit{TTTGCGGAACAATGGCAATTCATCAATCTGTATAATAATTTTTTTTTTTTTTTTTTTTTTTTTTCCAGAGCCC} \\ \hline
Handle\_B5\_2 &	\seqsplit{GAAACATGAAAGCTCAGTACCAGGCGAAAAATGCTGAACAAATTTTTTTTTTTTTTTTTTTTTTCCAGAGCCC} \\ \hline
Handle\_B5\_3 &	\seqsplit{TGTAGCATTCCAACGTTAGTAAATGAAGTGCCGCGCCACCCTTTTTTTTTTTTTTTTTTTTTTTCCAGAGCCC} \\ \hline
Handle\_C0\_0 &	\seqsplit{GCTCACAATTCCGTGAGCTAACTCACTGGAAGTAATGGTCAATTTTTTTTTTTTTTTTTTTTTTAGCGAACCC} \\ \hline
Handle\_C0\_1 &	\seqsplit{GGCCCTGAGAGAAGCAGGCGAAAATCATTGCGTAGAGGCGGTTTTTTTTTTTTTTTTTTTTTTTAGCGAACCC} \\ \hline
Handle\_C0\_2 &	\seqsplit{CTTAAACAGCTTATATATTCGGTCGCTTGATGGGGAACAAGATTTTTTTTTTTTTTTTTTTTTTAGCGAACCC} \\ \hline
Handle\_C0\_3 &	\seqsplit{TTTGCGGATGGCCAACTAAAGTACGGGCTTGCAGCTACAGAGTTTTTTTTTTTTTTTTTTTTTTAGCGAACCC} \\ \hline
Handle\_C1\_0 &	\seqsplit{ATCAAAATCATATATGTAAATGCTGAACAAACACTTGCTTCTTTTTTTTTTTTTTTTTTTTTTTGTTCGAGCG} \\ \hline
Handle\_C1\_1 &	\seqsplit{CAACGCTCAACAGCAGAGGCATTTTCAATCCAATGATAAATATTTTTTTTTTTTTTTTTTTTTTGTTCGAGCG} \\ \hline
Handle\_C1\_2 &	\seqsplit{AACGGGTATTAAGGAATCATTACCGCCAGTAATTCAACAATATTTTTTTTTTTTTTTTTTTTTTGTTCGAGCG} \\ \hline
Handle\_C1\_3 &	\seqsplit{TGATTGCTTTGAGCAAAAGAAGATGAAATAGCAGAGGTTTTGTTTTTTTTTTTTTTTTTTTTTTGTTCGAGCG} \\ \hline
Handle\_C2\_0 &	\seqsplit{ACTGTTGGGAAGCAGCTGGCGAAAGGATAGGTCAAGATCGCATTTTTTTTTTTTTTTTTTTTTTGGAGCAGGG} \\ \hline
Handle\_C2\_1 &	\seqsplit{ATAAATCATACATAAATCGGTTGTACTGTGCTGGCATGCCTGTTTTTTTTTTTTTTTTTTTTTTGGAGCAGGG} \\ \hline
Handle\_C2\_2 &	\seqsplit{GGTAGCTATTTTAGAGAATCGATGAAAACATTAAATGTGTAGTTTTTTTTTTTTTTTTTTTTTTGGAGCAGGG} \\ \hline
Handle\_C2\_3 &	\seqsplit{AGCTTTCATCAACGGATTGACCGTAAAATCGTATAATATTTTTTTTTTTTTTTTTTTTTTTTTTGGAGCAGGG} \\ \hline
Handle\_C3\_0 &	\seqsplit{AATAGCAATAGCACCAGAAGGAAACCTAAAGCCACTGGTAATTTTTTTTTTTTTTTTTTTTTTTGCCTCTCCC} \\ \hline
Handle\_C3\_1 &	\seqsplit{GACAGGAGGTTGAAACAAATAAATCCGCCCCCTCCGCCACCCTTTTTTTTTTTTTTTTTTTTTTGCCTCTCCC} \\ \hline
Handle\_C3\_2 &	\seqsplit{CAGAATCAAGTTTCGGCATTTTCGGTTAAATATATCACCAGTTTTTTTTTTTTTTTTTTTTTTTGCCTCTCCC} \\ \hline
Handle\_C3\_3 &	\seqsplit{TCATATGGTTTACGATTGAGGGAGGGAAACGCAATACATACATTTTTTTTTTTTTTTTTTTTTTGCCTCTCCC} \\ \hline
Handle\_C4\_0 &	\seqsplit{CAAATGCTTTAAAAAATCAGGTCTTTAAGAGCAGCCAGAGGGTTTTTTTTTTTTTTTTTTTTTTGGGCTCTGG} \\ \hline
Handle\_C4\_1 &	\seqsplit{AAACGAAAGAGGGCGAAACAAAGTACTGACTATATTCGAGCTTTTTTTTTTTTTTTTTTTTTTTGGGCTCTGG} \\ \hline
Handle\_C4\_2 &	\seqsplit{CTTCATCAAGAGAAATCAACGTAACAGAGATTTGTCAATCATTTTTTTTTTTTTTTTTTTTTTTGGGCTCTGG} \\ \hline
Handle\_C4\_3 &	\seqsplit{AAAGATTCATCAGGAATTACGAGGCATGCTCATCCTTATGCGTTTTTTTTTTTTTTTTTTTTTTGGGCTCTGG} \\ \hline
Handle\_C5\_0 &	\seqsplit{AGAGCCTAATTTGATTTTTTGTTTAAATCCTGAAATAAAGAATTTTTTTTTTTTTTTTTTTTTTGTGACACGC} \\ \hline
Handle\_C5\_1 &	\seqsplit{TTTGCGGAACAATGGCAATTCATCAATCTGTATAATAATTTTTTTTTTTTTTTTTTTTTTTTTTGTGACACGC} \\ \hline
Handle\_C5\_2 &	\seqsplit{GAAACATGAAAGCTCAGTACCAGGCGAAAAATGCTGAACAAATTTTTTTTTTTTTTTTTTTTTTGTGACACGC} \\ \hline
Handle\_C5\_3 &	\seqsplit{TGTAGCATTCCAACGTTAGTAAATGAAGTGCCGCGCCACCCTTTTTTTTTTTTTTTTTTTTTTTGTGACACGC} \\ \hline
Handle\_D0\_0 &	\seqsplit{GCTCACAATTCCGTGAGCTAACTCACTGGAAGTAATGGTCAATTTTTTTTTTTTTTTTTTTTTTGGTACAGCG} \\ \hline
Handle\_D0\_1 &	\seqsplit{GGCCCTGAGAGAAGCAGGCGAAAATCATTGCGTAGAGGCGGTTTTTTTTTTTTTTTTTTTTTTTGGTACAGCG} \\ \hline
Handle\_D0\_2 &	\seqsplit{CTTAAACAGCTTATATATTCGGTCGCTTGATGGGGAACAAGATTTTTTTTTTTTTTTTTTTTTTGGTACAGCG} \\ \hline
Handle\_D0\_3 &	\seqsplit{TTTGCGGATGGCCAACTAAAGTACGGGCTTGCAGCTACAGAGTTTTTTTTTTTTTTTTTTTTTTGGTACAGCG} \\ \hline
Handle\_D1\_0 &	\seqsplit{ATCAAAATCATATATGTAAATGCTGAACAAACACTTGCTTCTTTTTTTTTTTTTTTTTTTTTTTGGGTTCGCT} \\ \hline
Handle\_D1\_1 &	\seqsplit{CAACGCTCAACAGCAGAGGCATTTTCAATCCAATGATAAATATTTTTTTTTTTTTTTTTTTTTTGGGTTCGCT} \\ \hline
Handle\_D1\_2 &	\seqsplit{AACGGGTATTAAGGAATCATTACCGCCAGTAATTCAACAATATTTTTTTTTTTTTTTTTTTTTTGGGTTCGCT} \\ \hline
Handle\_D1\_3 &	\seqsplit{TGATTGCTTTGAGCAAAAGAAGATGAAATAGCAGAGGTTTTGTTTTTTTTTTTTTTTTTTTTTTGGGTTCGCT} \\ \hline
Handle\_D2\_0 &	\seqsplit{ACTGTTGGGAAGCAGCTGGCGAAAGGATAGGTCAAGATCGCATTTTTTTTTTTTTTTTTTTTTTGACTTGGCG} \\ \hline
Handle\_D2\_1 &	\seqsplit{ATAAATCATACATAAATCGGTTGTACTGTGCTGGCATGCCTGTTTTTTTTTTTTTTTTTTTTTTGACTTGGCG} \\ \hline
Handle\_D2\_2 &	\seqsplit{GGTAGCTATTTTAGAGAATCGATGAAAACATTAAATGTGTAGTTTTTTTTTTTTTTTTTTTTTTGACTTGGCG} \\ \hline
Handle\_D2\_3 &	\seqsplit{AGCTTTCATCAACGGATTGACCGTAAAATCGTATAATATTTTTTTTTTTTTTTTTTTTTTTTTTGACTTGGCG} \\ \hline
Handle\_D3\_0 &	\seqsplit{AATAGCAATAGCACCAGAAGGAAACCTAAAGCCACTGGTAATTTTTTTTTTTTTTTTTTTTTTTGTCGTCAGC} \\ \hline
Handle\_D3\_1 &	\seqsplit{GACAGGAGGTTGAAACAAATAAATCCGCCCCCTCCGCCACCCTTTTTTTTTTTTTTTTTTTTTTGTCGTCAGC} \\ \hline
Handle\_D3\_2 &	\seqsplit{CAGAATCAAGTTTCGGCATTTTCGGTTAAATATATCACCAGTTTTTTTTTTTTTTTTTTTTTTTGTCGTCAGC} \\ \hline
Handle\_D3\_3 &	\seqsplit{TCATATGGTTTACGATTGAGGGAGGGAAACGCAATACATACATTTTTTTTTTTTTTTTTTTTTTGTCGTCAGC} \\ \hline
Handle\_D4\_0 &	\seqsplit{CAAATGCTTTAAAAAATCAGGTCTTTAAGAGCAGCCAGAGGGTTTTTTTTTTTTTTTTTTTTTTGGGAGAGGC} \\ \hline
Handle\_D4\_1 &	\seqsplit{AAACGAAAGAGGGCGAAACAAAGTACTGACTATATTCGAGCTTTTTTTTTTTTTTTTTTTTTTTGGGAGAGGC} \\ \hline
Handle\_D4\_2 &	\seqsplit{CTTCATCAAGAGAAATCAACGTAACAGAGATTTGTCAATCATTTTTTTTTTTTTTTTTTTTTTTGGGAGAGGC} \\ \hline
Handle\_D4\_3 &	\seqsplit{AAAGATTCATCAGGAATTACGAGGCATGCTCATCCTTATGCGTTTTTTTTTTTTTTTTTTTTTTGGGAGAGGC} \\ \hline
Handle\_D5\_0 &	\seqsplit{AGAGCCTAATTTGATTTTTTGTTTAAATCCTGAAATAAAGAATTTTTTTTTTTTTTTTTTTTTTGTTGCAGCC} \\ \hline
Handle\_D5\_1 &	\seqsplit{TTTGCGGAACAATGGCAATTCATCAATCTGTATAATAATTTTTTTTTTTTTTTTTTTTTTTTTTGTTGCAGCC} \\ \hline
Handle\_D5\_2 &	\seqsplit{GAAACATGAAAGCTCAGTACCAGGCGAAAAATGCTGAACAAATTTTTTTTTTTTTTTTTTTTTTGTTGCAGCC} \\ \hline
Handle\_D5\_3 &	\seqsplit{TGTAGCATTCCAACGTTAGTAAATGAAGTGCCGCGCCACCCTTTTTTTTTTTTTTTTTTTTTTTGTTGCAGCC} \\ \hline

\caption{Handle sequence for the octahedral DNA origami} 
\label{tab:Octhandle2}
\end{longtable}
\end{spacing}
\clearpage

\bibliographystyle{old-mujstyl}

\bibliography{refsup,biblio}
